\documentclass[12pt,preprint]{aastex}
\usepackage{natbib}
\newcommand{\be}{\begin{equation}}
\newcommand{\ee}{\end{equation}}

\shorttitle{Fermi LAT Third Catalog}
\shortauthors{Ackermann et al.}

\begin{document}

 

\title{Fermi Large Area Telescope Third Source Catalog}

\slugcomment{Accepted for Publication in ApJS}

\author{
F.~Acero\altaffilmark{1}, 
M.~Ackermann\altaffilmark{2}, 
M.~Ajello\altaffilmark{3}, 
A.~Albert\altaffilmark{4}, 
W.~B.~Atwood\altaffilmark{5}, 
M.~Axelsson\altaffilmark{6,7}, 
L.~Baldini\altaffilmark{8,4}, 
J.~Ballet\altaffilmark{1,9}, 
G.~Barbiellini\altaffilmark{10,11}, 
D.~Bastieri\altaffilmark{12,13}
% ...
}
\altaffiltext{1}{Laboratoire AIM, CEA-IRFU/CNRS/Universit\'e Paris Diderot, Service d'Astrophysique, CEA Saclay, F-91191 Gif sur Yvette, France}
\altaffiltext{2}{Deutsches Elektronen Synchrotron DESY, D-15738 Zeuthen, Germany}
\altaffiltext{3}{Department of Physics and Astronomy, Clemson University, Kinard Lab of Physics, Clemson, SC 29634-0978, USA}
\altaffiltext{4}{W. W. Hansen Experimental Physics Laboratory, Kavli Institute for Particle Astrophysics and Cosmology, Department of Physics and SLAC National Accelerator Laboratory, Stanford University, Stanford, CA 94305, USA}
\altaffiltext{5}{Santa Cruz Institute for Particle Physics, Department of Physics and Department of Astronomy and Astrophysics, University of California at Santa Cruz, Santa Cruz, CA 95064, USA}
\altaffiltext{6}{Department of Physics, KTH Royal Institute of Technology, AlbaNova, SE-106 91 Stockholm, Sweden}
\altaffiltext{7}{Tokyo Metropolitan University, Department of Physics, Minami-osawa 1-1, Hachioji, Tokyo 192-0397, Japan}
\altaffiltext{8}{Universit\`a di Pisa and Istituto Nazionale di Fisica Nucleare, Sezione di Pisa I-56127 Pisa, Italy}
\altaffiltext{9}{email: jean.ballet@cea.fr}
\altaffiltext{10}{Istituto Nazionale di Fisica Nucleare, Sezione di Trieste, I-34127 Trieste, Italy}
\altaffiltext{11}{Dipartimento di Fisica, Universit\`a di Trieste, I-34127 Trieste, Italy}
\altaffiltext{12}{Istituto Nazionale di Fisica Nucleare, Sezione di Padova, I-35131 Padova, Italy}
\altaffiltext{13}{Dipartimento di Fisica e Astronomia ``G. Galilei'', Universit\`a di Padova, I-35131 Padova, Italy}
% ...

\begin{abstract}
We present the third {\it Fermi} Large Area Telescope source catalog (3FGL) of sources in the 100~MeV--300~GeV range. Based on the first four years of science data from the {\it Fermi Gamma-ray Space Telescope} mission, it is the deepest yet in this energy range.  Relative to the 2FGL catalog, the 3FGL catalog incorporates twice as much data as well as a number of analysis improvements, including improved calibrations at the event reconstruction level, an updated model for Galactic diffuse $\gamma$-ray emission, a refined procedure for source detection, and improved methods for associating LAT sources with potential counterparts at other wavelengths. The 3FGL catalog includes 3033 sources above $4\sigma$ significance, with source location regions, spectral properties, and monthly light curves for each.  Of these, 78 are flagged as potentially being due to imperfections in the model for Galactic diffuse emission.  Twenty-five sources are modeled explicitly as spatially extended, and overall 238 sources are considered as identified based on angular extent or correlated variability (periodic or otherwise) observed at other wavelengths.  For 1010 sources we have not found plausible counterparts at other wavelengths.  More than 1100 of the identified or associated sources are active galaxies of the blazar class; several other classes of non-blazar active galaxies are also represented in the 3FGL.  Pulsars represent the largest Galactic source class.  From source counts of Galactic sources we estimate the contribution of unresolved sources to the Galactic diffuse emission is $\sim$3\% at 1~GeV.
\end{abstract}


\section{Introduction}
\label{introduction}

This paper presents a catalog of high-energy $\gamma$-ray sources detected in the first four years of the {\it Fermi Gamma-ray Space Telescope} mission by the Large Area Telescope (LAT).  It is the successor to the LAT Bright Source List \citep[hereafter 0FGL,][]{LAT09_BSL}, the First {\it Fermi} LAT \citep[1FGL,][]{LAT10_1FGL} catalog, and the Second {\it Fermi} LAT \citep[2FGL,][]{LAT12_2FGL} catalog, which were based on 3 months, 11 months, and 2 years of flight data, respectively.  The 3FGL catalog both succeeds and complements the First {\it Fermi} LAT Catalog of Sources Above 10 GeV \citep[1FHL,][]{LAT13_1FHL}, which was based on 3 years of flight data but considered only sources detected above 10 GeV.  The new 3FGL catalog is the deepest yet in the 100~MeV--300~GeV energy range. The result of a dedicated effort for studying the Active Galactic Nuclei (AGN) population in the 3FGL catalog is published in an accompanying paper \citep[3LAC,][]{LAT14_3LAC}.

We have implemented a number of analysis refinements for the 3FGL catalog:


\begin{enumerate}
\item Pass 7 reprocessed data\footnote{See \url{http://fermi.gsfc.nasa.gov/ssc/data/analysis/documentation/Pass7REP\_usage.html}.} are now used (\S~\ref{LATData}).  The principal difference relative to the original Pass 7 data used for 2FGL is improved angular resolution above 3 GeV.  In addition, systematics of the instrument response functions (IRFs) are better characterized, and smaller.
\item This catalog employs a new model of the diffuse Galactic and isotropic emissions, developed for the 3FGL analysis (\S~\ref{DiffuseModel}).  The model has improved fidelity to the observations, especially for regions where the diffuse emission cannot be described using a spatial template derived from observations at other wavelengths.  In addition, the accuracy of the model is improved toward bright star-forming regions and at energies above 40~GeV generally.  The development of this model is described in a separate publication \citep{LAT14_DiffuseModel}.
\item We explicitly model twenty-five sources as extended emission regions (\S~\ref{catalog_extended}), up from twelve in 2FGL.  Each has an angular extent measured with LAT data.  Taking into account the finite sizes of the sources allows for more accurate flux and spectrum measurements for the extended sources as well as for nearby point sources.
\item We have further refined the method for characterizing and localizing source `seeds' evaluated for inclusion in the catalog (\S~\ref{catalog_detection}).  The improvements in this regard are most marked at low Galactic latitudes, where an iterative approach to finding seeds has improved the sensitivity of the catalog in the Galactic plane.
\item For studying the associations of LAT sources with counterparts at other wavelengths, we have updated several of the catalogs used for counterpart searches, and correspondingly recalibrated the association procedure.
\end{enumerate}

The exposure of the LAT is fairly uniform across the sky, but the brightness of the diffuse backgrounds, and hence the sensitivity for source detection, depends strongly on direction.  As for previous LAT source catalogs, for the 3FGL catalog sources are included based on the statistical significance of their detection considered over the entire time period of the analysis.  For this reason the 3FGL catalog is not a comprehensive catalog of transient $\gamma$-ray sources, however the catalog does include light curves on a monthly time scale for sources that meet the criteria for inclusion.
 
In \S~\ref{lat_and_background} we describe the LAT and the models for the diffuse  backgrounds, celestial and otherwise.  Section \ref{catalog_main} describes how the catalog is constructed, with emphasis on what has changed since the analysis for the 2FGL catalog.  The 3FGL catalog itself is presented in \S~\ref{3fgl_description}, along with a comparison to previous LAT catalogs. 
We discuss associations and identifications in \S~\ref{source_assoc_main} and Galactic source counts in \S~\ref{source_counts}.  The conclusions are presented in \S~\ref{conclusions}.  We provide appendices with technical details of the analysis and of the format of the electronic version of the catalog.

    \section{Instrument and Background}
\label{lat_and_background}

\subsection{The Large Area Telescope}
\label{LATDescription}

The LAT detects $\gamma$ rays in the energy range 20~MeV to more than 300~GeV, measuring their arrival times, energies, and directions.  The LAT is also an efficient detector of the intense background of charged particles from cosmic rays and trapped radiation at the orbit of the {\it Fermi} satellite.  Accounting for $\gamma$ rays lost in filtering charged particles from the data, the effective collecting area is $\sim$6500~cm$^2$ at 1~GeV (for the P7REP\_SOURCE\_V15 event selection used here; see below).  The live time 
is nearly 76\%, limited primarily by interruptions of data taking when $Fermi$ is passing through the South Atlantic Anomaly ($\sim$13\%) and readout dead-time fraction ($\sim$9\%).   The field of view of the LAT is 2.4~sr at 1~GeV.   The per-photon angular resolution (point-spread function, PSF, 68\% containment) is $\sim$5$\degr$ at 100~MeV, decreasing to $0\fdg8$ at 1~GeV (averaged over the acceptance of the LAT), varying with energy approximately as $E^{-0.8}$ and asymptoting at $\sim$$0\fdg2$ above 20~GeV. The tracking section of the LAT has 36 layers of silicon strip detectors interleaved with 16 layers of tungsten foil (12 thin layers, 0.03 radiation length, at the top or {\it Front} of the instrument, followed by 4 thick layers, 0.18 radiation length, in the {\it Back} section).  The silicon strips track charged particles, and the tungsten foils facilitate conversion of $\gamma$ rays to positron-electron pairs.  Beneath the tracker is a calorimeter composed of an 8-layer array of CsI crystals ($\sim$8.5 total radiation lengths) to determine the 
$\gamma$-ray energy.  A segmented charged-particle anticoincidence detector (plastic scintillators 
read out by photomultiplier tubes) around the tracker is used to reject charged-particle background events.  More information about the LAT is provided in  \citet{LAT09_instrument}, and the in-flight calibration of the LAT is described in \citet{LAT09__calib}, \citet{LAT_energyscale2012}, and \citet{LAT12__calib}.

\subsection{The LAT Data}
\label{LATData}

The data 
for the 3FGL catalog were taken during the period 2008 August 4 (15:43 UTC)--2012 July 31 (22:46 UTC) covering close to four years.
They are the public data available from the {\it Fermi} Science Support Center (FSSC).
Intervals around bright GRBs \citep[080916C, 090510, 090902B, 090926A, 110731A;][]{GRBCatalog} were excised. Solar flares became relatively frequent 
in 2011--12 (close to solar maximum) and were excised as well whenever they were bright enough to be detected over a month at the $3\sigma$ level. Solar flares last much longer than GRBs, so we were attentive not to reject too much time. Since the $\gamma$-ray emission is localized on the Sun, we kept intervals during which the Sun was at least $3^\circ$ below the Earth limb\footnote{This selection in FTOOLS notation is \texttt{ANGSEP(RA\_SUN,DEC\_SUN,RA\_ZENITH,DEC\_ZENITH) > 115}.} even during solar flares. The solar flares were detected over 3-hour intervals so the corresponding Good Time Intervals (GTI) are aligned to those 3-hour marks. Overall about two days were excised due to solar flares.
In order to reduce the contamination from the $\gamma$-ray-bright Earth limb (\S~\ref{EarthLimb}), times when the rocking angle of the spacecraft was larger than 52\degr, and events with zenith angles larger than 100\degr, were excised as well. The precise time intervals corresponding to selected events are recorded in the \texttt{GTI} extension of the 3FGL catalog (FITS version, App.~\ref{appendix_fits_format}).

\begin{figure*}[!ht]
\centering
\includegraphics[width=0.8\textwidth]{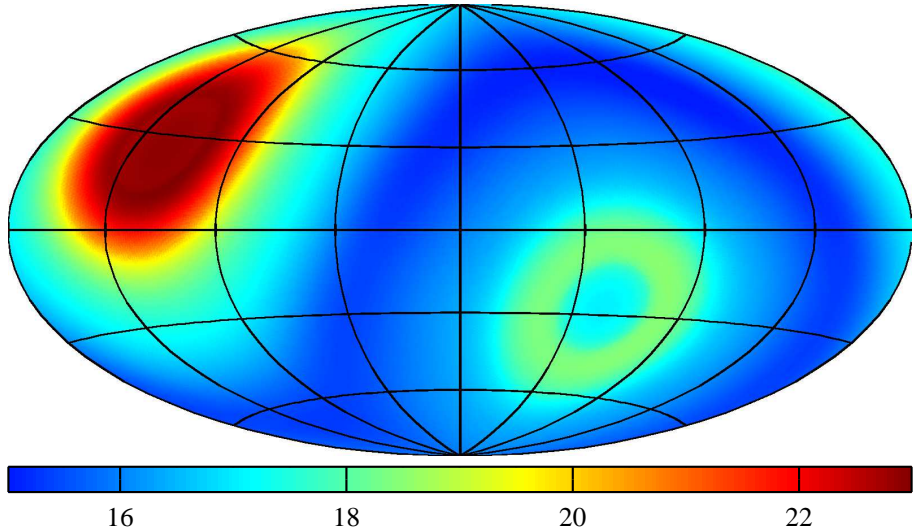}
\caption{The exposure at 1 GeV in Galactic coordinates and Hammer-Aitoff projection for the four-year period analyzed for the 3FGL catalog.  The units are equivalent on-axis observing time (in Ms).
}
\label{fig:exposure}
\end{figure*}

The rocking angle remained set at $50\degr$ after September 2009 (it was $35\degr$ for about 80\% of the time before that\footnote{See the LAT survey-mode history at \url{http://fermi.gsfc.nasa.gov/ssc/observations/types/allsky/}.}).  With the larger rocking angle the orbital plane is further off axis with the result that the survey is slightly non-uniform. The maximum exposure is reached at the North celestial pole. At 1~GeV it is 60\% larger than the minimum exposure, which is reached at the celestial equator (Figure~\ref{fig:exposure}).

In parallel with accumulating new data, developments on the instrument analysis side \citep{LAT13_P7repro} led to reprocessing all LAT data with new calibration constants, resulting in the Pass 7 reprocessed data that were used for 3FGL\footnote{Details about the performance of the LAT are available at \\ \url{http://www.slac.stanford.edu/exp/glast/groups/canda/lat\_Performance.htm}.}. The main advantage for the source catalog is that the reprocessing improved the PSF above 10 GeV by $\sim$30\%, improving the localization of hard sources (\S~\ref{catalog_detection}). We used the Source class event selection.

The lower bound of the energy range was left at 100~MeV, but the upper bound was raised to 300~GeV as in the 1FHL catalog. This is because as the source-to-background ratio decreases, the sensitivity curve \citep[Figure~18 of][1FGL]{LAT10_1FGL} shifts to higher energies.

    \subsection{Model for the Diffuse Gamma-Ray Background}
\label{DiffuseModel}

Models for the diffuse $\gamma$-ray backgrounds were updated for the 3FGL analysis, taking into account the new IRFs for Pass 7 reprocessed data and the improved statistics available with a 4-year data set, and also applying some refinements in the procedure for evaluating the models.  The primary components of the diffuse backgrounds are the diffuse $\gamma$-ray emission of the Milky Way and the approximately isotropic background consisting of emission from sub-threshold celestial sources plus residual charged particles misclassified as $\gamma$ rays.  In addition, we treat the `passive' emission of the Sun and Moon from cosmic-ray interactions with the solar atmosphere, solar radiation field, and the lunar lithosphere, as effectively a diffuse component, because the Sun and Moon move across the sky.  The residual Earth limb emission after the zenith angle selection (\S~\ref{LATData}) is also treated as effectively diffuse.  Each component of the background model for the 3FGL analysis is described in more detail below. 

\subsubsection{Diffuse emission of the Milky Way}

The diffuse $\gamma$-ray emission of the Milky Way originates in cosmic-ray interactions with interstellar gas and radiation.  As for 2FGL, for any given energy the model is primarily a linear combination of template maps derived from CO and H\,{\sc i} line survey data plus infrared maps of interstellar dust, which trace interstellar gas and in the model represent the $\gamma$-ray emission from pion decay and bremsstrahlung.  In addition we include in the model a template representing the intensity of emission from inverse Compton scattering of cosmic-ray electrons on the interstellar radiation field.  This component was calculated using the GALPROP code\footnote{See \url{http://galprop.stanford.edu}.} \citep{GALPROP1998}.  

For the 3FGL analysis we have made several improvements relative to 2FGL in modeling the diffuse emission.  The development of this new model is described by \citet{LAT14_DiffuseModel}\footnote{The model is available as \texttt{gll\_iem\_v05\_rev1.fit} from the FSSC.}. Here we briefly summarize the improvements.  For 3FGL
the representation of the gas traced uniquely by the infrared maps was improved in the vicinity of massive star-forming regions.  The overall model was fit to the LAT data iteratively, taking into account a preliminary version of the 3FGL source list, for the energy range 50 MeV--50 GeV.  Relative to the 2FGL model for Galactic diffuse emission, we have improved the extrapolation to lower and higher energies using the energy dependence of the $\gamma$-ray emissivity function.  We developed a new procedure to account for the structured celestial $\gamma$-ray emission that could not be fit using templates derived from observations at other wavelengths.  This residual component, also fit iteratively, was derived by deconvolving the residuals to take into account the effects of the PSF, filtering the result to reduce statistical fluctuations (removing structures on angular scales smaller than $\sim$2$\degr$).  The spectrum was modeled as inverse Compton emission from a population of cosmic-ray electrons with a spectral break at 965 MeV.  

\subsubsection{Isotropic background}

The isotropic diffuse background was derived from all-sky fits of the four-year data set using the Galactic diffuse emission model described above and a preliminary version of the 3FGL source list.  The diffuse background includes charged particles misclassified as $\gamma$ rays.  We implicitly assume that the acceptance for these residual charged particles is the same as for $\gamma$ rays in treating these diffuse background components together.  For the analysis we derived the contributions to the isotropic background separately for $Front$-converting and $Back$-converting events.  They are available as \texttt{iso\_source\_xxx\_v05.txt} from the FSSC, where \texttt{xxx} is \texttt{front} or \texttt{back}.

\subsubsection{Solar and lunar template}

The quiescent Sun and the Moon are fairly bright $\gamma$-ray sources \citep{LAT11_Sun,LAT11_Moon}.  The Sun moves in the ecliptic but the solar $\gamma$-ray emission is extended because of cosmic-ray interactions 
with the solar radiation field; detectable emission from inverse Compton scattering of cosmic-ray electrons on the radiation field of the Sun extends several degrees from the Sun \citep{LAT11_Sun}.  The Moon is not an extended source in this way but the lunar orbit is inclined somewhat relative 
to the ecliptic and the Moon moves through a larger fraction of the sky than the Sun.  Averaged over time, the $
\gamma$-ray emission from the Sun and Moon trace a region around the ecliptic.  We used models of their 
observed emission together with calculations of their motions and of the exposure of the observations by the LAT 
to make templates for the equivalent diffuse component for the 3FGL analysis using $gtsuntemp$ \citep{SST_2013}.  For the light curves (\S~\ref{catalog_variability}) we evaluated the equivalent diffuse components for the corresponding time intervals.

\subsubsection{Residual Earth limb template}
\label{EarthLimb}

The limb of the Earth is an intense source of $\gamma$ rays from cosmic-ray collisions with the upper atmosphere \citep{LAT09_Earth}.  At the $\sim$565~km altitude of the (nearly-circular) orbit of the LAT, the limb is $\sim$112$\degr$ from the zenith.  During survey-mode observations, which predominated in the first four years of the {\it Fermi} mission, the spacecraft was rocked toward the northern and southern orbital poles on alternate $\sim$90 minute orbits.  With these attitudes, the edge of the LAT field of view closest to the orbital poles generally subtended part of the Earth limb.  As described in \S~\ref{LATData}, we limited the data selection and exposure calculations to zenith angles less than 100$\degr$.  Because the Earth limb emission is so intense, and the tails of the LAT PSF are long \citep{LAT12__calib}, a residual component of limb emission remained in the data.  Over the course of a precession period of the orbit ($\sim$53~d), the residual glow fills out large `caps' around the celestial poles, with the angular radius determined by the sum of the orbital inclination ($25\fdg6$) and the angular distance of the zenith angle limit from the orbital pole (10$\degr$).  \citet{LAT14_DiffuseModel} describe how the map and spectrum of the residual component were derived.  The spectrum is well modeled as a steep power law in energy with index 4.25.  This is steep enough that the residual Earth limb emission contributes significantly only below 300 MeV.

\section{Construction of the Catalog}
\label{catalog_main}

The procedure used to construct the 3FGL catalog has a number of improvements relative to what was implemented for the 2FGL catalog.  In this section we review the procedure, with an emphasis on what is being done differently.
The significances (\S~\ref{catalog_significance}), spectral parameters (\S~\ref{catalog_spectral_shapes}) and fluxes (\S~\ref{catalog_flux_determination}) of all catalog sources were obtained using the standard $pyLikelihood$ framework (Python analog of $gtlike$) in the LAT Science Tools\footnote{See \url{http://fermi.gsfc.nasa.gov/ssc/data/analysis/documentation/Cicerone/}.} (version v9r32p5). The localization procedure (\S~\ref{catalog_detection}), which relies on $pointlike$, provided the source positions, the starting point for the spectral fitting, and a comparison for estimating the reliability of the results (\S~\ref{catalog_analysismethod}).
Throughout the text we use the Test Statistic $TS=2\Delta \log \mathcal{L}$ for quantifying how significantly a source emerges from the background, comparing the likelihood function $\mathcal{L}$ with and without that source.

    \subsection{Detection and Localization}
\label{catalog_detection}

This section describes the generation of a list of candidate sources, with locations and initial 
spectral fits, for processing by the standard LAT science analysis tools, 
especially $gtlike$ to compute the likelihood (\S~\ref{catalog_significance}). This initial stage uses 
instead $pointlike$ \citep{Kerr2010}. Compared with the $gtlike$-based analysis 
described in \S~\ref{catalog_significance} to \ref{catalog_limitations}, it uses 
the same data, exposure, and IRFs, 
but the partitioning of the sky, the 
computation of the likelihood function, and its optimization, are  
independent. Since this version of the computation of the likelihood function is used for localization, it needs to
represent a valid estimate of the probability of observing a point source with the
assumed spectral function. 

The process started with an initial set of sources from the 2FGL analysis; not 
just those reported in that catalog, but also including all candidates failing 
the significance threshold (i.e., with $TS < 25$). It also used the latest extended source list with 25 
entries (\S~\ref{catalog_extended}), and the three-source representation of the Crab (\S~\ref{catalog_spectral_shapes}). The same spectral models were considered for each source as in \S~\ref{catalog_spectral_shapes}, but the favored model (power law or curved) was not necessarily the same.

Many details of the processing were identical to the 2FGL procedure: using 
HEALPix\footnote{\url{http://healpix.sourceforge.net}.} \citep{Gorski2005} with 
$N_{\rm side} = 12$, to tile the sky, resulting in 1728 tiles of $\sim$25 deg$^2$ area; 
optimizing spectral parameters for the sources within each tile, for the data 
in a cone of 5$\degr$ radius about the center of the tile; and including the 
contributions of all sources within 10$\degr$ of the center.  The tiles are of 
course discrete, but the regions, which we refer to as RoIs, for Regions of 
Interest, are overlapping and not independent. The data were binned in energy 
(14 energy bands from 100~MeV to 316~GeV) and position, where the angular bin size 
(the bins also defined using HEALPix) was set to be small compared with the PSF 
for each energy, and event type. Separating the photons according to event type 
is important, especially for localization, since the $Front$-converting events 
have a factor of two narrower PSF core than the $Back$-converting events. Thus the parameter 
optimization was performed by maximizing the logarithm of the likelihood, expressed as a sum 
over each energy band and each of the two event types ($Front$, $Back$). The fits 
for each RoI, maximizing the likelihood as a function of the free 
parameters, were performed independently. Correlations between sources in 
neighboring RoIs were then accounted for by iterating all 1728 fits until the changes in the 
log likelihoods for all RoIs were less than 10.

After a set of iterations had converged, then the localization procedure was 
applied, and source positions updated for a new set of iterations. At this 
stage, new sources were occasionally added using the residual $TS$ procedure described below.
The detection and initial localization process resulted in 4029 candidate point sources
with $TS > 10$.

New features that are discussed below include an assessment of the 
reliability of each spectral fit and of the model as a whole in each RoI, a 
different approach to the normalization of the Galactic diffuse background 
component, and a method to unweight the likelihood to account for the effect of potential 
systematic errors in the Galactic diffuse on source spectra.

\subsubsection{Fit validation} 
\label{pointlike_validation}

An important criterion is that the spectral and 
spatial models for all sources are not only optimized, but that the predictions 
of the models are consistent with the data. Maximizing the likelihood as an 
estimator for spectral parameters and position is valid only if the likelihood, 
given a set of parameters, corresponds to the probability of observing the data. 
We have three measures. 

\begin{figure*}[!ht]
\centering
\includegraphics[width=\textwidth]{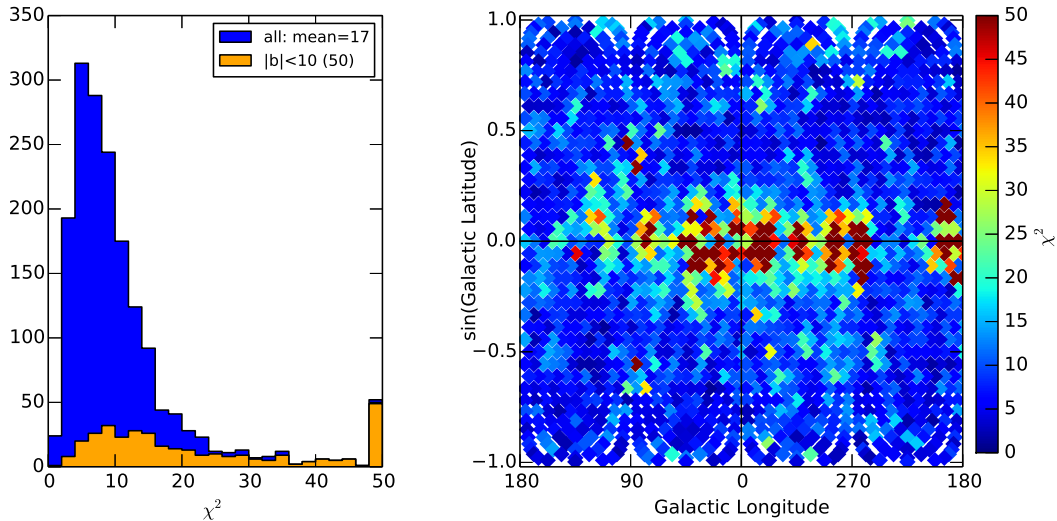}
\caption{Distributions of the $\chi^2$ measure of consistency of the measured spectrum 
of each RoI with the model (capped at 50). Left: histogram highlighting the low-latitude subset. 
Right: distribution of the values over the sky.}
\label{fig:fig_chisq}
\end{figure*}

The first compares the number of counts in each energy band, combining $Front$ and $Back$, for each of the 1728 
regions, defining a $\chi^2$-like measure as the sum of the squares of the 
deviations divided by the predicted number of counts. The number of counts is 
the expected variance for Poisson counting statistics. This 
measure is of course only a component of the likelihood, and depends only weakly 
on most of the point sources. That is, maximizing the likelihood does not 
necessarily minimize this quantity.  But it is important to check the reliability of the diffuse
model used, since this can distort the point source spectral fits.
Figure~\ref{fig:fig_chisq}  shows the distribution of that $\chi^2$-like measure and its values as a 
function of location on the sky. The number of 
degrees of freedom is 14 (the number of energy bands) minus the effective number of variables. The fact 
that the distribution peaks at $\simeq 9$ seems sensible. The $\sim$35 regions with $\chi^2 > 50$ 
indicate problems with the model.  Most are close to the 
Galactic plane, indicating difficulty with the component representing the Galactic diffuse emission. The few at high 
latitudes could be due to missing sources or, for very strong sources, inadequacy of 
the simple spectral models that we use.

\begin{figure}[!ht]
\centering
\includegraphics[width=0.4\textwidth]{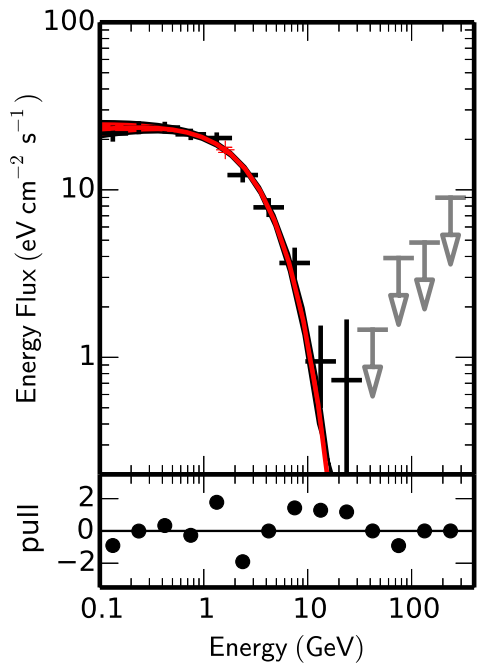}
\caption{The spectral energy distribution for a typical source, in this case PSR J1459$-$6053, 
as measured by the $pointlike$ analysis. The lower plot shows the $pulls$, defined as the 
square root of the difference $2\Delta \log \mathcal{L}$ between the fitted flux and the spectral model in each energy band, signed with the residual. The points with 
error bars reflect the 
dependence of the likelihood on the flux
for each energy band, combining $Front$ and $Back$, while the curve is the 
result of the fit to all the energy bands. 
}
\label{fig:PSR_J1459-6053_sed}
\end{figure}

The second measure is a check that the spectral model for each source is 
consistent with the data. The likelihood associated with a source is the product 
of the likelihoods for that source for each energy band, including the contributions
of nearby, overlapping sources, and the diffuse backgrounds.  The correlations induced by those are only relevant
for the lower energies, typically below 1~GeV. For this analysis, we keep these 
contributions fixed.
We form the spectral fit quality as $2\log(\mathcal{L}_{\rm bands}/\mathcal{L}_{\rm fit})$
where the flux for each band is optimized independently in $\mathcal{L}_{\rm bands}$ whereas the 
spectral model is applied in $\mathcal{L}_{\rm fit}$. The spectrum in Figure~\ref{fig:PSR_J1459-6053_sed} illustrates the concept.

\begin{figure*}[!ht]
\centering
\includegraphics[width=\textwidth]{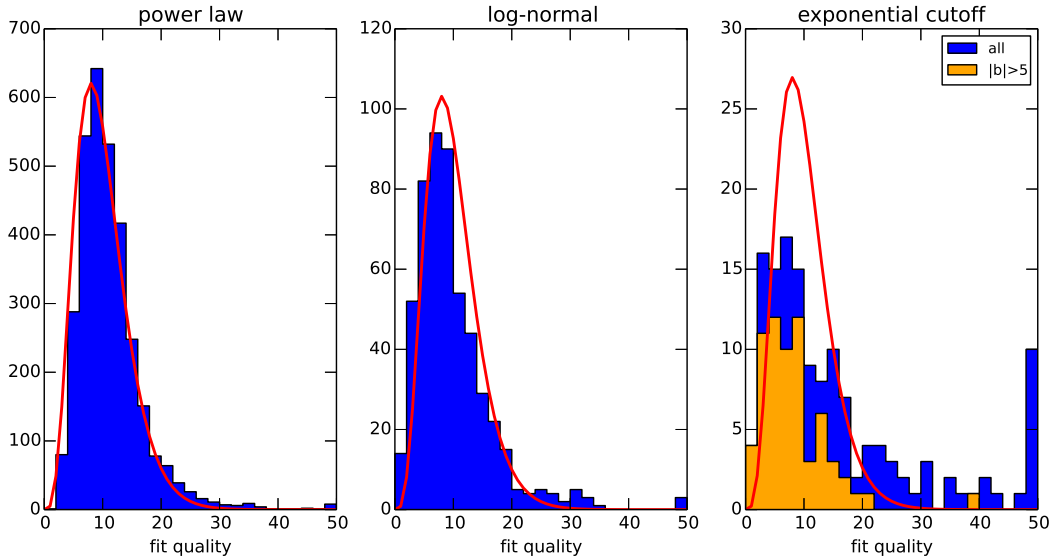}
\caption{Distributions of the spectral fit quality (capped at 50).
Left: sources fit with a power-law spectrum; 
Center: sources fit with a log-normal;
Right: sources fit with a power law with exponential cutoff.
All are overlaid with the $\chi^2$ distribution with 10 degrees of freedom.
}
\label{fig:fit_quality}
\end{figure*}

In Figure~\ref{fig:fit_quality}, we show
the distribution of the spectral fit quality 
for all preliminary spectra, with separate plots 
for the three different spectral functions (\S~\ref{catalog_spectral_shapes}): power law, log-normal, 
and power law with an exponential cutoff. The latter, applied almost exclusively 
to pulsars, is separated into sources in and out of the Galactic plane. It is seen that
sources in the plane often have poorer fits. All are compared with an example 
$\chi^2$ distribution with 10 degrees of freedom. There are 14 bands, and two to four 
parameters, but the higher-energy bands often do not contribute, so the number of degrees of freedom 
is not well defined and we use 10 for illustration only.

Finally, the localization process fits the logarithm of the likelihood as a function of 
position to a quadratic form, and checks the consistency with a $\chi^2$-like 
measure (\S~\ref{catalog_localization}).

\subsubsection{Galactic diffuse normalization and unweighting} 
The model that we used for the Galactic diffuse background is a global fit using 
the data, as described in \S~\ref{DiffuseModel}.  
For an individual RoI however, we found that we needed to adjust the 
normalizations for each band to fit the data. For the relatively broad energy 
bands, four per decade, used in the $pointlike$ fit 
we allow
the normalization for each band to vary, effectively ignoring the spectral 
prediction of the diffuse component analysis. So, for each of the 1728 RoIs, and 
for each of the eight energy bands below 10~GeV, we measured a normalization 
factor, which applies to both $Front$ and $Back$, by maximizing the likelihood 
with respect to it. A motivation for this procedure was that, for the lowest
energy bands, it often improved the fit consistencies of the spectral models of the 
sources in the same RoI. 

While the precision of the determination of the average contribution from the 
Galactic diffuse for an energy band is subject to only the statistics of the 
number of photons, the value of the Galactic diffuse intensity at the location of each source, that 
is, the angular distribution of the intensity, is subject to an additional systematic 
error. Since this intensity is strongly correlated with the measurement of the flux 
from the source itself, and the correlation can be very significant for weak sources, we have 
adopted an \textit{ad hoc}, but conservative procedure to account for the 
additional uncertainty by increasing the width of the log likelihood 
distribution from each energy band according to how sensitive it is to the 
Galactic diffuse contribution.  This is accomplished by dividing the log 
likelihood by max(1,$N_{\rm diff}$/1000) where $N_{\rm diff}$ is the predicted 
number of Galactic diffuse photons in the RoI. This has the effect of limiting the 
precision to the statistics of 1000 photons in the RoI and energy band, i.e. it 
unweights contributions from energy ranges for which the contribution from the 
diffuse component is relatively less well defined. 

\subsubsection{Localization}
\label{catalog_localization}

\begin{figure*}[!ht]
\centering
\includegraphics[width=0.45\textwidth]{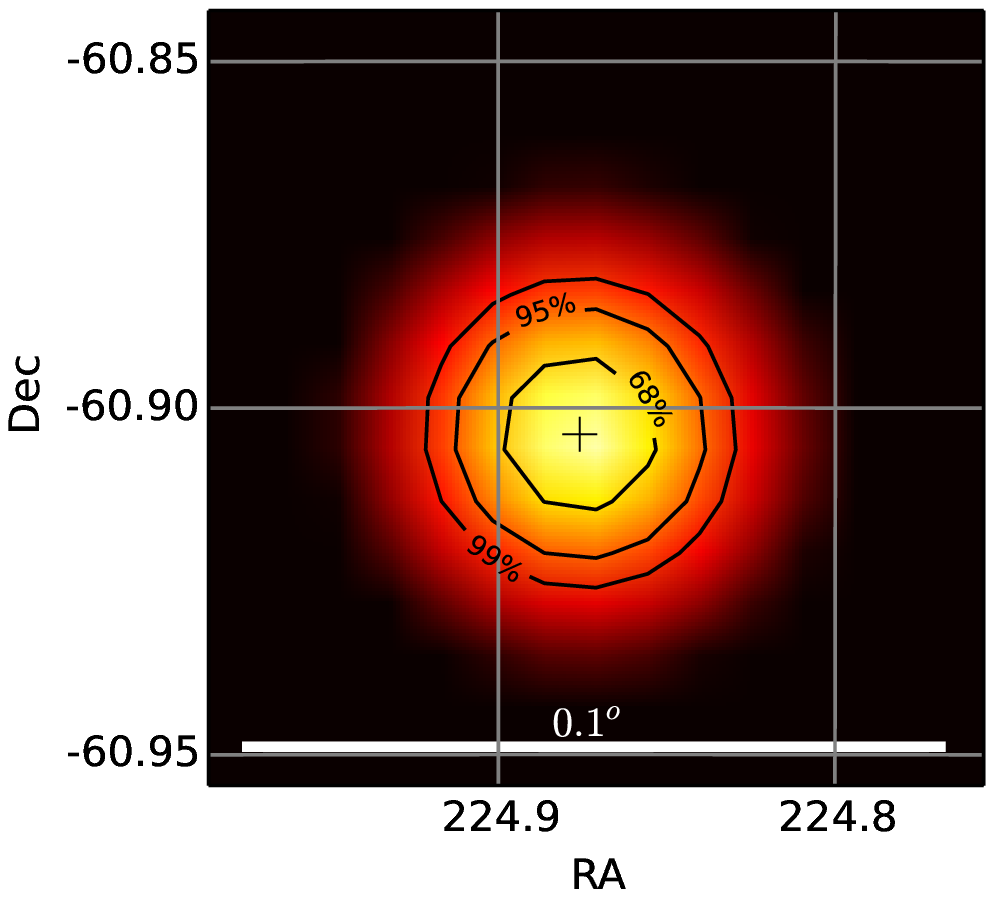}
\includegraphics[width=0.45\textwidth]{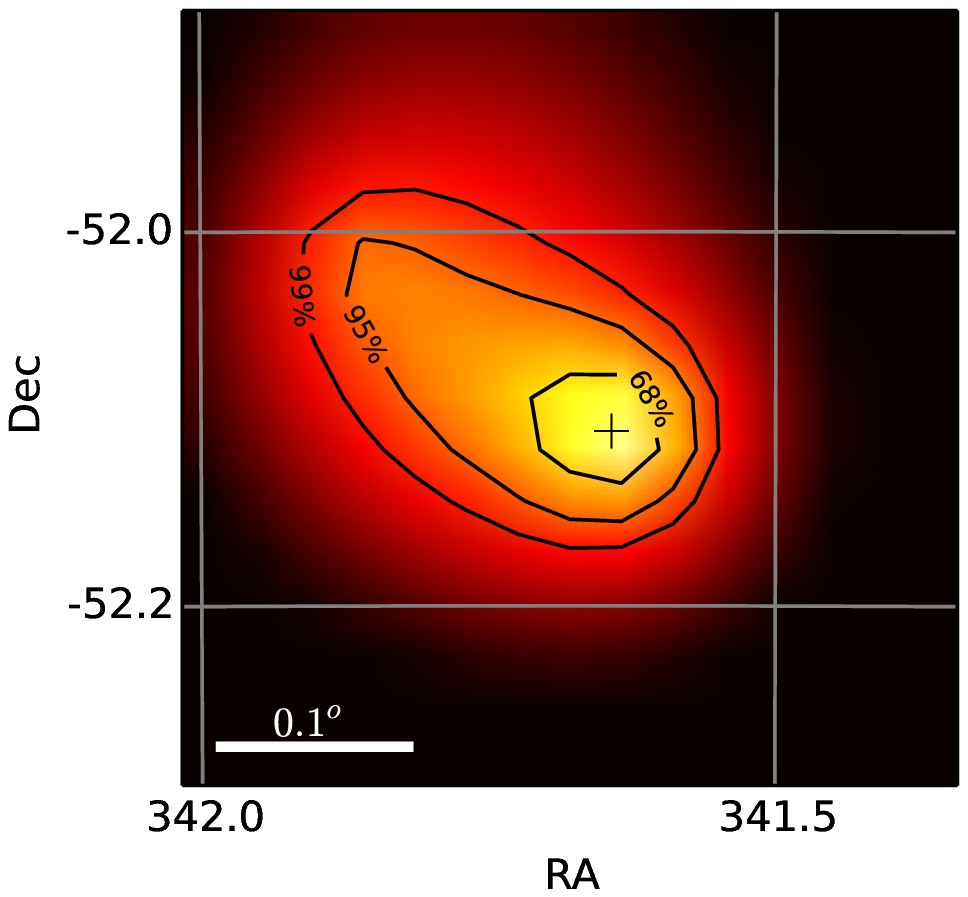}
\caption{Examples of localization $TS$ maps. The contours for 68\%, 95\%, and 99\%  containment are shown. The scale (in decimal degrees) is not the same in both plots.
Left: PSR J1459$-$6053, a good localization with LQ = 0.63.
Right: 3FGL J2246.7$-$5205, a bad localization with LQ = 14.} 
\label{fig:localization_example}
\end{figure*}

\begin{figure}[!ht]
\centering
\includegraphics[width=0.5\textwidth]{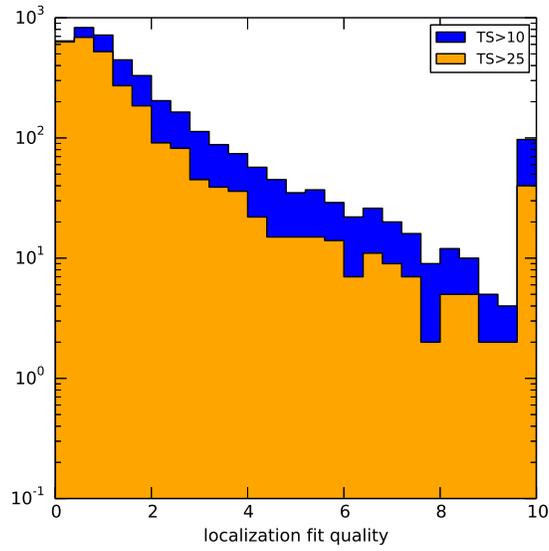}
\caption{The distribution, in the preliminary source list, of the localization quality LQ (capped at 10). }
\label{fig:localization_quality}
\end{figure}

The position of each source was determined by maximizing the likelihood with 
respect to its position only. That is, all other parameters are kept fixed. The possibility that 
a shifted position would affect the spectral models or positions of nearby sources is 
accounted for by iteration.
Ideally, the likelihood is the product of two 
Gaussians in two orthogonal angular variables. Thus the log likelihood is a 
quadratic form in any pair of angular variables, assuming small angles. 
We define LTS, for Localization Test Statistic, to be twice the log of the 
likelihood ratio of any position with respect to the maximum; the LTS evaluated 
for a grid of positions is called an $LTS$ map. We fit the distribution of LTS  
to a quadratic form to determine the uncertainty ellipse, the major and 
minor axes and orientation. We also define a measure, the localization quality 
(LQ), of how well the actual LTS distribution matches this expectation by 
reporting the sum of the squares of the deviations of eight points evaluated 
from the fit at a circle of radius corresponding to twice the geometric mean of 
the two Gaussian sigmas. 
Figure~\ref{fig:localization_example} shows examples  
of localization regions for point sources.
The distribution of the localization quality is shown in Figure~\ref{fig:localization_quality}. 

An important issue is how to treat apparently significant sources that do not 
have good localization fits, which we defined as LQ $>8$. An example is shown in 
Figure~\ref{fig:localization_example} (right). 
We flagged such sources (Flag 9 in 
Table~\ref{tab:flags}) and for them estimated the position and uncertainty by performing 
a moment analysis of the LTS function instead of fitting a quadratic form. 
Some sources that did not 
have a 
well-defined peak in the likelihood were discarded by hand, on the consideration that 
they were most likely related to residual diffuse emission. Another possibility 
is that two nearby sources produce a dumbbell-like shape; 
for some of these cases we added a new source by hand. A final selection demanding that
the semi-major radius ($1\sigma$) be less than $0\fdg25$ resulted in 3976 
candidate sources of which 142 were localized using the moment analysis.


As in 1FGL and 2FGL, we compared the localizations of the brightest sources with
associations with their true positions in the sky. This indicated that the absolute
precision is still the same, $\sim$$0\fdg005$ at the 95\% confidence level.
After the associations procedure (\S~\ref{source_assoc_automated}), 
we compared the distribution of distances to the high-confidence counterparts
(in units of the estimated $1\sigma$ errors) with a Rayleigh distribution, 
and noted that it was slightly broader, by a factor 1.05 (smaller than the 1.1
factor used in 1FGL and 2FGL).
Consequently, we multiplied all error estimates by 1.05 and added 0\fdg005 in 
quadrature to both 95\% ellipse axes. The resulting comparison with the Rayleigh
distribution is shown in Figure~3 of \citet[][3LAC]{LAT14_3LAC} and
indicates good agreement. 

\subsubsection{Detection of additional sources}

We used the $pointlike$ definition of likelihood itself to detect sources that 
needed to be added to the model of the sky. Using  HEALPix with $N_{\rm side} = 512$, we 
defined 3.2 M pixels in the sky, separated by $\simeq 0\fdg15$, 
then evaluated the improvement in the likelihood from adding a new point
source at the 
center of each, assuming a power-law spectrum with index 2.2. The $TS$ value for 
each attempt, assigned to the pixel, defines a residual $TS$ map of the sky. 
Next we performed a cluster analysis for all pixels with $TS > 10$, determining 
the number of pixels, the maximum $TS$, and the $TS$-weighted centroid. All such 
clusters with at least two pixels were added to a list of seeds. Then each seed 
was reanalyzed, now allowing the spectral index to vary, with a full 
optimization in the respective RoI, and then localized. The last step was to add 
all such refit seeds, if the fits to the spectrum and the position were 
successful, and $TS > 10$, as new sources, for a final optimization of the full 
sky. 

    \subsection{Significance and Thresholding}
\label{catalog_significance}

The framework for this stage of the analysis is inherited from the 2FGL catalog. It splits the sky into RoIs, varying typically half a dozen sources near the center of the RoI at the same time. There were 840 RoIs for 3FGL, listed in the \texttt{ROIs} extension of the catalog (App.~\ref{appendix_diffuse_params}). The global best fit is reached iteratively, injecting the spectra of sources in the outer parts of the RoI from the previous step.
In that approach the diffuse emission model (\S~\ref{DiffuseModel}) is taken from the global templates (including the spectrum, unlike what is done with $pointlike$ in \S~\ref{catalog_detection}) but it is modulated in each RoI by three parameters: normalization and small corrective slope of the Galactic component and normalization of the isotropic component. 
Appendix~\ref{appendix_diffuse_params} shows how those parameters vary over the sky.

Among more than 4000 seeds coming from the localization stage, we keep only sources at $TS > 25$, corresponding to a significance of just over $4\sigma$ evaluated from the $\chi^2$ distribution with 4 degrees of freedom \citep[position and spectral parameters,][]{mattox96}. The model for the current RoI is readjusted after removing each seed below threshold, so that the final model fits the full data. The low-energy flux of the seeds below threshold (a fraction of which are real sources) can be absorbed by neighboring sources closer than the PSF radius. There is no pair of seeds closer than $0\fdg1$, so the neighbors are unaffected at high energy. The fixed sources outside the core of the RoI are not tested and therefore not removed during the last fit of an RoI. Since the $TS$ threshold at the previous step was set to 16, seeds with $16 < TS < 25$ still populate the outer parts of the RoI, preventing the background level to rise (bullet 5 below).

\begin{figure*}
   \centering
   \begin{tabular}{cc} \hspace{-0.5cm}
  \includegraphics[bb=126 262 460 514,clip,width=0.5\textwidth]{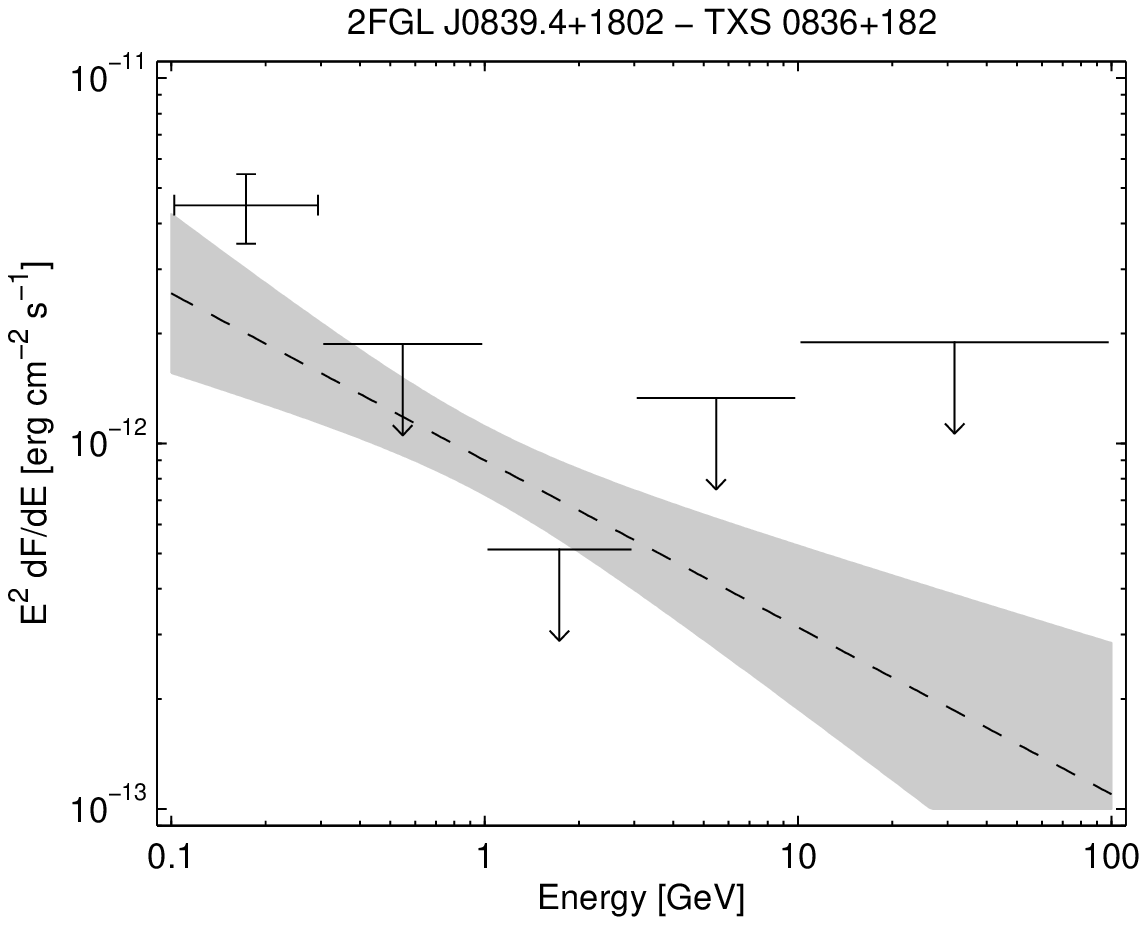} &  \hspace{-0.5cm}
   \includegraphics[width=0.5\textwidth]{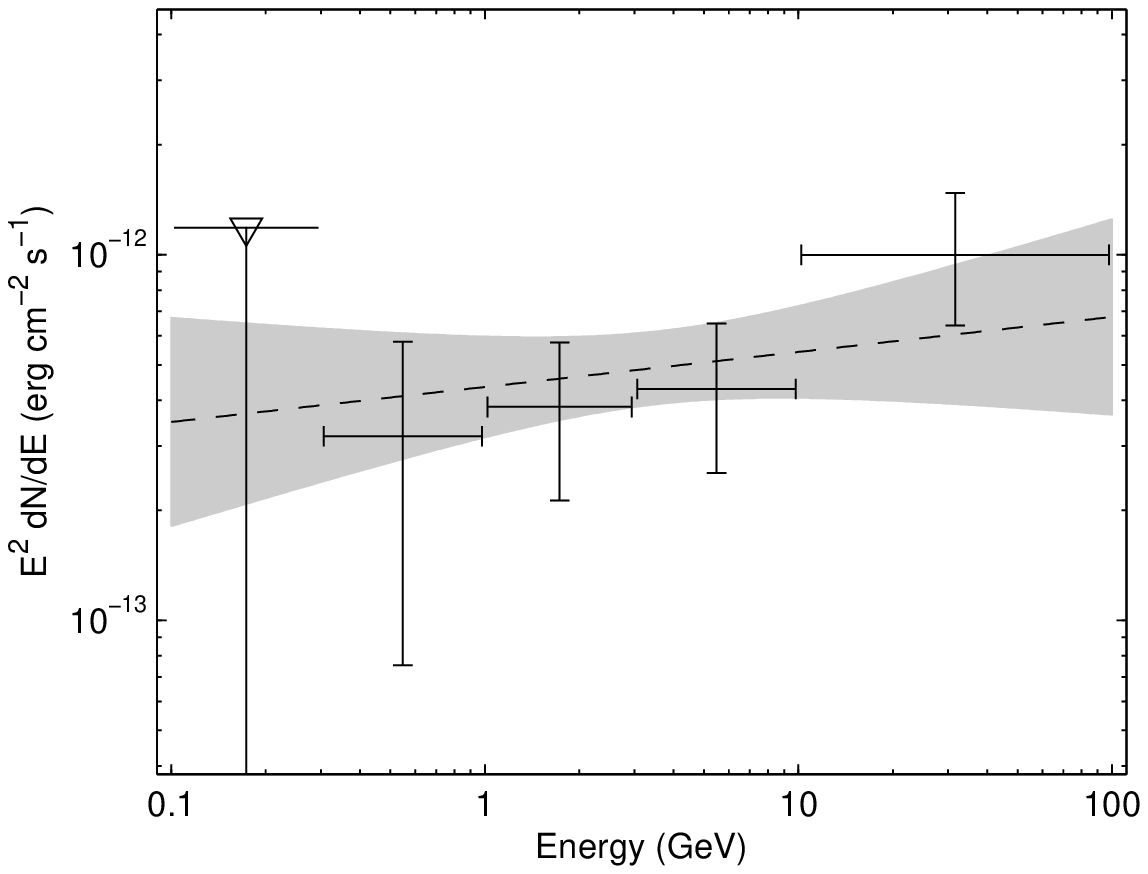}
   \end{tabular}
   \caption{Spectral energy distribution of the same source (the BL Lac TXS 0836+182) as 2FGL J0839.4+1802 (left) and 3FGL J0839.6+1803 (right). This source was flagged as `possibly due to the Sun' (Flag 11 in Table~\ref{tab:flags}) in 2FGL. Entering the Sun and Moon into the background model has reduced to zero the low energy signal that drove the 2FGL fit, resulting in a hard source in 3FGL. The dashed line is the best fit over the full energy range (\S~\ref{catalog_significance}) and the gray-shaded area is the statistical uncertainty around the best fit (for a given spectral form). The vertical scale is not the same in the left and right plots. Note that the 300~MeV to 1~GeV point has (asymmetric) error bars in 3FGL as explained in \S~\ref{catalog_flux_determination} even though its significance is less than $2\sigma$. Upper limits (indicated by a downward triangle in 3FGL and a downward arrow in 2FGL) are at 95\% confidence level.}
   \label{fig:SEDSun}
\end{figure*}

We introduced a number of improvements with respect to 2FGL (by decreasing order of importance):
\begin{enumerate}
\item After 2FGL was completed we understood that it was important to account for the different instrumental backgrounds in $Front$ and $Back$ events (\S~\ref{DiffuseModel}). Implicitly assuming that they were equal as in 2FGL resulted in lower $TS$ (fewer sources) and tended to underestimate the low-energy flux. The impact is largest at high latitude. We used different isotropic spectral templates for $Front$ and $Back$ events, but a common renormalization parameter. We also used different $Front$ and $Back$ models of the Earth limb. The same distinction was introduced for computing the fluxes per energy band (\S~\ref{catalog_flux_determination}) and per month (\S~\ref{catalog_variability}).
\item Another effect discovered after 2FGL was a slight inconsistency (8\% at 100~MeV) between the $Front$ and $Back$ effective areas. This affected mostly the Galactic plane, where the strong interstellar emission makes up 90\% of the events. That effect created opposite low-energy residuals in $Front$ and $Back$ which did not compensate each other because of the differing PSF. It was corrected empirically in the P7REP\_SOURCE\_V15 version of the IRFs \citep{LAT13_P7repro}.
\item We put in place an automatic iteration procedure at the next-to-last step of the process checking that the all-sky result is stable (2FGL used a fixed number of five iterations), similar to what was done for localization in 2FGL. Quantitatively, we iterated an RoI and its neighbors until $\log \mathcal{L}$ did not change by more than 10. In practice this changes nothing at high latitude, but improves convergence in the Galactic plane. Fifteen iterations were required to reach full convergence. That iteration procedure was run twice, allowing sources to switch to a curved spectral shape (\S~\ref{catalog_spectral_shapes}) after the first convergence.
\item The software issue which prevented using unbinned likelihood in 2FGL was solved. We took advantage of that by using unbinned likelihood at high energy where keeping track of the exact direction of each event helps. At low energy we used binned likelihood in order to cap the memory and CPU resources. The dividing energy was set to 3~GeV, resulting in data cubes (below 3~GeV) and event lists (above 3~GeV) of approximately equal size. Both data sets were split between $Front$ and $Back$. This was implemented in the $SummedLikelihood$ framework of $pyLikelihood$. In binned mode, the pixel size was set to $0\fdg2$ and $0\fdg3$ for $Front$ and $Back$ events, respectively (at 3~GeV the full width at half maximum of the PSF is $0\fdg25$ and $0\fdg38$, respectively). The energy binning was set to 10 bins per decade as in 2FGL. In the exposure maps for unbinned mode, the pixel size was set to $0\fdg1$ (even though the exposure varies very slowly, this is required to model precisely the edge of the field of view).
\item We changed the criterion for including sources outside the RoI in the model. We replaced the flat $7\degr$ distance threshold by a threshold on contributed counts (predicted from the model at the previous step). We kept all sources contributing more than 2\% of the counts per square degree in the RoI. This is a good compromise between reliability and memory/CPU requirements, and accounts for bright sources even far outside the RoI (at 100~MeV the 95\% containment radius for $Back$ events is $14\degr$). Compared to 2FGL, that new procedure affects mostly high latitudes (where the sources make up a larger fraction of the diffuse emission). Because it brings more low-energy events from outside in the model, it tends to reduce the fitted level of the low-energy diffuse emission, resulting in slightly brighter and softer source spectra.
\item The fits are now performed up to 300~GeV, and the overal significances (\texttt{Signif\_Avg}) as well as the spectral parameters refer to the full 100~MeV to 300~GeV band.
\item We introduced explicitly the model of the Sun and Moon contributions (\S~\ref{DiffuseModel}), without any adjustment or free parameter in the likelihood analysis. The success of that procedure is illustrated in Figures~\ref{fig:SEDSun} and \ref{fig:LCSun}.
\item For homogeneity (so that the result does not depend on which spectral model we start from) the $TS > 25$ threshold was always applied to the power-law model, even if the best-fit model was curved. There are 21 sources in 2FGL with $TS - TS_{\rm curve} < 25$ which would not have made it with this criterion (see \S~\ref{catalog_spectral_shapes} for the definition of $TS_{\rm curve}$).
\end{enumerate}

    \subsection{Spectral Shapes}
\label{catalog_spectral_shapes}

The spectral representation of sources was mostly the same as in 2FGL.
We introduced an additional parameter modeling a super- or subexponentially cutoff power law, as in the pulsar catalog \citep{LAT13_2PC}. However this was applied only to the brightest pulsars (PSR J0835$-$4510 in Vela, J0633+1746, J1709$-$4429, J1836+5925, J0007+7303). The global fit with nearby sources was too unstable for the fainter ones, which were left with a simple exponentially cutoff power law. The subexponentially cutoff power law was also adopted for the brightest blazar 3C 454.3\footnote{That is only a mathematical model, it should not be interpreted in a physical sense since it is an average over many different states of that very variable object.}. The fit was very significantly better than with either a log-normal or a broken power law shape. Even though bright sources are not a scientific objective of a catalog, avoiding low-energy spectral residuals (which translate into spatial residuals because of the broad PSF) is important for nearby sources.

Therefore the spectral representations which can be found in 3FGL are:
\begin{itemize}
\item a log-normal representation (\texttt{LogParabola} in the tables) for all significantly curved spectra except pulsars and 3C 454.3:
\begin{equation}
\frac{{\rm d}N}{{\rm d}E} = K \left (\frac{E}{E_0}\right )^{-\alpha -
\beta\log(E/E_0)}
\label{eq:logparabola}
\end{equation}
where $\log$ is the natural logarithm. The reference energy $E_0$ is set to \texttt{Pivot\_Energy} in the tables. The parameters $K$, $\alpha$ (spectral slope at $E_0$) and the curvature $\beta$ appear as \texttt{Flux\_Density}, \texttt{Spectral\_Index} and \texttt{beta} in the tables, respectively. No negative $\beta$ (spectrum curved upwards) was found. The maximum allowed $\beta$ was set to 1 as in 2FGL.
\item an exponentially cutoff power law for all significantly curved pulsars and a super- or subexponentially cutoff power law for the bright pulsars and 3C 454.3 (\texttt{PLExpCutoff} or \texttt{PLSuperExpCutoff} in the tables, depending on whether $b$ was fixed to 1 or left free):
\begin{equation}
\frac{{\rm d}N}{{\rm d}E} = K \left (\frac{E}{E_0}\right )^{-\Gamma} \exp \left (\left (\frac{E_0}{E_c}\right )^b-\left (\frac{E}{E_c}\right )^b\right )
\label{eq:expcutoff}
\end{equation}
where the reference energy $E_0$ is set to \texttt{Pivot\_Energy} in the tables and the parameters $K$, $\Gamma$ (low-energy spectral slope), $E_c$ (cutoff energy) and $b$ (exponential index) appear as \texttt{Flux\_Density}, \texttt{Spectral\_Index}, \texttt{Cutoff} and \texttt{Exp\_Index} in the tables, respectively. Note that this is not the way that spectral shape appears in the Science Tools (no $(E_0/E_c)^b$ term in the exponential), so the error on $K$ in the tables was obtained from the covariance matrix. The minimum $\Gamma$ was set to 0.5 (in 2FGL it was set to 0).
\item a simple power-law form for all sources not significantly curved. 
\end{itemize}
As in 2FGL, a source is considered significantly curved if $TS_{\rm curve} > 16$ where $TS_{\rm curve} = 2 (\log \mathcal{L}$(curved spectrum)$ - \log \mathcal{L}$(power-law)). The curved spectrum is \texttt{PLExpCutoff} (or \texttt{PLSuperExpCutoff}) for pulsars and 3C 454.3, \texttt{LogParabola} for all other sources. The curvature significance is reported as \texttt{Signif\_Curve} (see \S~\ref{catalog_flux_determination}).

Another difference with 2FGL is that the complex spectrum of the Crab was represented as three components:
\begin{itemize}
\item a \texttt{PLExpCutoff} shape for the pulsar, with free $K$, $\Gamma$ and $E_c$.
\item a soft power-law shape for the synchrotron emission of the nebula, with free $K$ and $\Gamma$ since the synchrotron emission is variable \citep{LAT11_CrabFlares}. The synchrotron component is called 3FGL J0534.5+2201s.
\item a hard power-law shape for the inverse Compton emission of the nebula, with parameters fixed to those found in \citet{LAT10_Crab}. That component does not vary, and leaving it free made the fit unstable. It is called 3FGL J0534.5+2201i.
\end{itemize}

In 2FGL, two sources (MSH 15$-$52 and Vela X) spatially coincident with pulsars had trouble converging and their spectra were fixed to the result of the dedicated analysis \citep{LAT10_PSR1509, LAT10_VelaX}.
In 3FGL the spectra of five sources were fixed for the same reason: the same two, the Inverse Compton component of the Crab Nebula, the Cygnus X cocoon \citep{LAT11_CygCocoon} and the $\gamma$-Cygni supernova remnant.
The spatial template of $\gamma$-Cygni was taken from \citet{LAT12_extended} as in 1FHL. We did not switch to the more complex spatial template used in \citet{LAT11_CygCocoon} but the spectral template was obtained from a reanalysis of the Cygnus region including the Cygnus X cocoon (L. Tibaldo, private communication).

Overall in 3FGL six sources (the five brightest pulsars and 3C 454.3) were fit as \texttt{PLSuperExpCutoff} (with $b$ of Eq.~\ref{eq:expcutoff} $< 1$), 110 pulsars were fit as \texttt{PLExpCutoff}, 395 sources were fit as \texttt{LogParabola} and the rest (including the five fixed sources) were represented as power laws.

    \subsection{Extended Sources}
\label{catalog_extended}

As for the 2FGL and 1FHL catalogs, we explicitly model as spatially extended those LAT sources that have been shown in dedicated analyses to be resolved by the LAT.  Twelve extended sources were entered in the 2FGL catalog. That number grew to 22 in the 1FHL catalog. The spatial templates were based on dedicated analysis of each source region and have been normalized to contain the entire flux from the source ($>99\%$ of the flux for unlimited spatial distributions such as 2-D Gaussians). The spectral form chosen for each source is the best adapted among those used in the catalog analysis (see \S~\ref{catalog_spectral_shapes}). Three more extended sources have been reported since then and were included in the same way in the 3FGL analysis\footnote{The templates and spectral models are available through the {\it Fermi} Science Support Center.}.

The catalog process does not involve looking for new extended sources or testing possible extension of sources detected as point-like. This was last done comprehensively by \citet{LAT12_extended} based on 1FGL. The extended sources published since then were the result of focussed studies so there most likely remain unreported faint extended sources in the Fermi-LAT data set. The process does not attempt to refit the spatial shape of known extended sources either.

The extended sources include twelve supernova remnants (SNRs), nine pulsar wind nebulae (PWNe) or candidates, the Cygnus X cocoon, the Large and Small Magellanic Clouds (LMC and SMC), and the lobes of the radio galaxy Centaurus A.  Below we provide notes on new sources and changes since 2FGL:  
\begin{itemize}
\item HB 21 is an SNR recently reported as a LAT source \citep{Reichardt13_HB21}. We added it to the list, using the simple disk template and \texttt{LogParabola} spectral shape derived by \citet{LAT13_HB21}.
\item HESS J1303$-$631 and HESS J1841$-$055 are two H.E.S.S. sources (most likely PWNe) recently reported as faint hard LAT sources by \citet{LAT13_TeVPWN}. We added them to the list, using the original H.E.S.S. template rather than the best spatial fit to the LAT data, in keeping with the spectral analysis in that paper.
\item We changed the spectral representation of the LMC and the Cygnus Loop from \texttt{PLExpCutoff} to \texttt{LogParabola}, which fits the data better. The curvature of the fainter SMC spectrum is not significant; therefore it was fit as a power law.
\end{itemize}

In general, we did not allow any point source inside the extended templates, even when the $TS$ maps indicated that adding new seeds would improve the fit. Most likely (pending a dedicated reanalysis) those additional seeds were simply residuals due to the fact that the very simple geometrical representations that we adopted are not precise enough, rather than independent point sources. We preferred not splitting the source flux into pieces. The only exceptions are 3FGL J1823.2$-$1339 within HESS J1825$-$137, 3FGL J2053.9+2922 inside the Cygnus Loop, 3FGL J0524.5$-$6937 inside the LMC, and sources inside the Cygnus X cocoon. The first one is as significant as the extended source and was a 2FGL source already. The next two are well localized over large extended sources and show a very hard spectrum, so they do not impact the spectral characteristics of the extended sources. The Cygnus X cocoon was fixed (\S~\ref{catalog_spectral_shapes}) and allowing point sources on top of it was necessary to reach a reasonable representation of the region.

Table~\ref{tbl:extended} lists the source name, spatial template description, spectral form and the reference for the dedicated analysis. These sources are tabulated with the point sources, with the only distinction being that no position uncertainties are reported and their names end in \texttt{e} (see \S~\ref{catalog_description}).  Unidentified point sources inside extended ones are marked by ``xxx field'' in the \texttt{ASSOC2} column of the catalog.

\begin{deluxetable}{lllcll}
\tabletypesize{\scriptsize}
\tablecaption{Extended Sources Modeled in the 3FGL Analysis
\label{tbl:extended}}
\tablewidth{0pt}
\tablehead{

\colhead{3FGL Name}&
\colhead{Extended Source}&
\colhead{Spatial Form}&
\colhead{Extent (deg)}&
\colhead{Spectral Form}&
\colhead{Reference}
}

\startdata
J0059.0$-$7242e & SMC & 2D Gaussian & 0.9 & \texttt{PowerLaw} & \citet{LAT10_SMC} \\
J0526.6$-$6825e & LMC & 2D Gaussian\tablenotemark{a} & 1.2, 0.2 & \texttt{LogParabola} & \citet{LAT10_LMC} \\
J0540.3+2756e & S 147 & Map & \nodata & \texttt{PowerLaw} & \citet{LAT12_S147} \\
J0617.2+2234e & IC 443 & 2D Gaussian & 0.26 & \texttt{LogParabola} & \citet{LAT10_IC443} \\
J0822.6$-$4250e & Puppis A & Disk & 0.37 & \texttt{PowerLaw} & \citet{LAT12_extended} \\
J0833.1$-$4511e & Vela X & Disk & 0.88 & \texttt{PowerLaw} & \citet{LAT10_VelaX} \\
J0852.7$-$4631e & Vela Junior & Disk & 1.12 & \texttt{PowerLaw} & \citet{LAT11_RXJ0852} \\
J1303.0$-$6312e & HESS J1303$-$631 & 2D Gaussian & 0.16 & \texttt{PowerLaw} & \citet{HESS05_J1303} \\
J1324.0$-$4330e & Centaurus A (lobes) & Map & \nodata & \texttt{PowerLaw} & \citet{LAT10_CenAlobes} \\
J1514.0$-$5915e & MSH 15$-$52 & Disk & 0.25 & \texttt{PowerLaw} & \citet{LAT10_PSR1509} \\
J1615.3$-$5146e & HESS J1614$-$518 & Disk & 0.42 & \texttt{PowerLaw} & \citet{LAT12_extended} \\
J1616.2$-$5054e & HESS J1616$-$508 & Disk & 0.32 & \texttt{PowerLaw} & \citet{LAT12_extended} \\
J1633.0$-$4746e & HESS J1632$-$478 & Disk & 0.35 & \texttt{PowerLaw} & \citet{LAT12_extended} \\
J1713.5$-$3945e & RX J1713.7$-$3946 & Map & \nodata & \texttt{PowerLaw} & \citet{LAT11_RXJ1713} \\
J1801.3$-$2326e & W28 & Disk & 0.39 & \texttt{LogParabola} & \citet{LAT10_W28} \\
J1805.6$-$2136e & W30 & Disk & 0.37 & \texttt{LogParabola} & \citet{LAT12_W30} \\
J1824.5$-$1351e & HESS J1825$-$137 & 2D Gaussian & 0.56 & \texttt{LogParabola} & \citet{LAT11_J1825} \\
J1836.5$-$0655e & HESS J1837$-$069 & Disk & 0.33 & \texttt{PowerLaw} & \citet{LAT12_extended} \\
J1840.9$-$0532e & HESS J1841$-$055 & 2D Gaussian\tablenotemark{b}  & (0.41, 0.25) & \texttt{PowerLaw} & \citet{HESS08_Unid} \\
J1855.9+0121e & W44 & Ring\tablenotemark{b} & (0.22, 0.14), (0.30, 0.19) & \texttt{LogParabola} & \citet{LAT10_W44} \\
J1923.2+1408e & W51C & Disk\tablenotemark{b} & (0.40, 0.25) & \texttt{LogParabola} & \citet{LAT09_W51C} \\
J2021.0+4031e & $\gamma$-Cygni & Disk & 0.63 & \texttt{PowerLaw} & \citet{LAT12_extended} \\
J2028.6+4110e & Cygnus X cocoon & 2D Gaussian & 2.0 & \texttt{PowerLaw} & \citet{LAT11_CygCocoon} \\
J2045.2+5026e & HB 21 & Disk & 1.19 & \texttt{LogParabola} & \citet{LAT13_HB21} \\
J2051.0+3040e & Cygnus Loop & Ring & 0.7, 1.6 & \texttt{LogParabola} & \citet{LAT11_CygnusLoop} \\
\enddata

\tablenotetext{a}{Combination of two 2D Gaussian spatial templates.}
\tablenotetext{b}{The shape is elliptical; each pair of parameters $(a, b)$ represents the semi-major $(a)$ and semi-minor $(b)$ axes.}

\tablecomments{~List of all sources that have been modeled as extended sources. The Extent column indicates the radius for Disk sources, the dispersion for Gaussian sources, and the inner and outer radii for Ring sources.}

\end{deluxetable}

    \subsection{Flux Determination}
\label{catalog_flux_determination}

The source photon fluxes are reported in the same five energy bands (100 to 300~MeV; 300~MeV to 1~GeV; 1 to 3~GeV; 3 to 10~GeV; 10 to 100~GeV) as in 2FGL. The fluxes were obtained by freezing the spectral index to that obtained in the fit over the full range and adjusting the normalization in each spectral band. For the curved spectra (\S~\ref{catalog_spectral_shapes}) the spectral index in a band was set to the local spectral slope at the logarithmic mid-point of the band $\sqrt{E_n E_{n+1}}$, restricted to be in the interval [0,5].
The photon flux between 1 and 100~GeV as well as the energy flux between 100~MeV and 100~GeV ($F_{35}$ and $S_{25}$ in Table~\ref{tab:desc}; the subscript $ij$ indicates the energy range as 10$^i$--10$^j$ MeV), are derived from the full-band analysis assuming the best spectral shape, and their uncertainties from the covariance matrix. Even though the full analysis is carried out up to 300~GeV in 3FGL, we have not changed the energy range over which we quote fluxes so that they can be easily compared with past fluxes. The photon flux above 100~GeV is negligible anyway and the energy flux above 100~GeV is not precisely measured (even for hard sources).

\begin{figure}
\plotone{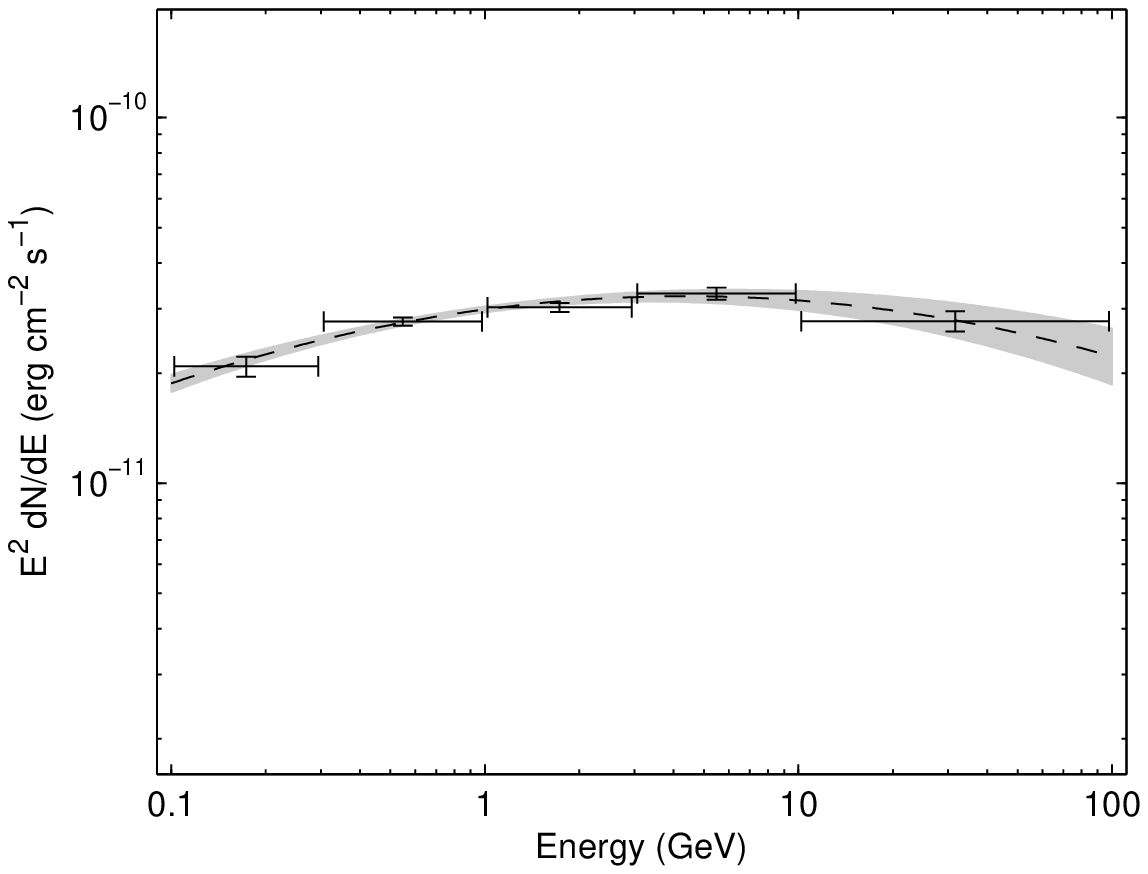}
\caption{Spectral energy distribution of 3FGL J0222.6+4301 (3C 66A) fitted by a \texttt{LogParabola} spectrum with $\beta = 0.039 \pm 0.007$ but \texttt{Signif\_Curve} (defined just before Eq.~\ref{eq:systratio}) = 2.81. The curvature is statistically significant but a power law cannot be excluded given the range of the systematic errors on effective area.}
\label{fig:SEDCurve}
\end{figure}

\begin{figure}
\plotone{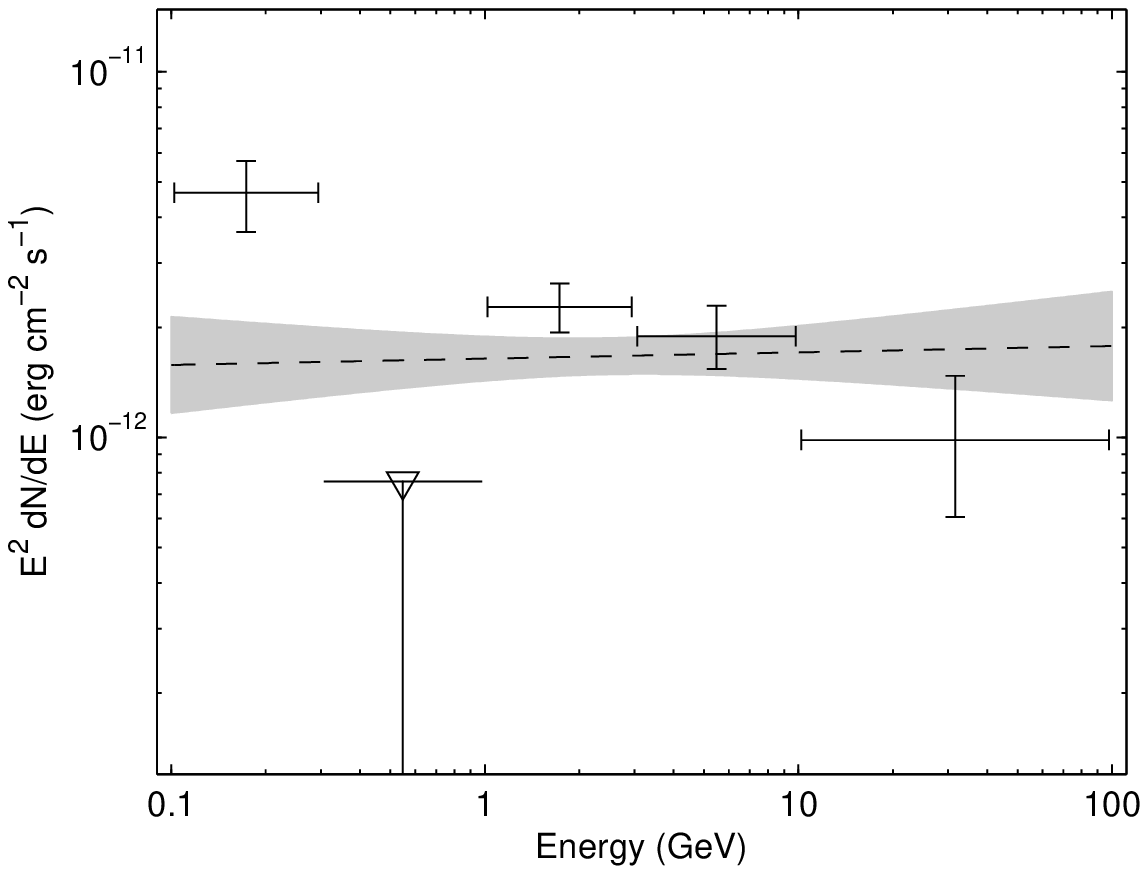}
\caption{Spectral energy distribution of 3FGL J1226.9$-$1329 (PMN J1226$-$1328) flagged with bad spectral fit quality (Flag 10 in Table~\ref{tab:flags}). The first two points deviate from the power-law fit. This source is within 1\fdg3 of the much brighter pulsar PSR J1231$-$1411 so it is confused with it (within $r_{68}$) below 600~MeV.}
\label{fig:SEDQuality}
\end{figure}

Improvements with respect to the 2FGL analysis are:
\begin{itemize}
\item We used binned likelihood in the first three bands (up to 3~GeV) and unbinned likelihood in the last two bands, distinguishing $Front$ and $Back$ events. The pixel sizes in each band in binned mode were $0\fdg3$ and $0\fdg5$, $0\fdg2$ and $0\fdg3$, $0\fdg1$ and $0\fdg15$ where in each band, the first value is for $Front$, the second one for $Back$. This reduces error bars by 10--15\% compared to mixing $Front$ and $Back$ events as in 2FGL.
\item Following what was done in the 1FHL catalog, the errors on the fluxes of moderately faint sources ($TS \ge 1$ in the band) were computed as $1\sigma$ errors with MINOS in the Minuit\footnote{\url{http://lcgapp.cern.ch/project/cls/work-packages/mathlibs/minuit/home.html}.} package. This was done whenever the relative error on flux in the quadratic approximation (from the covariance matrix) was larger than 10\%. Both errors (lower and upper) are reported in the FITS table (App.~\ref{appendix_fits_format}). The lower error is reported with a minus sign (when the error comes from the quadratic approximation, the lower error is simply minus the upper error). The upper limits $F_i^{UL}$ for very faint sources ($TS < 1$) were computed as in 2FGL, using the Bayesian method \citep{helene83} at 95\% of the posterior probability. The upper error is then reported as $0.5(F_i^{UL}-F_i^{BF})$ where $F_i^{BF}$ is the best-fit flux, and the lower error is set to NULL.
\item The same iteration procedure described in \S~\ref{catalog_significance} was put in place for the fluxes per energy band using a more stringent criterion ($\Delta \log \mathcal{L} < 3$). Convergence was fast at high energy (little cross-talk between sources). It was a little slower at low energy (6 iterations in the first band) but much faster than the full-band fit because no spectral adjustment was involved.
\item We report as \texttt{nuFnuxxx\_yyy} the Spectral Energy Distribution (SED) in the band defined by xxx to yyy MeV, which can be directly overlaid on an SED plot. The SED was obtained by dividing the energy flux in the band by the band width in natural logarithm log(yyy/xxx). Since the fit is performed on the flux only (no spectral freedom in each band), the relative error on the SED is the same as that on the corresponding flux.
\end{itemize}

As in 2FGL we report in 3FGL a curvature significance \texttt{Signif\_Curve} = $\sqrt{TS_{\rm curve} \; R_{\rm syst}}$ (in $\sigma$ units) after approximately accounting for systematic uncertainties on effective area via
\begin{equation}
\label{eq:systratio}
R_{\rm syst} = \frac{\sum_i (F_i - F_i^{\rm PL})^2 / (\sigma_i^2 + (f_i^{\rm rel} F_i^{\rm fit})^2)}{\sum_i (F_i - F_i^{\rm PL})^2 / \sigma_i^2}
\end{equation}
where $i$ runs over all bands, $F_i^{\rm PL}$ is the flux predicted by the power-law model and $F_i^{\rm fit}$ is the flux predicted by the best-fit (curved) model in that band from the spectral fit to the full band. $f_i^{\rm rel}$ reflects the systematic uncertainty on effective area (\S~\ref{catalog_limitations}). The values were set to 0.1, 0.05, 0.05, 0.05, 0.1 in our five bands (the fourth one went down from 0.08 in 2FGL, thanks to improved calibration).
Eq.~\ref{eq:systratio} is not exactly the same formula used for 2FGL. In 2FGL $F_i^{\rm fit}$ would have been replaced by $F_i^{\rm PL}$. The disadvantage of the previous estimate was that it capped \texttt{Signif\_Curve} to rather low values (below 15) resulting in a small dynamic range because the largest relative systematic errors are in the two extreme bands and in those bands the power-law fit can run way above the points (because the spectra are curved downwards). Using the curved fit (closer to the points) to estimate the systematic errors is a more reasonable procedure, and recovers a larger dynamic range (up to 85 in 3FGL).

As in 2FGL we consider that only sources with \texttt{Signif\_Curve} $>$ 4 are significantly curved (at the $4\sigma$ level). When $R_{\rm syst}$ is small (bright source) it can happen that $TS_{\rm curve} > 16$ (triggering a curved model following \S~\ref{catalog_spectral_shapes}) but \texttt{Signif\_Curve} $<$ 4. The 43 such sources with \texttt{LogParabola} spectra (and 2 pulsars with \texttt{PLExpCutoff} spectra) but \texttt{Signif\_Curve} $<$ 4 could be power laws within systematic errors. Nevertheless we do not go back to power-law spectra for those sources because they are better fit with curved models and power-law models would result in negative low-energy residuals which might affect nearby sources. One of them is illustrated in Figure~\ref{fig:SEDCurve}. All are bright sources with modest curvature.

Spectral fit quality (for Flag 10 in Table~\ref{tab:flags}) is computed as in Eq.~3 of \citet[][2FGL]{LAT12_2FGL} rather than as in \S~\ref{pointlike_validation}. Among the 42 sources flagged because of a too large spectral fit quality, most show deviations at low energy and are in confused regions or close to a brighter neighbor, as in Figure~\ref{fig:SEDQuality}.

Spectral plots for all 3FGL sources overlaying the best model on the individual SED points are available from the FSSC.

    \subsection{Variability}
\label{catalog_variability}

\begin{figure}
\plotone{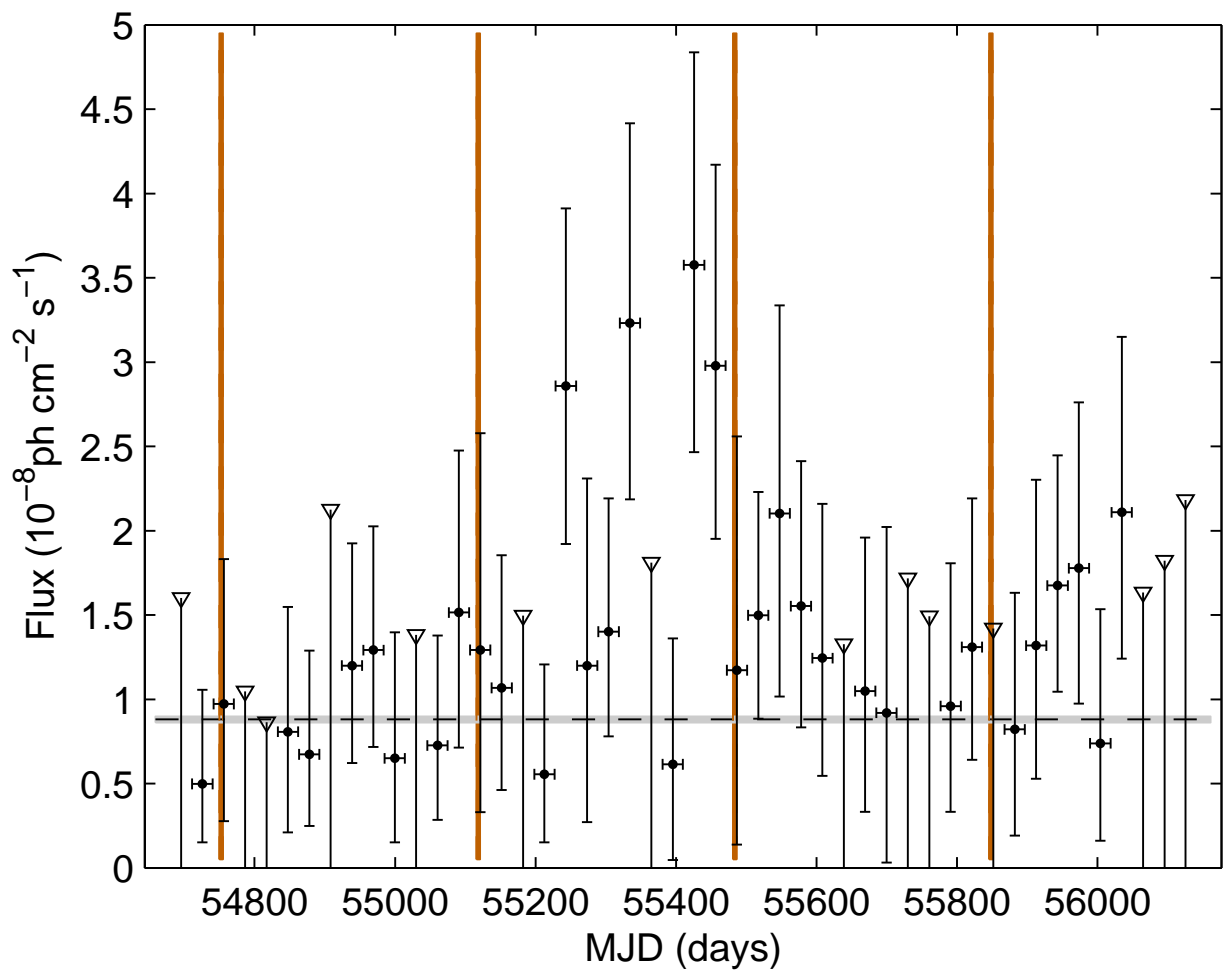}
\caption{Light curve of 3FGL J1315.7$-$0732 (NVSS J131552$-$073301) in the ecliptic plane. That source is significantly variable. The flares do not correspond to the times when the Sun passed through the region (vertical orange bands). The only effect of the Sun passage is somewhat larger error bars. The gray-shaded horizontal area materializes the systematic uncertainty of 2\%. Upper limits (indicated by a downward triangle) are at 95\% confidence level.}
\label{fig:LCSun}
\end{figure}

\begin{figure}
\plotone{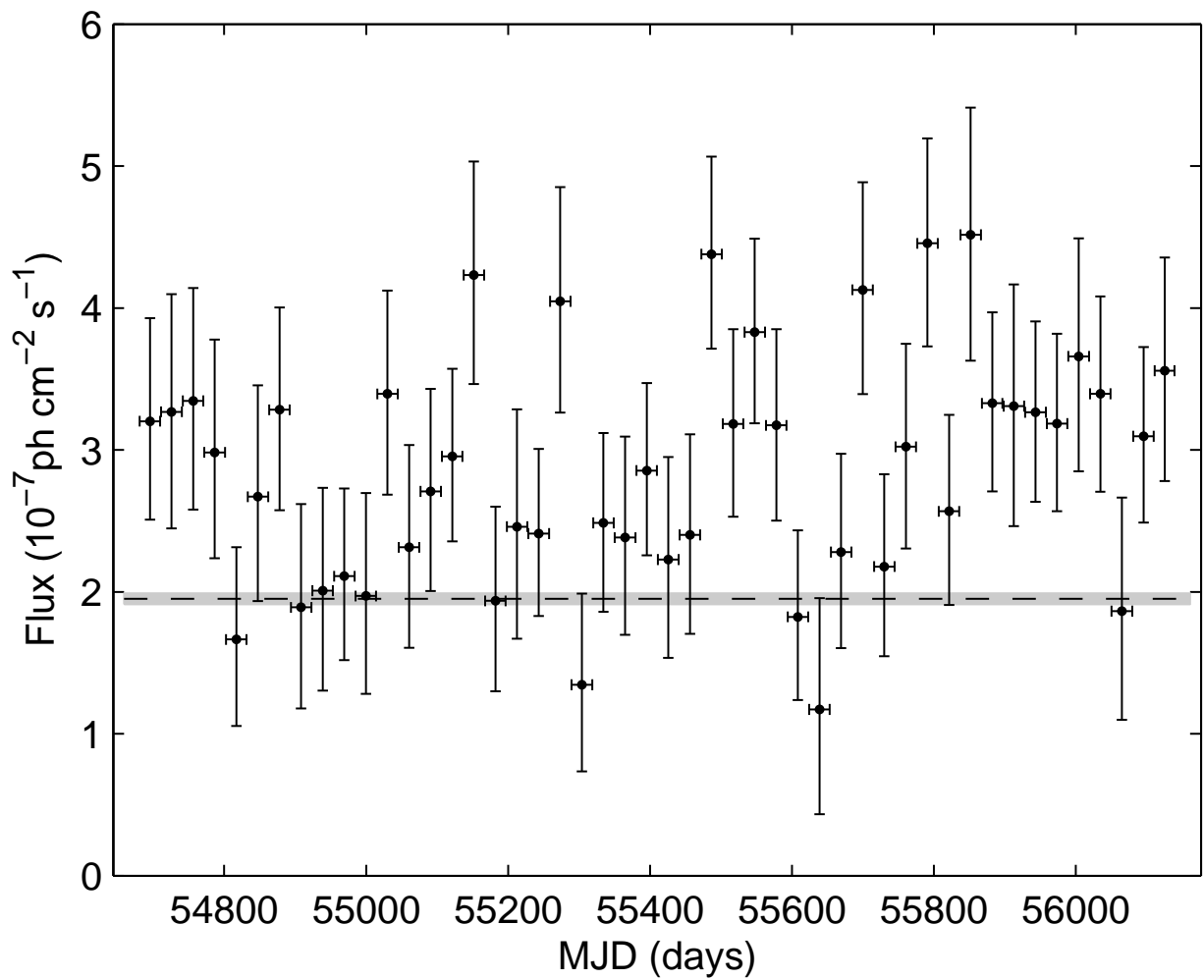}
\caption{Light curve of 3FGL J1616.2$-$5054e (HESS J1616$-$508). That is an extended source that should not be variable. Indeed the monthly fluxes are compatible with a constant, but not with the flux extracted over the full four years (dashed line with gray-shaded uncertainty). That inconsistency is due to a remaining difference between binned and unbinned likelihood fits affecting only extended sources.}
\label{fig:LCExtended}
\end{figure}

\begin{figure}
\plotone{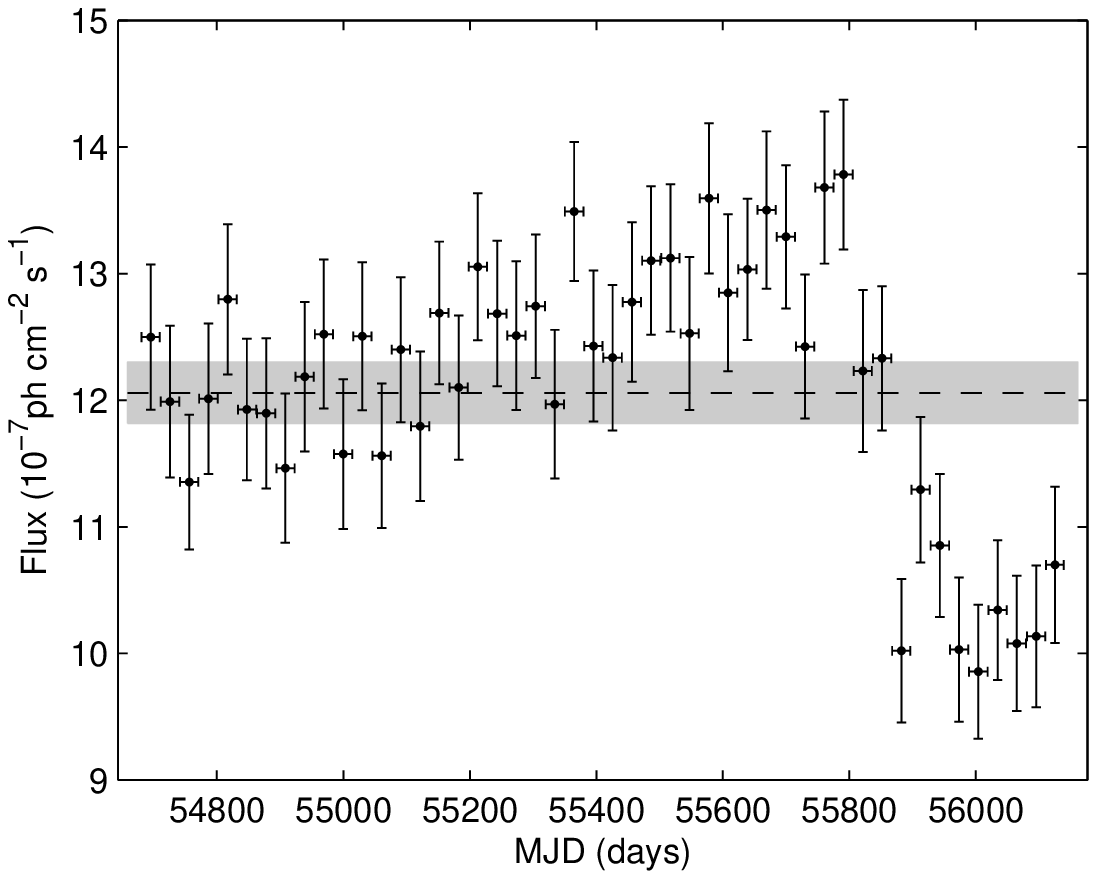}
\caption{Light curve of 3FGL J2021.5+4026 (PSR J2021+4026 in the $\gamma$ Cygni SNR). The variability of that pulsar is easily detected by the automatic procedure. The vertical scale does not start at 0.}
\label{fig:LCPulsar}
\end{figure}

The light curves were computed over the same (1-month) intervals as in 1FGL and 2FGL (there are now 48 points). The first 23 intervals correspond exactly to 2FGL.  The fluxes in each bin were obtained by freezing the spectral parameters to those obtained in the fit over the full range and adjusting the normalization. We used unbinned likelihood over the full energy range for the light curves. Over short intervals it does not incur a large CPU or memory penalty and it preserves the full information. We used a different isotropic and Earth limb model for $Front$ and $Back$ events, as in the main fit (\S~\ref{catalog_significance}). We also used a different Sun/Moon model for each month (the Sun is obviously at a different place in the sky each month). That improvement, together with our removing the solar flares, effectively mitigated the peaks that we noted in the 2FGL light curves due to the Sun passage near the source (Flag 11 in Table~\ref{tab:flags}). We have not noted any obvious Sun-related peak in the 3FGL light curves (Figure~\ref{fig:LCSun}).

As in the band fluxes calculation (\S~\ref{catalog_flux_determination}) the errors on the monthly fluxes of moderately faint sources ($TS > 1$) were computed as lower and upper $1\sigma$ errors with MINOS in Minuit. Both errors (lower and upper) are reported in the FITS table (Table~\ref{tab:columns}) so the \texttt{Unc\_Flux\_History} column is a $2 \times 48$ array. This allowed providing more information in the light curve plots\footnote{These plots are available from the FSSC.} by keeping points with error bars whenever $TS > 1$ (the lower error does not reach 0). When $TS < 1$ the 95\% upper limit is converted into an upper error in the same way as in 2FGL and the band fluxes calculation.

We noted an inconsistency between the light curve and the flux from the main fit (over the full interval) in several extended sources, whereby the average of the light curve appears distinctly above the flux from the main fit. It is particularly obvious in Cen~A lobes, HESS J1616$-$508 (Figure~\ref{fig:LCExtended}), S 147, W28, and W30. We traced the problem to the fact that we used unbinned likelihood over the whole energy range for the light curves, but binned likelihood for the main fit below 3~GeV. 
We have not found any evidence that this affects the point sources. Since we do not expect variability in extended sources, we left this inconsistency in the catalog as a known feature.

The variability indicator \texttt{Variability\_Index} is the same as in 2FGL, with the same relative systematic error of 2\%. Variability is considered probable when \texttt{Variability\_Index} exceeds the threshold of 72.44 corresponding to 99\% confidence in a $\chi^2$ distribution with 47 degrees of freedom.

The Crab nebula and pulsar are a particularly difficult case. The nebula is very variable \citep{AGILE11_CrabFlares,LAT11_CrabFlares} while the pulsar has no detected variability. So we would have liked the synchrotron component to absorb the full variability in 3FGL. It does not turn out this way, however, because the spectrum of the nebula becomes much harder during flares. This is not accounted for in the variability analysis (the spectral slopes are fixed to that in the full interval). As a result, the pulsar component also increases during the nebular flares and the pulsar becomes formally variable. We stress here that it is only a feature of our automatic analysis and is in no way a real detection of variability in the Crab pulsar.
Besides the Crab, we detect the (real) variability of PSR J2021+4026 \citep[Figure~\ref{fig:LCPulsar},][]{LAT13_variablePSR}. The only other formally variable pulsar is PSR J1732$-$3131 just above threshold. Since this is one in 137 pulsars, it is compatible with a chance occurrence at the 99\% confidence level.

    \subsection{Limitations and Systematic Uncertainties}
\label{catalog_limitations}

\subsubsection{Source confusion}
\label{catalog_confusion}

\begin{figure}[ht]
\epsscale{1.0}
\plotone{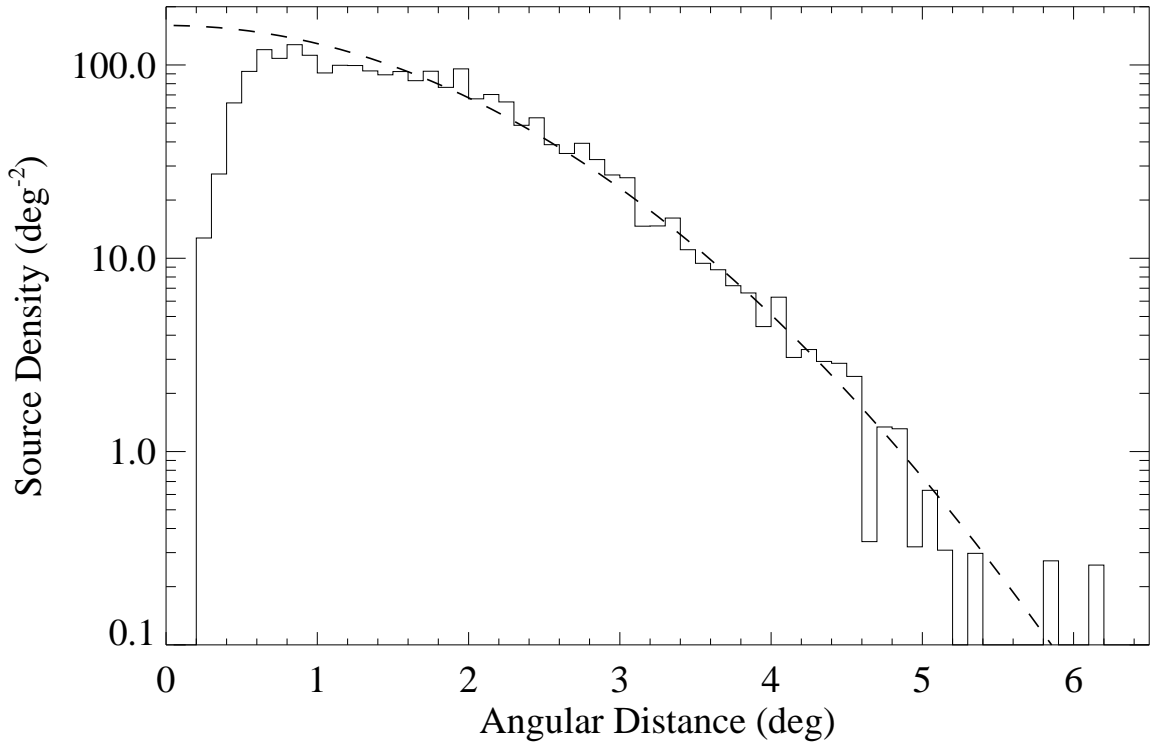}
\caption{Distribution of nearest-neighbor distances for 3FGL sources at $|b| > 10\degr$.  The dashed curve was derived as described in \citet[][1FGL]{LAT10_1FGL}.  It is the distribution expected if sources could be detected at arbitrarily small angular separations.  The dashed curve is normalized to match the total observed number of sources for separations $> 0\fdg8$ (2035). This corresponds to an expected true number of sources (extrapolated down to 0 separation) of 2336 at $|b| > 10\degr$.
}
\label{fig:nearest}
\end{figure}

As for the 1FGL and 2FGL catalogs we investigated source confusion by comparing the actual distribution of angular separations between 3FGL sources with what would be expected for a population of sources that could be detected independently regardless how small their angular separations.  The formalism is defined in \citet[][1FGL]{LAT10_1FGL}.  We considered the region of the sky above $|b| = 10\degr$, within which the average angular separation of 3FGL sources is $2\fdg2$.  The distribution of nearest-neighbor distances is shown in Figure~\ref{fig:nearest} along with the distribution expected if the source detection efficiency did not decrease for closely-spaced sources.   The observed density of nearest-neighbor starts to fall below the expected curve at about $0\fdg8$ angular separation.  The implied number of missing closely-spaced sources is $\sim$140, or about 6\% of the estimated true source count in the region.  For the 2FGL catalog the fraction was only 3.3\%. This indicates that even though the PSF improved after the Pass 7 reprocessing, the larger number of detected sources (2193 vs. 1319) is now pushing the LAT catalog into the confusion limit even outside the Galactic plane. Because the confusion process goes as the square of the source density, the expected number of sources above the detection threshold within $0\fdg5$ of another one (most of which are not resolved) has increased by a factor of 3 between 2FGL and 3FGL.

The consequence of source confusion is not only losing a fraction of sources. It can also lead to ``composite'' $\gamma$-ray sources merging the characteristics of two very nearby astronomical objects. An example is the unassociated 3FGL J0536.4$-$3347, located between two bright blazars. Its spectrum is relatively soft, similar to that expected from the FSRQ BZQ J0536$-$3401, 14\arcmin\ away. Its location, however, is closer (4\arcmin) to the BL Lac BZB J0536$-$3343 because that one dominates at high energy where the $Fermi$ PSF is best. That issue is discussed in more detail in the 3LAC paper.

\subsubsection{Instrument response functions}
\label{catalog_IRFs}

The systematic uncertainties on effective area have improved since 2FGL, going from P7SOURCE\_V6 to P7REP\_SOURCE\_V15. They are now estimated to be 5\% between 316~MeV and 10~GeV, increasing to 10\% at 100~MeV and 15\% at 1~TeV (see the caveats page at the FSSC), following the methods described by \citet{LAT12__calib}.
As in previous LAT catalogs, we have not included those uncertainties in any of the error columns, because they apply uniformly to all sources. They must be kept in mind when working with absolute numbers, but comparisons between sources can be carried out at better precision.
The systematic uncertainties on effective area have been included in the curvature significance (\S~\ref{catalog_flux_determination}) and a systematic uncertainty of 2\% on the stability of monthly flux measurements (measured directly on the bright pulsars) has been included in the variability index (\S~\ref{catalog_variability}).

\subsubsection{Diffuse emission model}
\label{catalog_diffusemodel}

\begin{table*}
\begin{tabular}{l|l|l|l|l|l}
Selection & Quantity & \multicolumn{2}{c}{Diffuse model (\S~\ref{catalog_diffusemodel})} & \multicolumn{2}{c}{Analysis method (\S~\ref{catalog_analysismethod})} \\
 &  & Bias & Scatter & Bias & Scatter \\
\hline
Galactic & Eflux (174) & $+1.88\sigma$ (+21\%) & $3.40\sigma$ (42\%) & $-0.47\sigma$ ($-$7\%) & $1.93\sigma$ (27\%) \\
Ridge & Index (88) & $+1.44\sigma$ (+0.14) & $1.81\sigma$ (0.37) & $-0.08\sigma$ ($-$0.01) & $2.40\sigma$ (0.21) \\
\hline
Galactic & Eflux (662) & $+0.51\sigma$ (+7\%) & $2.19\sigma$ (32\%) & $-0.66\sigma$ ($-$12\%) & $1.26\sigma$ (23\%) \\
Plane & Index (470) & $+0.34\sigma$ (+0.04) & $1.54\sigma$ (0.21) & $-0.44\sigma$ ($-$0.06) & $1.15\sigma$ (0.15) \\
\hline
High & Eflux (2193) & $+0.07\sigma$ (+1\%) & $0.98\sigma$ (15\%) & $-0.42\sigma$ ($-$7\%) & $0.74\sigma$ (13\%) \\
Latitude & Index (1960) & $+0.23\sigma$ (+0.03) & $0.73\sigma$ (0.10) & $-0.34\sigma$ ($-$0.05) & $0.73\sigma$ (0.10) \\
\hline
\end{tabular}
\caption{~The table gives the bias and the scatter induced by changing one of two important elements in the analysis chain, first in units of the statistical error (i.e., on $(A_i^{\rm alt}-A_i)/\sigma_i$), then in absolute terms (i.e., on $A_i^{alt}-A_i$), where $A_i$ is either the log of the energy flux between 100~MeV and 100~GeV or the spectral index in the standard analysis, $A_i^{\rm alt}$ is the same quantity in the alternative analysis and $\sigma_i$ the statistical uncertainty on $A_i$. The spectral index comparison is restricted to pure power-law sources. The Galactic Ridge is defined as $|b| < 2\degr$ and $| l | < 60\degr$. High Latitude is defined as $|b| \ge 10\degr$. The Galactic Plane is everything else (i.e., it does not include the Galactic Ridge). The number of sources in each selection is given in parentheses after the quantity.}
\label{tbl:alternative}
\end{table*}

The model of diffuse emission is the main source of uncertainties for faint sources. Contrary to the effective area, it does not affect all sources equally: its effects are smaller outside the Galactic plane where the diffuse emission is fainter and varying on larger angular scales.
It is also less of a concern in the high-energy bands ($>$ 3~GeV) where the core of the PSF is narrow enough that the sources dominate the background under the PSF.
But it is a serious concern inside the Galactic plane in the low-energy bands ($<$ 1~GeV) and particularly inside the Galactic ridge ($|l| < 60\degr$) where the diffuse emission is strongest and very structured, following the molecular cloud distribution.
It is not easy to assess precisely how large the uncertainties are, because they relate to uncertainties in the distributions of interstellar gas, the interstellar radiation field, and cosmic rays, which depend in detail on position on the sky.

For an assessment we have tried re-extracting the source spectra using one of the eight alternative interstellar emission models described in \citet{LAT13_DiffuseSystematics}, namely the one obtained with optically thin H\,{\sc i}, an SNR cosmic-ray source distribution and a 4 kpc halo, adapted to the P7REP IRFs. For computational reasons we have not used all eight alternative models, but that one should be representative. In each RoI we left free the normalization of each component of the model contributing (with its normalization set to 1) more than 3\% of the total counts in the RoI. The isotropic normalization was also left free, but the inverse Compton, Loop I and {\it Fermi} bubbles components were fixed (too large scale to be fitted inside a single RoI). That approach (independent components) differs enough from the standard diffuse model that it can provide a stronger test than comparing with the previous generation diffuse model, as we did for 2FGL. Nevertheless both models still rely on nearly the same set of H\,{\sc i} and CO maps of the gas in the interstellar medium, so they are not as independent as we would like.

The results show that the systematic uncertainty more or less follows the statistical one, i.e., it is larger for fainter sources in relative terms.
We list the induced biases and scatters of flux and spectral index in Table~\ref{tbl:alternative}.
We have not increased the flux and index errors in the catalog itself accordingly because this alternative model does not fit the data as well as the newer one. The fit quality is nearly everywhere worse, except near the Carina region where we know the standard model does not fit the data very well (App.~\ref{appendix_diffuse_params}).
From that point of view we may expect these estimates of the systematic uncertainties to be upper limits.
So we regard the values as qualitative estimates.
In the Galactic plane (and even worse in the Galactic ridge) the systematic uncertainties coming from the diffuse model are larger than the statistical ones. In the Galactic ridge, even the bias is larger than the statistical uncertainty. The effect is larger than what we estimated for 2FGL (even though the diffuse model has improved), partly because the exposure is twice as deep and partly because the new alternative model is further from the standard one.
Outside the Galactic plane the systematic uncertainty due to the diffuse model remains less than the statistical one, and the bias is negligible.

The same comparison also allows flagging outliers as suspect (\S~\ref{catalog_analysis_flags}). 119 sources received Flag 1 (Table~\ref{tab:flags}) because they ended up with $TS < 25$ with the alternative model, and 118 received Flag 3, indicating that their photon or energy fluxes changed by more than $3\sigma$.
That uncertainty also appears in Flag 4 whereby we flag all sources with source-to-background ratio less than 10\% in all bands in which they are statistically significant.

\subsubsection{Analysis method}
\label{catalog_analysismethod}

The check presented in this section is new to 3FGL. As explained in \S~\ref{catalog_detection} the $pointlike$-based method used to detect and localize sources also provides an estimate of the source spectra and significance. Therefore we use it to estimate systematic errors due to the analysis itself. Many aspects differ between the two methods: the code, the RoIs, the Earth limb representation. The alternative method does not remove faint sources (with $TS < 25$) from the model. The diffuse model is the same spatially but it was rescaled spectrally in each energy bin. The $pointlike$-based method also rescales $\log \mathcal{L}$ in order to play down the energy bins in which the source-to-background ratio is low.

The procedure to compare the results is the same as in \S~\ref{catalog_diffusemodel}. We list the induced biases and scatters of flux and spectral index in Table~\ref{tbl:alternative}.
In general, the effect of changing the analysis procedure is less than changing the diffuse model. Outside the Galactic ridge (and even outside the Galactic plane), we observe a negative bias on flux and index (i.e. fainter harder sources with the $pointlike$ pipeline) close to half the statistical error.
That effect is probably the result of removing the sources below threshold in the standard method. This favors absorbing the flux of faint neighbors at low energy (where the PSF is broad), resulting in somewhat brighter and softer sources.

A total of 118 sources received Flag 1 ($TS < 25$ with $pointlike$), and 101 received Flag 3 (flux changed by more than $3\sigma$).
Only 25 (Flag 1) and 19 (Flag 3) sources are flagged from both the diffuse model and the analysis method comparisons. In other words, the 3FGL catalog is more or less half way between the result from $pointlike$ and the result with the alternative diffuse model. Comparing the lists from $pointlike$ and the alternative diffuse model would result in 202 sources with Flag 1 and 209 with Flag 3.

    \subsection{Sources Toward Local Interstellar Clouds and the Galactic Ridge}
\label{catalog_ism}

As we did for the 2FGL catalog, we carefully evaluated which sources are potentially artifacts due to systematic uncertainties in modeling the Galactic diffuse emission.  The procedure, described in more detail in the 2FGL paper, flags unassociated sources with moderate $TS$ and spectral index $\Gamma > 2$, corresponding to features in individual gas components. For 3FGL we did not consider sources that have very curved spectra to be artifacts.
Very soft sources with power-law spectra are instead more likely to be problematic.  Sources considered to be potential artifacts are assigned an analysis flag in the catalog (\S~\ref{catalog_analysis_flags}).  We also append \texttt{c} to the source names.

Relative to the 2FGL catalog, far fewer \texttt{c} sources are flagged here (78 here vs. 162 for 2FGL) despite the much greater number of sources overall in the 3FGL catalog.  Away from the Galactic plane, the reduction of \texttt{c} sources is primarily due to improvement of the representation of the dark gas component of the Galactic diffuse emission model in the vicinity of massive star-forming regions (\S~\ref{DiffuseModel}).  At low latitudes, the reduction primarily is due to relaxing the criterion on unassociated sources with very curved spectra.

\begin{figure*}
   \centering
   \includegraphics[width=6in]{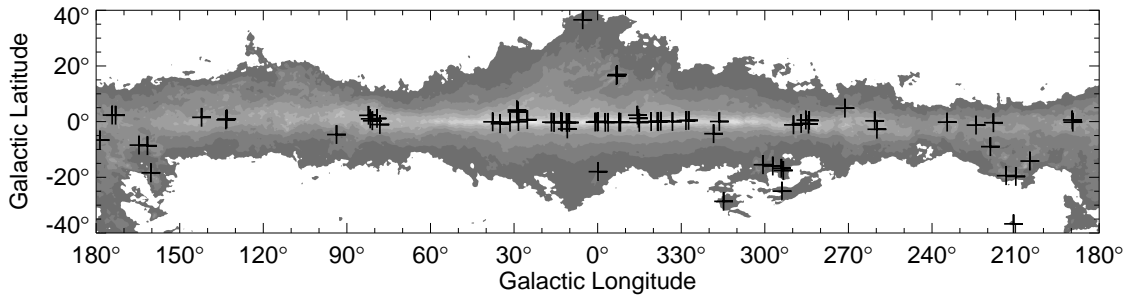}
   \caption{Locations of the \texttt{c} sources in the 3FGL catalog overlaid on a grayscale representation of the model
   for the Galactic diffuse $\gamma$-ray emission used for the 3FGL analysis (see \S~\ref{DiffuseModel}).  The plotted
   symbols are centered on the locations of the sources.  The model diffuse intensity is shown for 1~GeV and the spacing of the levels is logarithmic from 1\% to 100\% of the peak intensity.}

   \label{fig:csources}
\end{figure*}

Figure~\ref{fig:csources} shows the locations of the \texttt{c} sources for 3FGL.  The majority are close to the Galactic plane, where the diffuse $\gamma$-ray emission is brightest and very structured.  Clusters are apparent in regions where spiral arms of the Milky Way are viewed essentially tangentially, in particular the Cygnus ($l \sim 80\degr$) and Carina ($l \sim 285\degr$) regions where the systematic uncertainties of the Galactic diffuse emission model are especially large.  None of the \texttt{c} sources is identified (\S~\ref{source_assoc_firm}) and 63 ($\sim$80\%) have no firm association with a counterpart at other wavelengths, a much larger fraction than the overall average ($\sim$30\%) for 3FGL (Table~\ref{tab:classes}).

    \subsection{Analysis Flags}
\label{catalog_analysis_flags}

\begin{deluxetable}{cl}

\tablecaption{Definitions of the Analysis Flags \label{tab:flags}}
\tablehead{
\colhead{Flag\tablenotemark{a}} & 
\colhead{Meaning}
}

\startdata
  1  & Source with $TS > 35$ which went to $TS < 25$ when changing the diffuse model \\
     & (\S~\ref{catalog_diffusemodel}) or the analysis method (\S~\ref{catalog_analysismethod}). Sources with $TS \le 35$ are not flagged \\
     & with this bit because  normal statistical fluctuations can push them to $TS < 25$. \\
  2  & Not used. \\
  3  & Flux ($>$ 1~GeV) or energy flux ($>$ 100~MeV) changed by more than $3\sigma$ when \\
     & changing the diffuse model or the analysis method. Requires also that the flux \\ 
     & change by more than 35\% (to not flag strong sources). \\
  4  & Source-to-background ratio less than 10\% in highest band in which $TS > 25$. \\
     & Background is integrated over $\pi r_{68}^2$ or 1 square degree, whichever is smaller.\\
  5  & Closer than $\theta_{\rm ref}$ from a brighter neighbor.  $\theta_{\rm ref}$ is defined in the highest band in \\
     & which source $TS > 25$, or the band with highest $TS$ if all are $< 25$. $\theta_{\rm ref}$ is set \\
     & to $2\fdg17$ (FWHM) below 300~MeV, $1\fdg38$ between 300~MeV and 1~GeV, $0\fdg87$ \\
     & between 1~GeV and 3~GeV, $0\fdg67$ between 3 and 10~GeV and $0\fdg45$ above \\
     & 10~GeV ($2 \, r_{68}$). \\
  6  & On top of an interstellar gas clump or small-scale defect in the model of \\
     & diffuse emission; equivalent to the \texttt{c} designator in the source name (\S~\ref{catalog_ism}). \\
  7  & Unstable position determination; result from {\it gtfindsrc} outside the 95\% ellipse \\
     & from {\it pointlike}. \\
  8  & Not used. \\
  9  & Localization Quality $>$ 8 in {\it pointlike} (\S~\ref{catalog_detection}) or long axis of 95\% ellipse $> 0\fdg25$. \\
 10  & Spectral Fit Quality $> 16.3$ \citep[Eq. 3 of][2FGL]{LAT12_2FGL}. \\
 11  & Possibly due to the Sun (\S~\ref{catalog_variability}). \\
 12  & Highly curved spectrum; \texttt{LogParabola} $\beta$ fixed to 1 or \texttt{PLExpCutoff} \\
     & \texttt{Spectral\_Index} fixed to 0.5 (see \S~\ref{catalog_spectral_shapes}). \\
 \enddata
 
 \tablenotetext{a}{In the FITS version the values are encoded as individual bits in a single column, with Flag $n$ having value $2^{(n-1)}$.  For information about the FITS version of the table see Table~\ref{tab:columns} in  App.\ref{appendix_fits_format}.}

\end{deluxetable}

As in 2FGL we identified a number of conditions
that should be considered cautionary regarding the reality of a source or the magnitude of the systematic uncertainties of its measured properties. 
They are described in Table~\ref{tab:flags}.

Each flag has the same definition as for the 2FGL catalog, except for Flag 7, which was unused in that catalog.

Flags 1 to 12 have similar intent as in 2FGL, but differ in detail:
\begin{itemize}
\item Flags 1 and 3 are now applied not only when a source is sensitive to changing the diffuse model (\S~\ref{catalog_diffusemodel}) but also to the analysis method (\S~\ref{catalog_analysismethod}).
\item Flag 2 is not used. We didn't go so far as to rerun the full detection and localization procedure (\S~\ref{catalog_detection}) with the alternative diffuse model. Assessing the changes in source characteristics is normally enough.
\item For Flag 4, we lowered the threshold for flagging the source-to-background ratio to 10\%, recognizing that the uncertainties in the interstellar emission model are now reduced (App.~\ref{appendix_diffuse_params}).
\item We reinstated Flag 7 (comparison between $pointlike$ and $gtfindsrc$ localizations) which was not used in 2FGL because of an inconsistency in the unbinned likelihood results. It indicates sources for which the source locations derived from $pointlike$ (\S~\ref{catalog_localization}) and $gtfindsrc$ are inconsistent at the 95\% confidence level. $gtfindsrc$ was applied only above 3~GeV due to computing time constraints. This is appropriate for most sources (because the PSF is much better at high energy) but does not allow testing the localization of soft sources.
\item Flag 8 has been merged into Flag 9. Both tested localization reliability.
\item Flag 11 is deprecated because we put in place an explicit time-dependent model for the Sun and Moon emission (\S~\ref{DiffuseModel}).
\end{itemize}


\section{The 3FGL Catalog}
\label{3fgl_description}

We present a basic description of the 3FGL catalog in \S~\ref{catalog_description}, including a listing of the main table contents and some of the primary properties of the sources in the catalog.  We present a detailed comparison of the 3FGL catalog with the 2FGL catalog in \S~\ref{fgl_comparison}.

\begin{figure*}
   \centering
      \includegraphics[width=0.99\textwidth]{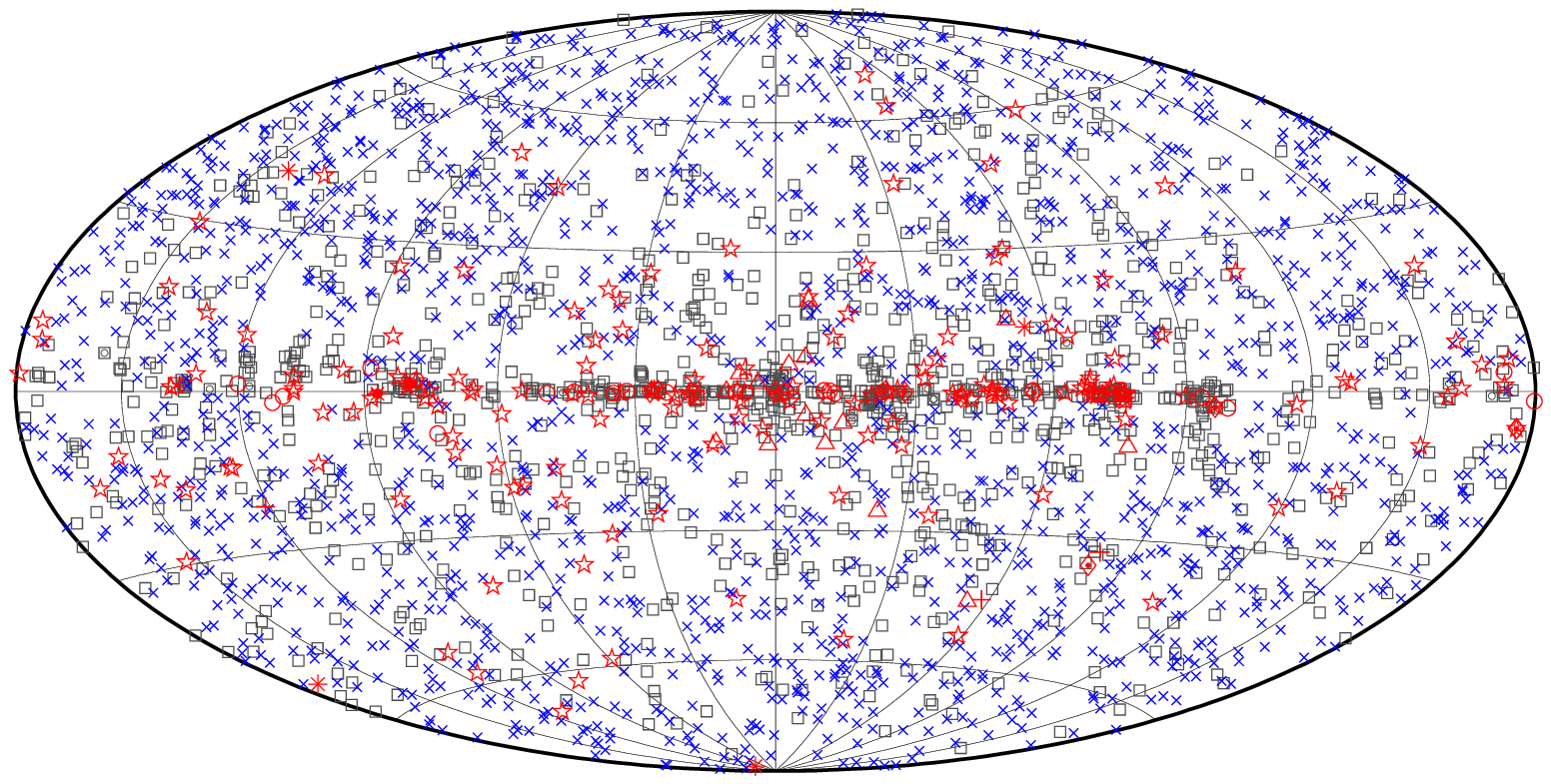}
      \includegraphics[width=0.80\textwidth]{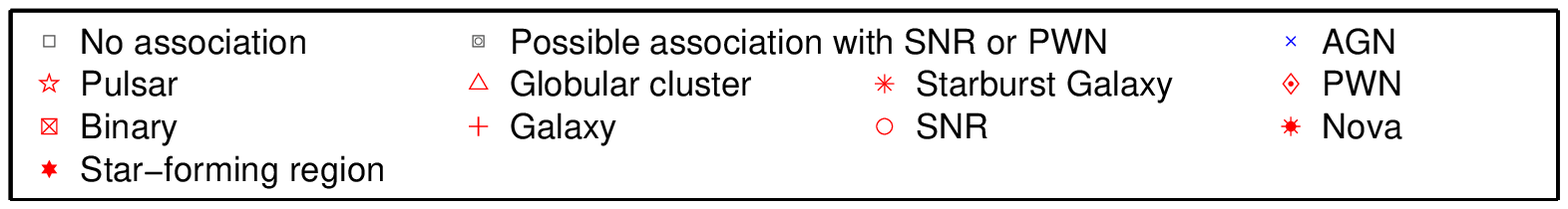}
      \includegraphics[width=0.99\textwidth]{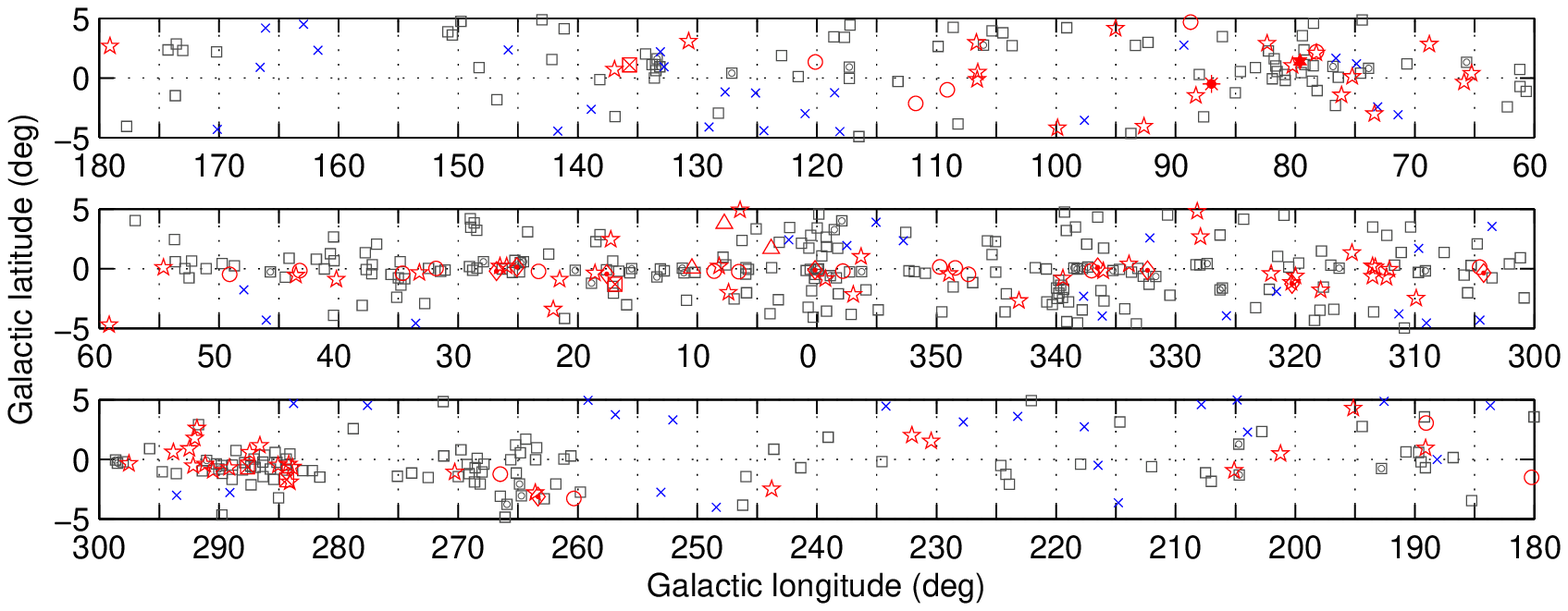}
   \caption{Full sky map (top) and blow-up of the inner Galactic region
  (bottom) showing sources by source class (see
  Table~\ref{tab:classes}). All AGN classes are plotted with the same symbol for simplicity.}
   \label{fig:map_id_assoc}
\end{figure*}

\subsection{Catalog Description}
\label{catalog_description}

Table~\ref{tab:sources} is the catalog, with information for each of the 3033 sources\footnote{Table~\ref{tab:sources} has 3034 entries because the PWN component of the Crab nebula is represented by two cospatial sources (\S~\ref{catalog_spectral_shapes}).}; see Table~\ref{tab:desc} for descriptions of the columns.  The source designation is \texttt{3FGL JHHMM.m+DDMM} where the \texttt{3} indicates that this is the third LAT catalog, \texttt{FGL} represents {\it Fermi} Gamma-ray LAT.  Sources close to the Galactic ridge and some nearby interstellar cloud complexes are assigned names of the form \texttt{3FGL JHHMM.m+DDMMc}, where the \texttt{c} indicates that caution should be used in interpreting or analyzing these sources.  Errors in the model of interstellar diffuse emission, or an unusually high density of sources, are likely to affect the measured properties or even existence of these 78 sources (see \S~\ref{catalog_ism}).  In addition a set of analysis flags has been defined to indicate sources with unusual or potentially problematic characteristics (see \S~\ref{catalog_analysis_flags}).  The \texttt{c} designator is encoded as one of these flags.  An additional 572 sources have one or more of the other analysis flags set.  The 25 sources that were modeled as extended for 3FGL (\S~\ref{catalog_extended}) are singled out by an \texttt{e} appended to their names.

The designations of the classes that we use to categorize the 3FGL sources are listed in Table~\ref{tab:classes} along with the numbers of sources assigned to each class. Figure~\ref{fig:map_id_assoc} illustrates where the source classes are in the sky. We distinguish between associated and identified sources, with associations depending primarily on close positional correspondence (see \S~\ref{source_assoc_automated}) and identifications requiring measurement of correlated variability at other wavelengths or characterization of the 3FGL source by its angular extent (see \S~\ref{source_assoc_firm}).  In the cases of multiple associations with a 3FGL source, we adopt the single association that is statistically most likely to be true if it is above the confidence threshold (see \S~\ref{source_assoc_automated}).
Sources associated with SNRs are often also associated with PWNe and pulsars, and the SNRs themselves are often not point-like.  We do not attempt to distinguish among the possible classifications and instead list in Table~\ref{tab:snrext} plausible associations of each class for unidentified 3FGL sources found to be positionally associated with SNRs\footnote{Four sources positionally associated with SNRs were also found to be associated with blazars.  We cannot quantitatively compare association probabilities between the blazar and the (spatially extended) SNR classes.  In the 3FGL catalog, we list only the blazar associations for them.  The sources and SNR associations are 3FGL J0217.3+6209 (G137.2+01.3), 3FGL J0223.5+6313 (G132.7+01.3), 3FGL J0526.0+4253 (G166.0+04.3), and 3FGL J0215.6+3709 (G074.9+01.2).}.  The Crab pulsar and PWN are represented by a total of three entries, two of which (designated \texttt{i} and \texttt{s}) represent spectral components of the PWN (see \S~\ref{source_assoc_firm}).  We consider these three entries to represent two sources. 

The photon flux for 1--100~GeV ($F_{35}$) and the energy flux for 100~MeV to 100~GeV in Table~\ref{tab:sources} are evaluated from the fit to the full band (see \S~\ref{catalog_flux_determination}). 
We do not present the integrated photon flux for 100~MeV to 100~GeV (see \S~\ref{catalog_flux_determination}). Table~\ref{tab:fluxes} presents the fluxes in individual bands as defined in \S~\ref{catalog_flux_determination}.

\begin{deluxetable}{lrrrrrrrrrrrrrcccclccp{3cm}}
\setlength{\tabcolsep}{0.02in}
\tabletypesize{\scriptsize}
\rotate
\tablewidth{0pt}
\tablecaption{LAT 4-year Catalog\label{tab:sources}}
\tablehead{
\colhead{Name 3FGL} &
\colhead{R.A.} &
\colhead{Decl.} &
\colhead{$l$} &
\colhead{$b$} &
\colhead{$\theta_{\rm 1}$} &
\colhead{$\theta_{\rm 2}$} &
\colhead{$\phi$} &
\colhead{$\sigma$} &
\colhead{$F_{35}$} &
\colhead{$\Delta F_{35}$} &
\colhead{$S_{25}$} &
\colhead{$\Delta S_{25}$} &
\colhead{$\Gamma_{25}$} &
\colhead{$\Delta \Gamma_{25}$} &
\colhead{Mod} &
\colhead{Var} &
\colhead{Flags} &
\colhead{$\gamma$-ray Assoc.} &
\colhead{TeV} &
\colhead{Class\tablenotemark{a}} &
\colhead{ID or Assoc.}
}
\startdata
 J0000.1+6545 &   0.038 & 65.752 & 117.694 & 3.403 & 0.102 & 0.078 & 41 &      6.8 &     1.0 &   0.2 &   13.6 &    2.1 & 2.41 & 0.08 & PL & \nodata & 3 & 2FGL J2359.6+6543c        & \nodata & \nodata & \nodata\\
 J0000.2$-$3738 &   0.061 & $-$37.648 & 345.411 & $-$74.947 & 0.073 & 0.068 & $-$89 &      5.1 &     0.2 &   0.1 &    2.4 &    0.7 & 1.87 & 0.18 & PL & \nodata & \nodata & \nodata & \nodata & \nodata & \nodata\\
 J0001.0+6314 &   0.254 & 63.244 & 117.293 & 0.926 & 0.248 & 0.160 & $-$65 &      6.2 &     0.6 &   0.1 &   13.0 &    1.9 & 2.73 & 0.11 & PL & \nodata & 3,4,5 & 2FGL J2358.9+6325         & \nodata & spp & \nodata\\
 J0001.2$-$0748 &   0.321 & $-$7.816 &  89.022 & $-$67.324 & 0.082 & 0.070 & $-$19 &     11.3 &     0.7 &   0.1 &    7.8 &    0.9 & 2.15 & 0.09 & PL & \nodata & \nodata & 2FGL J0000.9$-$0748         & \nodata & bll   & PMN J0001$-$0746            \\
  &  &  &  &  &  &  &  &  &  &  &  &  &  &  &  &  &  & 1FGL J0000.9$-$0745         &  & \\
 J0001.4+2120 &   0.361 & 21.338 & 107.665 & $-$40.047 & 0.211 & 0.188 & $-$33 &     11.4 &     0.3 &   0.1 &    8.1 &    0.8 & 2.78 & \nodata & LP & T & \nodata & 3EG J2359+2041                & \nodata & fsrq  & TXS 2358+209              \\
 J0001.6+3535 &   0.404 & 35.590 & 111.661 & $-$26.188 & 0.213 & 0.167 & 8 &      4.2 &     0.3 &   0.1 &    3.4 &    0.8 & 2.35 & 0.19 & PL & \nodata & 4 & \nodata & \nodata & \nodata & \nodata\\
 J0002.0$-$6722 &   0.524 & $-$67.370 & 310.139 & $-$49.062 & 0.102 & 0.086 & 69 &      5.9 &     0.3 &   0.1 &    3.3 &    0.8 & 1.95 & 0.16 & PL & \nodata & \nodata & \nodata & \nodata & \nodata & \nodata\\
 J0002.2$-$4152 &   0.562 & $-$41.883 & 334.070 & $-$72.143 & 0.217 & 0.140 & 68 &      5.2 &     0.3 &   0.1 &    3.0 &    0.7 & 2.09 & 0.19 & PL & \nodata & \nodata & 2FGL J0001.7$-$4159         & \nodata & bcu   & 1RXS J000135.5$-$415519     \\
  &  &  &  &  &  &  &  &  &  &  &  &  &  &  &  &  &  & 1FGL J0001.9$-$4158         &  & \\
 J0002.6+6218 &   0.674 & 62.301 & 117.302 & $-$0.037 & 0.061 & 0.054 & $-$55 &     18.0 &     2.8 &   0.2 &   18.4 &    1.7 & 2.35 & \nodata & LP & \nodata & \nodata & 2FGL J0002.7+6220         & \nodata & \nodata & \nodata\\
 J0003.2$-$5246 &   0.815 & $-$52.777 & 318.976 & $-$62.825 & 0.070 & 0.061 & $-$44 &      5.7 &     0.3 &   0.1 &    3.0 &    0.8 & 1.90 & 0.17 & PL & \nodata & \nodata & \nodata & \nodata & bcu   & RBS 0006                  \\
 J0003.4+3100 &   0.858 & 31.008 & 110.964 & $-$30.745 & 0.181 & 0.163 & 13 &      6.3 &     0.3 &   0.1 &    4.9 &    0.8 & 2.55 & 0.13 & PL & \nodata & \nodata & \nodata & \nodata & \nodata & \nodata\\
 J0003.5+5721 &   0.890 & 57.360 & 116.486 & $-$4.912 & 0.089 & 0.072 & 1 &      5.4 &     0.5 &   0.1 &    5.4 &    1.1 & 2.18 & 0.13 & PL & \nodata & \nodata & \nodata & \nodata & \nodata & \nodata\\
\enddata
\tablenotetext{a}{See Table~\ref{tab:classes} for class designators.}
\tablecomments{This table is published in its entirety in the electronic edition of the Astrophysical Journal Supplements. A portion is shown here for guidance regarding its form and content.}
\end{deluxetable}

\begin{deluxetable}{ll}
\setlength{\tabcolsep}{0.04in}
\tabletypesize{\scriptsize}
\tablecaption{LAT Third Catalog Description\label{tab:desc}}
\tablehead{
\colhead{Column} &
\colhead{Description} 
}
\startdata
Name &   \texttt{3FGL JHHMM.m+DDMM[c/e/i/s]}, constructed according to IAU Specifications for Nomenclature;  \texttt{m} is decimal\\
  &  minutes of R.A.; in the name, R.A. and Decl. are truncated at 0.1 decimal minutes and 1\arcmin, respectively; \\
  &  \texttt{c} indicates that based on the region of the sky the source is considered to be potentially confused \\
  & with Galactic diffuse emission; \texttt{e} indicates a source that was modeled as spatially extended (see \S~\ref{catalog_extended}); \\
  & the two spectral components of the Crab PWN are designated \texttt{i} and \texttt{s} \\
R.A. & Right Ascension, J2000, deg, 3 decimal places  \\
Decl. & Declination, J2000, deg, 3 decimal places \\
 $l$  & Galactic Longitude, deg, 3 decimal places \\
 $b$  & Galactic Latitude, deg, 3 decimal places \\
 $\theta_1$ & Semimajor radius of 95\% confidence region, deg, 3 decimal places\\
 $\theta_2$ & Semiminor radius of 95\% confidence region, deg, 3 decimal places\\
 $\phi$ & Position angle of 95\% confidence region, deg. East of North, 0 decimal places\\
$\sigma$ & Significance derived from likelihood Test Statistic for 100~MeV--300~GeV analysis, 1 decimal place \\
 $F_{35}$ &  Photon flux for 1~GeV--100~GeV, 10$^{-9}$ ph cm$^{-2}$ s$^{-1}$, summed over 3 bands, 1 decimal place \\
$\Delta F_{35}$ & $1\sigma$ uncertainty on  $F_{35}$, same units and precision \\
$S_{25}$  &  Energy flux for 100~MeV--100~GeV, 10$^{-12}$ erg cm$^{-2}$ s$^{-1}$, from power-law fit, 1 decimal place \\
 $\Delta S_{25}$ & $1\sigma$ uncertainty on $S_{25}$, same units and precision \\
$\Gamma_{25}$  & Photon number power-law index, 100~MeV--100~GeV, 2 decimal places \\
 $\Delta \Gamma_{25}$ & $1\sigma$ uncertainty of photon number power-law index, 100~MeV--100~GeV, 2 decimal places \\
Mod. &  \texttt{PL} indicates power-law fit to the energy spectrum;  \texttt{LP} indicates log-parabola fit to the energy spectrum;\\
    &     \texttt{EC} indicates power-law with exponential cutoff fit to the energy spectrum \\
 Var. &  \texttt{T} indicates $<$~1\% chance of being a steady source; see note in text  \\
Flags & See Table~\ref{tab:flags} for definitions of the flag numbers  \\
$\gamma$-ray Assoc.  & Positional associations with 0FGL, 1FGL, 2FGL, 3EG, EGR, or 1AGL sources \\
TeV & Positional association with a TeVCat source,  \texttt{P} for unresolved angular size,  \texttt{E} for extended \\
 Class & Like `ID' in 3EG catalog, but with more detail (see Table \ref{tab:classes}).  Capital letters indicate firm identifications;\\
  &  lower-case letters indicate associations \\
 ID or Assoc.  & Designator of identified or associated source \\
\enddata
\end{deluxetable}

\begin{deluxetable}{lcrcr}
\setlength{\tabcolsep}{0.04in}
\tablewidth{0pt}
\tabletypesize{\scriptsize}
\tablecaption{LAT 3FGL Source Classes \label{tab:classes}}
\tablehead{
\colhead{Description} & 
\multicolumn{2}{c}{Identified} &
\multicolumn{2}{c}{Associated} \\
& 
\colhead{Designator} &
\colhead{Number} &
\colhead{Designator} &
\colhead{Number}
}
\startdata
Pulsar, identified by pulsations & PSR & 143 & \nodata & \nodata \\
Pulsar, no pulsations seen in LAT yet & \nodata & \nodata & psr & 24 \\
Pulsar wind nebula & PWN & 9 & pwn & 2 \\
Supernova remnant & SNR & 12 & snr & 11 \\
Supernova remnant / Pulsar wind nebula & \nodata & \nodata & spp  & 49 \\
Globular cluster & GLC & 0 & glc & 15 \\
High-mass binary & HMB & 3 & hmb & 0 \\
Binary & BIN & 1 & bin & 0 \\
Nova & NOV & 1 & nov & 0 \\
Star-forming region & SFR & 1 & sfr & 0 \\
Compact Steep Spectrum Quasar & CSS & 0 & css & 1 \\
BL Lac type of blazar & BLL & 18 & bll & 642 \\ 
FSRQ type of blazar & FSRQ & 38 & fsrq & 446 \\
Non-blazar active galaxy & AGN & 0 & agn & 3 \\ 
Radio galaxy & RDG & 3 & rdg & 12 \\
Seyfert galaxy & SEY & 0 & sey & 1 \\
Blazar candidate of uncertain type &  BCU & 5 & bcu & 568 \\ 
Normal galaxy (or part) & GAL & 2 & gal &  1 \\
Starburst galaxy & SBG & 0 & sbg & 4 \\
Narrow line Seyfert 1 &NLSY1 & 2 & nlsy1 & 3 \\
Soft spectrum radio quasar &SSRQ & 0 & ssrq & 3 \\
Total & \nodata & 238 & \nodata & 1785 \\
\hline
Unassociated & \nodata & \nodata & \nodata & 1010\ 
\enddata
\tablecomments{The designation `spp' indicates potential association with SNR or PWN (see Table~\ref{tab:snrext}).  Designations shown in capital letters are firm identifications; lower case letters indicate associations. In the case of AGN, many of the associations have high confidence.  
Among the pulsars, those with names beginning with LAT were discovered with the LAT.}
\end{deluxetable}

\begin{deluxetable}{llllc}
\setlength{\tabcolsep}{0.04in}
\tabletypesize{\scriptsize}
\tablecaption{Potential Associations for Sources Near SNRs\label{tab:snrext}}
\tablehead{
\colhead{Name 3FGL} &
\colhead{SNR name} &
\colhead{PWN name} &
\colhead{TeV name} &
\colhead{Common name}
}
\startdata
 J0001.0+6314 & G116.5+01.1 &  \nodata  &  \nodata  & \nodata \\
 J0128.4+6257 & G127.1+00.5 &  \nodata  &  \nodata  & R5 \\
 J0220.1+6202c & G132.7+01.3 &  \nodata  &  \nodata  & HB3 \\
 J0224.0+6235 & G132.7+01.3 &  \nodata  &  \nodata  & HB3 \\
 J0500.3+5237 & G156.2+05.7 &  \nodata  &  \nodata  & \nodata \\
 J0610.6+1728 & G192.8$-$01.1 &  \nodata  &  \nodata  & PKS 0607+17 \\
 J0631.6+0644 & G205.5+00.5 &  \nodata  &  \nodata  & Monoceros Loop \\
 J0640.9+0752 & G205.5+00.5 &  \nodata  &  \nodata  & Monoceros Loop \\
 J0838.1$-$4615\tablenotemark{a} & G263.9$-$03.3 &  \nodata  &  \nodata  & Vela  \\
 J0839.1$-$4739 & G263.9$-$03.3 &  \nodata  &  \nodata  & Vela  \\
 J0843.1$-$4546 & G263.9$-$03.3 &  \nodata  &  \nodata  & Vela  \\
 J1101.9$-$6053 & G290.1$-$00.8 &  \nodata  &  \nodata  & MSH 11$-$61A \\
 J1111.9$-$6038 & G291.0$-$00.1 & G291.0$-$0.1 &  \nodata  & MSH 11$-$62 \\
 J1209.1$-$5224 & G296.5+10.0 &  \nodata  &  \nodata  & PKS 1209$-$51 \\
 J1212.2$-$6251 & G298.5$-$00.3 &  \nodata  &  \nodata  & \nodata \\
 J1214.0$-$6236 & G298.6$-$00.0 &  \nodata  &  \nodata  & \nodata \\
 J1345.1$-$6224 & G308.8$-$00.1 &  \nodata  &  \nodata  & \nodata \\
 J1441.5$-$5955c & G316.3$-$00.0 &  \nodata  &  \nodata  & MSH 14$-$57 \\
 J1549.1$-$5347c & G327.4+00.4 &  \nodata  &  \nodata  & \nodata \\
 J1551.1$-$5610 & G326.3$-$01.8 &  \nodata  &  \nodata  & Kes 25 \\
 J1552.9$-$5610 & G326.3$-$01.8 &  \nodata  &  \nodata  & Kes 25 \\
 J1615.3$-$5146e &  \nodata  & HESS J1614$-$518 & \nodata \\
 J1628.9$-$4852 & G335.2+00.1 &  \nodata  &  \nodata  & \nodata \\
 J1636.2$-$4709c & G337.2+00.1 &  \nodata  & HESS J1634$-$472 & \nodata \\
 J1638.6$-$4654\tablenotemark{a} & G337.8$-$00.1 &  \nodata  &  \nodata  & Kes 41 \\
 J1640.4$-$4634c & G338.3$-$00.0 &  \nodata  & HESS J1640$-$465 & \nodata \\
 J1641.1$-$4619c & G338.5+00.1 &  \nodata  & HESS J1641$-$463 & \nodata \\
 J1645.9$-$5420 & G332.5$-$05.6 &  \nodata  &  \nodata  & \nodata \\
 J1722.9$-$4529 & G343.0$-$06.0 &  \nodata  &  \nodata  & RCW 114 \\
 J1725.1$-$2832 & G358.0+03.8 &  \nodata  &  \nodata  & \nodata \\
 J1728.0$-$4606 & G343.0$-$06.0 &  \nodata  &  \nodata  & RCW 114 \\
 J1729.5$-$2824 & G358.0+03.8 &  \nodata  &  \nodata  & \nodata \\
 J1737.3$-$3214c & G356.3$-$00.3 &  \nodata  &  \nodata  & \nodata \\
 J1745.1$-$3011 & G359.1$-$00.5 &  \nodata  & HESS J1745$-$303 & \nodata \\
 J1745.6$-$2859c & G000.0+00.0 & G359.95$-$0.04 & Galactic Centre & Sgr A East \\
 J1810.1$-$1910 & G011.1+00.1 &  \nodata  & HESS J1809$-$193 & \nodata \\
 J1811.3$-$1927c & G011.2$-$00.3 &  \nodata  & HESS J1809$-$193 & \nodata \\
 J1817.2$-$1739 & G013.3$-$01.3 &  \nodata  &  \nodata  & \nodata \\
 J1818.7$-$1528\tablenotemark{a} & G015.4+00.1 &  \nodata  &  \nodata  & \nodata \\
 J1828.4$-$1121\tablenotemark{a} & G020.0$-$00.2 &  \nodata  &  \nodata  & \nodata \\
 J1829.7$-$1304 & G018.9$-$01.1 &  \nodata  &  \nodata  & \nodata \\
 J1833.9$-$0711\tablenotemark{a} & G024.7+00.6 &  \nodata  &  \nodata  & \nodata \\
 J1834.6$-$0659 & G024.7+00.6 &  \nodata  &  \nodata  & \nodata \\
 J1840.1$-$0412 & G027.8+00.6 &  \nodata  &  \nodata  & \nodata \\
 J1915.9+1112 & G045.7$-$00.4 &  \nodata  &  \nodata  & \nodata \\
 J1951.6+2926 & G065.7+01.2 &  \nodata  &  \nodata  & \nodata \\
 J2014.4+3606 & G073.9+00.9 &  \nodata  &  \nodata  & \nodata \\
 J2022.2+3840 & G076.9+01.0 &  \nodata  &  \nodata  & \nodata \\
 J2225.8+6045 & G106.3+02.7 &  \nodata  & G106.3+2.7 & \nodata \\
\enddata
\tablenotetext{a}{These sources have been found to be significantly variable, i.e., \texttt{Variability\_Index} $> 72.44$ (\S~\ref{catalog_variability}), which would be unexpected for physical associations with SNRs or PWNe.}
\tablecomments{These sources are classified as spp in Table~\ref{tab:sources}.  They may be pulsars rather than the SNR or PWN named.  Four additional 3FGL sources are associated with both an SNR and a blazar.  For these the catalog lists the blazar associations; see text.}
\end{deluxetable}

\begin{deluxetable}{lrrrrrrrrrr}
\setlength{\tabcolsep}{0.077in}
\tablecolumns{11}
\tabletypesize{\scriptsize}
\tablecaption{LAT 4-year Catalog:  Spectral Information\label{tab:fluxes}}
\tablehead{
\colhead{} & \multicolumn{2}{c}{0.1--0.3 GeV} & \multicolumn{2}{c}{0.3--1 GeV} & \multicolumn{2}{c}{1--3 GeV} & \multicolumn{2}{c}{3--10 GeV} & \multicolumn{2}{c}{10--100 GeV} \\ \cline{2-3} \cline{4-5} \cline{6-7} \cline{8-9} \cline{10-11} \\
\colhead{Name 3FGL} &
\colhead{$F_{\rm 1}$\tablenotemark{a}} &
\colhead{$\sqrt{TS_{\rm 1}}$} &
\colhead{$F_{\rm 2}$\tablenotemark{a}} &
\colhead{$\sqrt{TS_{\rm 2}}$} &
\colhead{$F_{\rm 3}$\tablenotemark{b}} &
\colhead{$\sqrt{TS_{\rm 3}}$} &
\colhead{$F_{\rm 4}$\tablenotemark{c}} &
\colhead{$\sqrt{TS_{\rm 4}}$} &
\colhead{$F_{\rm 5}$\tablenotemark{c}} &
\colhead{$\sqrt{TS_{\rm 5}}$} 
}
\startdata
 J0000.1+6545  & $   1.81^{+   0.82}_{-   0.84}$ &    2.2 & $   0.69^{+   0.14}_{-   0.14}$ &    5.3 & $   1.24^{+   0.23}_{-   0.23}$ &    6.0 & $   0.58^{+   0.49}_{-   0.40}$ &    1.5 & $   0.28^{+   0.19}_{-   0.15}$ &    2.4\\
 J0000.2$-$3738  & $   0.01^{+   0.20}_{-   0.01}$ &    0.0 & $   0.01^{+   0.03}_{-   0.01}$ &    0.6 & $   0.15^{+   0.07}_{-   0.06}$ &    2.9 & $   0.75^{+   0.32}_{-   0.26}$ &    4.2 & $   0.16^{+   0.15}_{-   0.09}$ &    3.5\\
 J0001.0+6314  & $   2.91^{+   0.74}_{-   0.71}$ &    4.0 & $   0.47^{+   0.11}_{-   0.11}$ &    4.3 & $   0.30^{+   0.18}_{-   0.18}$ &    1.7 & $   1.16^{+   0.48}_{-   0.42}$ &    3.3 & $   0.02^{+   0.13}_{-   0.02}$ &    0.4\\
 J0001.2$-$0748  & $   0.41^{+   0.26}_{-   0.26}$ &    1.6 & $   0.21^{+   0.05}_{-   0.04}$ &    5.3 & $   0.52^{+   0.11}_{-   0.10}$ &    7.0 & $   2.11^{+   0.51}_{-   0.45}$ &    8.0 & $   0.10^{+   0.15}_{-   0.08}$ &    1.6\\
 J0001.4+2120  & $   1.52^{+   0.24}_{-   0.24}$ &    6.8 & $   0.36^{+   0.05}_{-   0.05}$ &    7.6 & $   0.36^{+   0.10}_{-   0.09}$ &    4.5 & $   0.00^{+   0.19}_{-   0.00}$ &    0.0 & $   0.00^{+   0.11}_{-   0.00}$ &    0.0\\
 J0001.6+3535  & $   0.95^{+   0.27}_{-   0.27}$ &    3.6 & $   0.04^{+   0.04}_{-   0.04}$ &    0.9 & $   0.17^{+   0.09}_{-   0.08}$ &    2.5 & $   0.40^{+   0.31}_{-   0.23}$ &    2.0 & $   0.23^{+   0.19}_{-   0.13}$ &    2.3\\
 J0002.0$-$6722  & $   0.05^{+   0.28}_{-   0.05}$ &    0.2 & $   0.06^{+   0.03}_{-   0.03}$ &    2.0 & $   0.20^{+   0.08}_{-   0.07}$ &    3.4 & $   0.83^{+   0.32}_{-   0.27}$ &    4.8 & $   0.28^{+   0.19}_{-   0.14}$ &    3.1\\
 J0002.2$-$4152  & $   0.53^{+   0.22}_{-   0.22}$ &    2.5 & $   0.06^{+   0.03}_{-   0.03}$ &    2.1 & $   0.09^{+   0.07}_{-   0.06}$ &    1.5 & $   0.98^{+   0.36}_{-   0.31}$ &    4.7 & $   0.26^{+   0.19}_{-   0.14}$ &    2.6\\
 J0002.6+6218  & $   1.05^{+   0.40}_{-   0.40}$ &    2.6 & $   0.80^{+   0.09}_{-   0.09}$ &    9.4 & $   2.44^{+   0.23}_{-   0.23}$ &   13.3 & $   3.58^{+   0.69}_{-   0.64}$ &    7.4 & $   0.00^{+   0.15}_{-   0.00}$ &    0.0\\
 J0003.2$-$5246  & $   0.66^{+   0.38}_{-   0.38}$ &    1.8 & $   0.00^{+   0.03}_{-   0.00}$ &    0.0 & $   0.18^{+   0.08}_{-   0.07}$ &    3.1 & $   0.79^{+   0.36}_{-   0.28}$ &    3.7 & $   0.30^{+   0.19}_{-   0.14}$ &    4.5\\
 J0003.4+3100  & $   0.86^{+   0.27}_{-   0.25}$ &    3.4 & $   0.15^{+   0.04}_{-   0.04}$ &    3.6 & $   0.36^{+   0.10}_{-   0.09}$ &    4.9 & $   0.20^{+   0.26}_{-   0.19}$ &    1.1 & $   0.01^{+   0.14}_{-   0.01}$ &    0.1\\
 J0003.5+5721  & $   0.00^{+   0.31}_{-   0.00}$ &    0.0 & $   0.23^{+   0.07}_{-   0.07}$ &    3.5 & $   0.49^{+   0.14}_{-   0.14}$ &    4.0 & $   0.92^{+   0.41}_{-   0.35}$ &    3.3 & $   0.24^{+   0.16}_{-   0.12}$ &    3.0\\
 J0003.8$-$1151  & $   0.01^{+   0.21}_{-   0.01}$ &    0.0 & $   0.05^{+   0.04}_{-   0.03}$ &    1.5 & $   0.23^{+   0.08}_{-   0.07}$ &    3.8 & $   0.47^{+   0.30}_{-   0.24}$ &    2.5 & $   0.18^{+   0.16}_{-   0.11}$ &    2.5\\
 J0004.2+6757  & $   0.59^{+   1.16}_{-   0.59}$ &    0.5 & $   0.56^{+   0.13}_{-   0.13}$ &    4.4 & $   0.59^{+   0.19}_{-   0.18}$ &    3.4 & $   1.15^{+   0.47}_{-   0.42}$ &    3.2 & $   0.09^{+   0.11}_{-   0.07}$ &    1.4\\
 J0004.2+0843  & $   0.23^{+   0.25}_{-   0.23}$ &    0.8 & $   0.03^{+   0.04}_{-   0.03}$ &    0.7 & $   0.12^{+   0.09}_{-   0.08}$ &    1.5 & $   1.16^{+   0.41}_{-   0.35}$ &    5.3 & $   0.08^{+   0.12}_{-   0.06}$ &    2.0\\
\enddata
\tablecomments{This table is published in its entirety in the electronic edition of the Astrophysical Journal Supplements.  A portion is shown here for guidance regarding its form and content.}
\tablenotetext{a}{In units of $10^{-8}$ photons cm$^{-2}$ s$^{-1}$}
\tablenotetext{b}{In units of $10^{-9}$ photons cm$^{-2}$ s$^{-1}$}
\tablenotetext{c}{In units of $10^{-10}$ photons cm$^{-2}$ s$^{-1}$}
\end{deluxetable}

\subsection{Comparison with 0FGL, 1FGL, 2FGL and 1FHL}
\label{fgl_comparison}


\subsubsection{General comparison}



We compare the 3FGL with previous catalogs released by the LAT collaboration. These are listed in Table~\ref{tab:latcat}.

\begin{deluxetable}{lcccccc}
\tablewidth{0pc} 
\setlength{\tabcolsep}{0.18cm}
\tabletypesize{\scriptsize}
\tablecaption{ Statistics of Sources in LAT Catalogs \label{tab:latcat}}
\tablehead{ 
\colhead{Category} &
\colhead{0FGL\tablenotemark{a}} &
 \colhead{1FGL } &
 \colhead{2FGL } &
 \colhead{1FHL\tablenotemark{b}} &
 \colhead{3FGL} &
}
\startdata 
Total & 205 & 1451 & 1873 & 514 &	3033 \\
High-Latitude sources & 132 & 1043 & 1319 & 399 & 2193 \\
Low-Latitude sources & 73 & 408 & 554 & 115 & 841 \\
`Lost' sources\tablenotemark{(c)} ~in 3FGL& 12 & 310 & 300 & 17 & - \\
\enddata 
\tablenotetext{a}{0FGL, the LAT Bright Source List, has a lower energy limit of 200 MeV and a significance threshold $TS > 100$.}
\tablenotetext{b}{1FHL is a catalog for the energy range $>$10 GeV.}
\tablenotetext{c}{Sources without a counterpart in 3FGL catalog, at the level of overlapping 95\% source location confidence contours. These sources are discussed in Table~\ref{tab:012fgl_3fgl}.}
\end{deluxetable} 


Associations among 3FGL and 0FGL/1FGL/2FGL and 1FHL sources are based on the following relation:
\be
\Delta  \le\  d_{x,a} = \sqrt{ \theta_{x,a}^2 +\theta_{x,{\rm 3FGL}}^2}
\label{eq:compnFGL}
 \ee

\noindent where  $\Delta$ is the angular distance between the sources, $d_{x}$ is defined in terms of the semi-major axis of the  $x$\% confidence error ellipse for the position of each source, e.g., the 95\% confidence error for the automatic source association procedure (\S~\ref{source_assoc_automated}) and `a' is alternatively 0FGL, 1FGL, 2FGL, and 1FHL. 
In total, 1720 3FGL sources were automatically associated with entries in either the 0FGL, 1FGL, 2FGL, or 1FHL catalogs. The statistics of the association results is reported in Table~\ref{tab:assoc_results}.


\begin{figure}[!ht]
\plotone{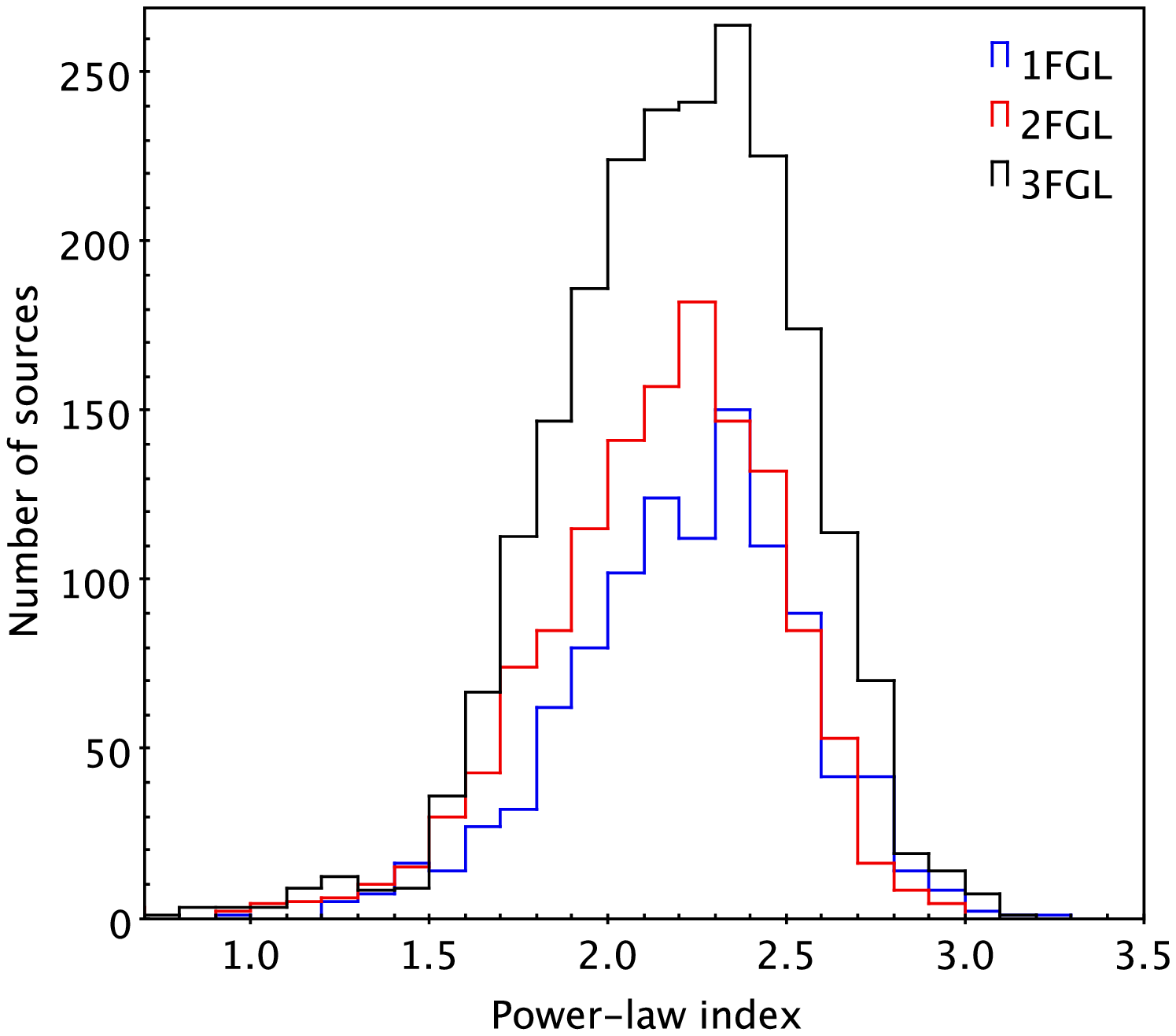}
\caption{Distributions of the spectral index for the high-latitude sources ($|b|>10\degr$) in 1FGL (1043 sources, blue line), 2FGL (1173 sources, red line) 3FGL (1960 sources, black line) catalogs. 2FGL and 3FGL samples include only power-law spectrum type.}
\label{fig:fig_spec1}
\end{figure}

In the 3FGL analysis the spectral fits are made using  power-law, power-law with an exponential cutoff, or log-parabola models (\S~\ref{catalog_spectral_shapes}). For each 2FGL and 3FGL source we also evaluated the spectral index ($\Gamma$) of the best power-law fit and this enables a comparison of the spectral characteristics of the 1FGL, 2FGL, and 3FGL sources. Figure~\ref{fig:fig_spec1} shows the distributions of the power-law indices of the sources at high Galactic latitude and only those with a power-law spectral type in the 1FGL, 2FGL and 3FGL catalogs, to avoid possible confusion from more complex features. The three distributions are very similar, with an average $\Gamma_{\rm 1FGL} = 2.23\pm0.01$, average $\Gamma_{\rm 2FGL} = 2.21\pm0.01$, and average $\Gamma_{\rm 3FGL} = 2.19\pm0.01$. 
However, the peaks of the three distributions are not exactly coincident; also, they have different skewnesses.  
The small differences in the power-law index distributions could be related to slightly different systematic uncertainties in the effective area between the instrument response functions P7REP\_SOURCE\_V15, P7SOURCE\_V6 and P6\_V3\_DIFFUSE used respectively for 3FGL, 2FGL, and 1FGL.

\begin{figure}[!ht]
\plotone{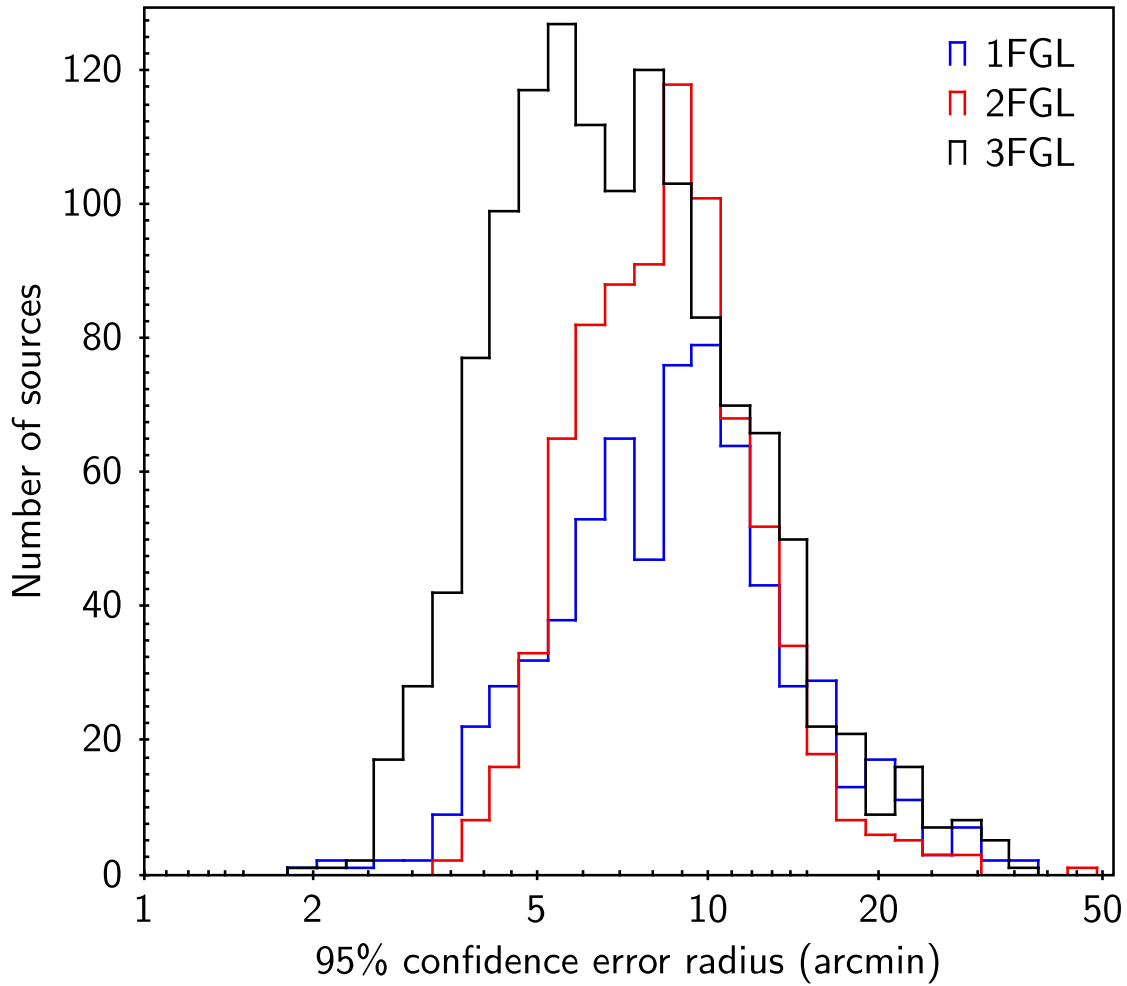}
\caption{Distributions of the 95\% confidence error radii for high-latitude sources ($|b|>10\degr$) with $25<TS<100$ in 1FGL (blue line), 2FGL (red line) and 3FGL (black line), illustrating the improvement of localizations for sources of equivalent detection significances.
}
\label{fig:fig_conf95_HL_ts100}
\end{figure}

We have compared the distribution of the 95\% confidence error radii of the 1FGL, 2FGL, and 3FGL sources at high Galactic latitude. 
The distribution of 95\% confidence error radius for those sources with $25<TS<100$ in any of the 1FGL, 2FGL, and 3FGL catalogs (Figure~\ref{fig:fig_conf95_HL_ts100}) shows the localization improvement for a given range of source detection significances.  We evaluated the 95\% confidence error radius as the geometric mean of the semi-major and semi-minor axes of the 95\% confidence error ellipse.


\begin{figure}[!ht]
\plotone{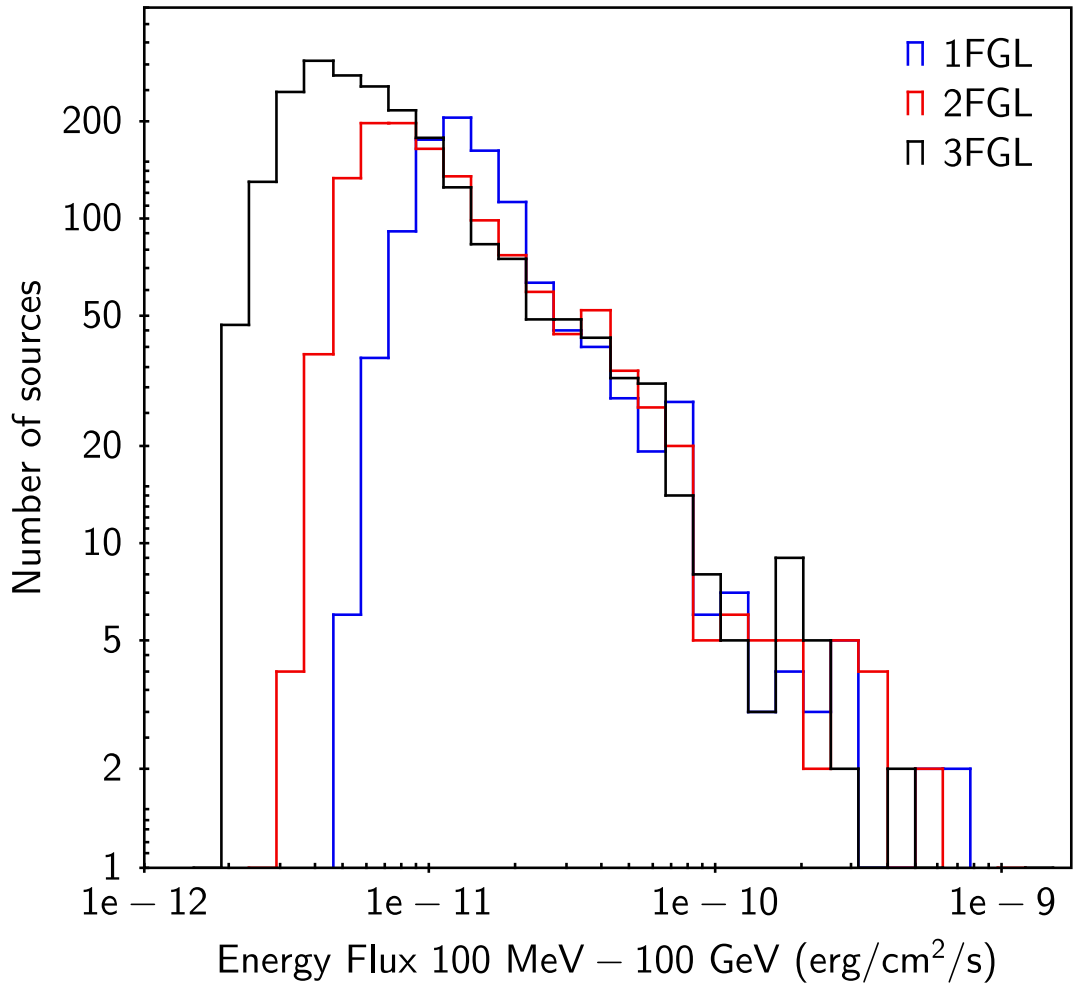}
\caption{Distributions of the energy flux for 1FGL (blue line), 2FGL (red line), and 3FGL (black line) sources at high Galactic latitude ($|b|>10\degr$).}
\label{fig:fig_enflux_123_HL}
\end{figure}

Figure~\ref{fig:fig_enflux_123_HL} shows the energy flux distribution in 1FGL, 2FGL, and 3FGL. Comparing the current flux threshold with those published in previous LAT Catalog papers we see that in 3FGL the threshold is down to $\simeq 3 \times 10^{-12}$ erg cm$^{-2}$ s$^{-1}$, from $\simeq 5 \times 10^{-12}$ erg cm$^{-2}$ s$^{-1}$ in 2FGL and $\simeq 8 \times 10^{-12}$ erg cm$^{-2}$ s$^{-1}$ in 1FGL. Above that flux the 2FGL and 3FGL distributions are entirely compatible.

However the 1FGL distribution shows a distinct bump between 1 and $2 \times 10^{-11}$ erg cm$^{-2}$ s$^{-1}$. That accumulation of fluxes was clearly incorrect. We attribute it primarily to overestimating significances and fluxes due to the unbinned likelihood bias in the 1FGL analysis, and also to the less accurate procedure then used to extract source flux  (see discussion in the 2FGL paper).

\subsubsection{Comparison of individual sources}



\begin{figure}[!ht]
\plotone{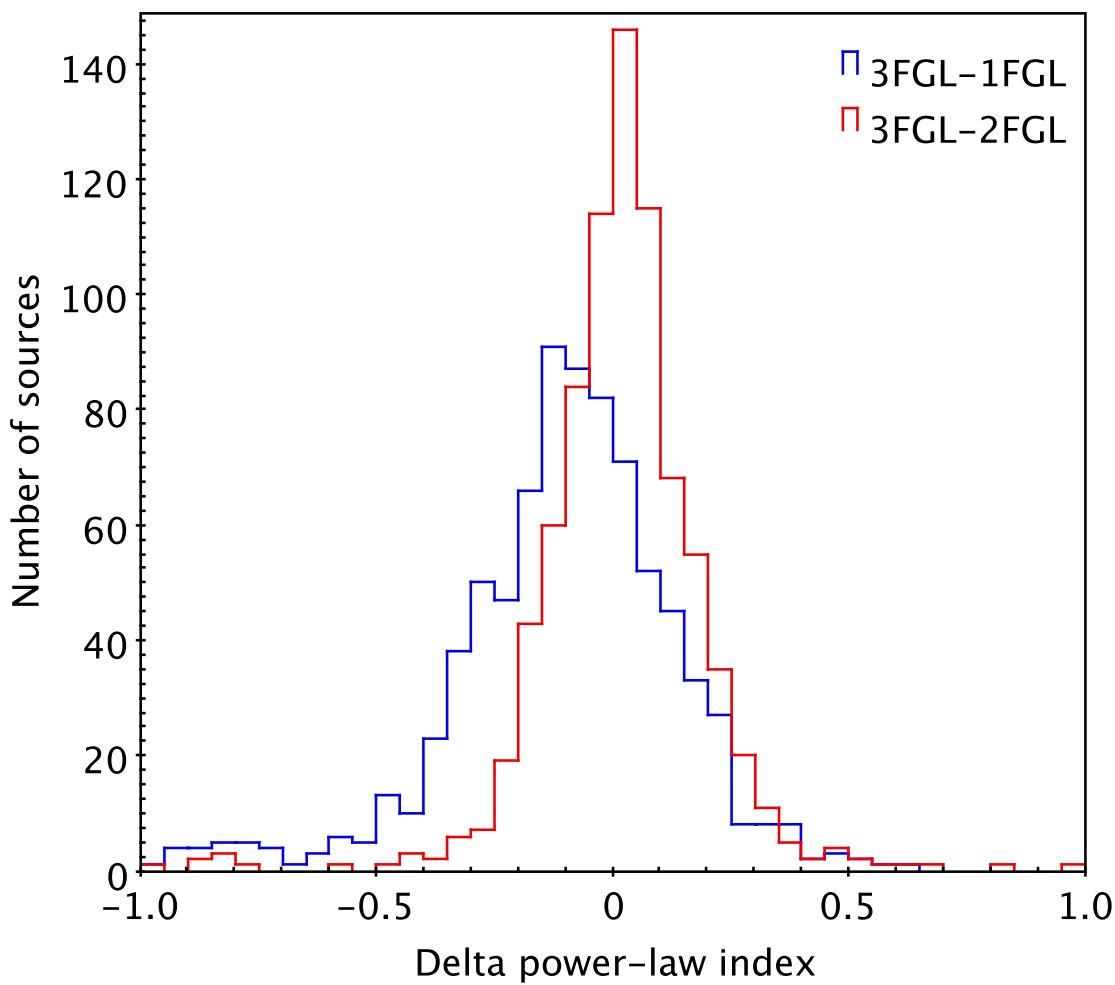}
\caption{Distribution of the differences $\Gamma_{\rm 3FGL} - \Gamma_{\rm 1FGL}$ (blue line) and $\Gamma_{\rm 3FGL} - \Gamma_{\rm 2FGL}$ (red line) for the 621 sources at high latitude ($|b|>10\degr$) in common among the 1FGL, 2FGL and 3FGL catalogs. For the 2FGL and 3FGL samples only power-law spectrum type sources have been considered.}
\label{fig:fig_spec}
\end{figure}

Figure~\ref{fig:fig_spec} shows the distribution of the differences $\Gamma_{\rm 3FGL}-\Gamma_{\rm 2FGL}$ and $\Gamma_{\rm 3FGL}-\Gamma_{\rm 1FGL}$ for the 621 high-latitude sources with power-law spectrum type in common among the three catalogs. The average of the 3FGL -- 2FGL distribution is $0.04\pm0.01$, with the 3FGL sources  slightly softer than the 2FGL ones, while the average of the 3FGL -- 1FGL distribution is $-0.04\pm0.01$, with the 3FGL sources  slightly harder than the 1FGL ones.  


\begin{figure}[!ht]
\plotone{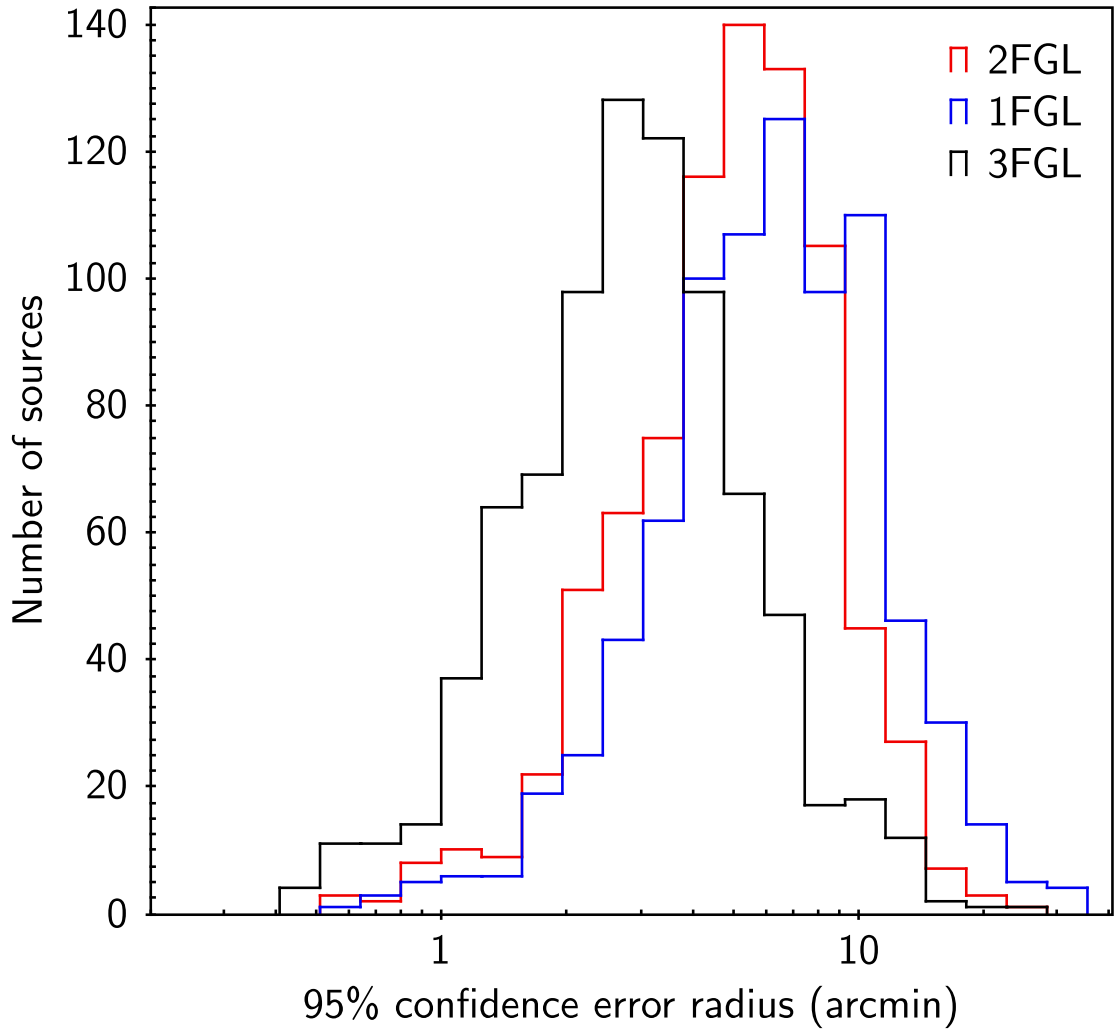}
\caption{Distributions of the 95\% confidence error radius for high-latitude sources ($|b|>10\degr$) in common among 1FGL (blue line), 2FGL (red line), and 3FGL (black line), illustrating the improvement of localizations for sources of equivalent detection significances.
}
\label{fig:fig_conf95_HL_123}
\end{figure}

When comparing the distribution of 95\% confidence error radius for the sources in common among all the LAT catalogs, we see that for 3FGL this parameter extends to lower values than for the earlier catalogs, showing that the localization has improved, thanks to improvements in the 3FGL analysis and increased statistics over the longer integration period for 3FGL (Figure~\ref{fig:fig_conf95_HL_123}).

\subsubsection{Possible causes for losing sources}








In the remainder of this section we describe a variety of reasons why the `lost'  0FGL, 1FGL, 2FGL, and 1FHL sources might not appear in the 3FGL catalog. 
Table~\ref{tab:012fgl_3fgl} shows the statistics of the `lost' sources.

\begin{deluxetable}{lccccc}
\rotate
\tablewidth{0pc} 
\setlength{\tabcolsep}{0.18cm}
\tabletypesize{\scriptsize}
\tablecaption{Statistics of `Lost' 0FGL, 1FGL, 2FGL, and 1FHL Sources
\label{tab:012fgl_3fgl}}
\tablehead{ 
\colhead{} &
\colhead{0FGL not in 3FGL} &
 \colhead{1FGL not in 3FGL} &
 \colhead{2FGL not in 3FGL} &
 \colhead{1FHL not in 3FGL} &
}
\startdata 
All & 12 & 310 & 300 & 17	\\
With flags & - & 131 & 211 & - \\
Name-FGL \texttt{c} \tablenotemark{(a)}& - & 104 & 87 & - \\
AGN & 1 & 22 & 27 & 1 \\
PSR & 0 & 1 & 3 & 0 \\
Unassociated & 11 & 264 & 234 & 16 \\
Within 1\degr~of a 3FGL \texttt{e} \tablenotemark{(b)}& 3 & 27 & 33 & 4 \\
\hline
 & & sources in other FGL catalogs & & \\
 \hline
 0FGL & - & 5 & 5 & 0 \\
 1FGL & 4 & - & 56 & 1 \\
 2FGL & 3 & 67 & - & 1 \\
 1FHL & 0 & 2 & 8 & - \\
 Not in any other {\it Fermi} catalog &  7 & 237 &  237 &  15 \\
\enddata 
\tablenotetext{a}{\texttt{c} indicates that based on the region of the sky the source is considered to be potentially confused with Galactic diffuse emission.}
\tablenotetext{b}{\texttt{e} indicates a source that was modeled as spatially extended.}
\end{deluxetable} 

We have also produced tables with all the `lost' sources for each previous LAT catalog. The first rows of the `lost' $\gamma$-ray source table for the 2FGL catalog are listed in Table~\ref{tab:2-3fgl}, only reported for guidance.\footnote{The full table of lost 2FGL sources and similar tables for lost 0FGL, lost 1FGL, and lost 1FHL sources are available only in the electronic version.}
In the last columns of Table~\ref{tab:2-3fgl} we assigned to each source one or more flags corresponding to possible causes for it to be lost and which we will discuss in the following paragraphs. In many cases, no one reason can be singled out. 

Lost sources from previous LAT catalogs are in general equally distributed over all latitudes, with a slight excess at low latitudes for 2FGL `lost' sources. In fact about 10\% of the 2FGL `lost' sources are at low Galactic latitude compared to a 6\% of high-latitude `lost' 2FGL sources.
We remind the reader that at low latitudes the Galactic diffuse emission is most intense and improvements in the model for the diffuse emission would be expected to have the most influence (\S~\ref{DiffuseModel}). The sources in common among 3FGL and the previous LAT catalogs are primarily outside the Galactic plane, as are the sources newly detected in 3FGL.
Most of the `lost' sources were also listed as unassociated in the previous FGL catalogs.
Among the former associated `lost' sources, most of them were associated with AGN and a few with pulsars. For sources of the AGN type their absence from the 3FGL catalog can be due to their intrinsic variability. A faint source which flared during the first year, allowing it to be detected in 0FGL, can be diluted and become undetectable in a longer time interval.

Most of the `lost' sources have analysis flags or the \texttt{c} designator in 1FGL and 2FGL names, indicating that these sources were already flagged as influenced by the diffuse emission and recognized as potentially problematic or possibly spurious.


Some other 1FGL, 2FGL, and 1FHL sources do not have counterparts in the 3FGL catalog because they have been resolved into two or more 3FGL sources or candidate source seeds. We flag them with `S' (split) in  the `Flag' column of Table~\ref{tab:2-3fgl}. In some cases only one of the seeds  reached $TS>25$ and so was included in the 3FGL list. 


Several other possible causes of `lost' sources are evident: (1) the 3FGL $\gamma$-ray centroid has shifted with respect to the previous FGL catalogs, preventing the matching; (2) statistical threshold effects, i.e. their $TS$ has dropped below 25.
Additional considerations include variability and (generally small) effects from the different event selections used for the analyses (P7REP\_SOURCE\_V15 for 3FGL, P7CLEAN\_V6 for 1FHL, P7SOURCE\_V6 for 2FGL and P6\_V3\_DIFFUSE for 0FGL); different Galactic diffuse emission models; different analysis procedures (unbinned likelihood analysis for 0FGL and 1FGL, binned likelihood analysis for 2FGL and 1FHL, and a combination of binned and unbinned for 3FGL). We analyze those causes in more detail for 2FGL in \S~\ref{compare_stepbystep}. We stress that these differences are often not negligible.



\begin{figure}[!ht]
\plotone{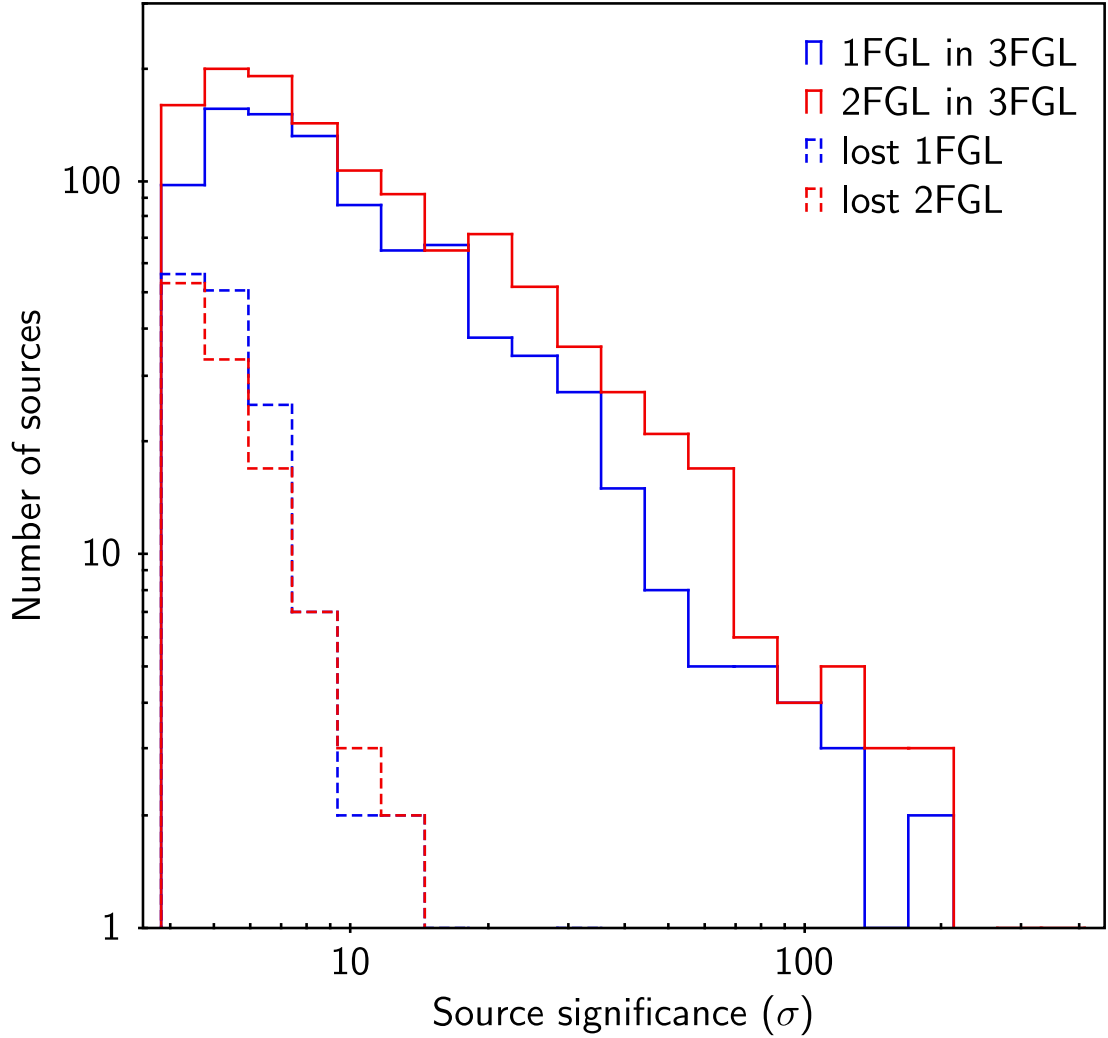}
\caption{Distributions of the significances of `lost' 1FGL and `lost' 2FGL sources compared to the 1FGL and 2FGL sources which are associated to 3FGL sources. All sources at high Galactic latitudes ($|b|>10\degr$) are included. (2FGL sources associated to 3FGL sources: solid red line, `lost' 2FGL sources: dashed red line, 1FGL sources associated to 3FGL sources: solid blue line, `lost' 1FGL sources: dashed blue line). 
}
\label{fig:fig_sign-avg_InOut}
\end{figure}

A comparison of the source significances of the `lost' sources with those in the 3FGL catalog shows that (Figure~\ref{fig:fig_sign-avg_InOut}) in the latter we have not lost highly-significant sources.
The peaks of the source significance distributions for all the sources of the FGL catalogs (not shown in the Figure~\ref{fig:fig_sign-avg_InOut}) have shifted from 4--6$\sigma$ for 1FGL to 4--5$\sigma$ for 2FGL and 3FGL. 

\begin{figure}[!ht]
\plotone{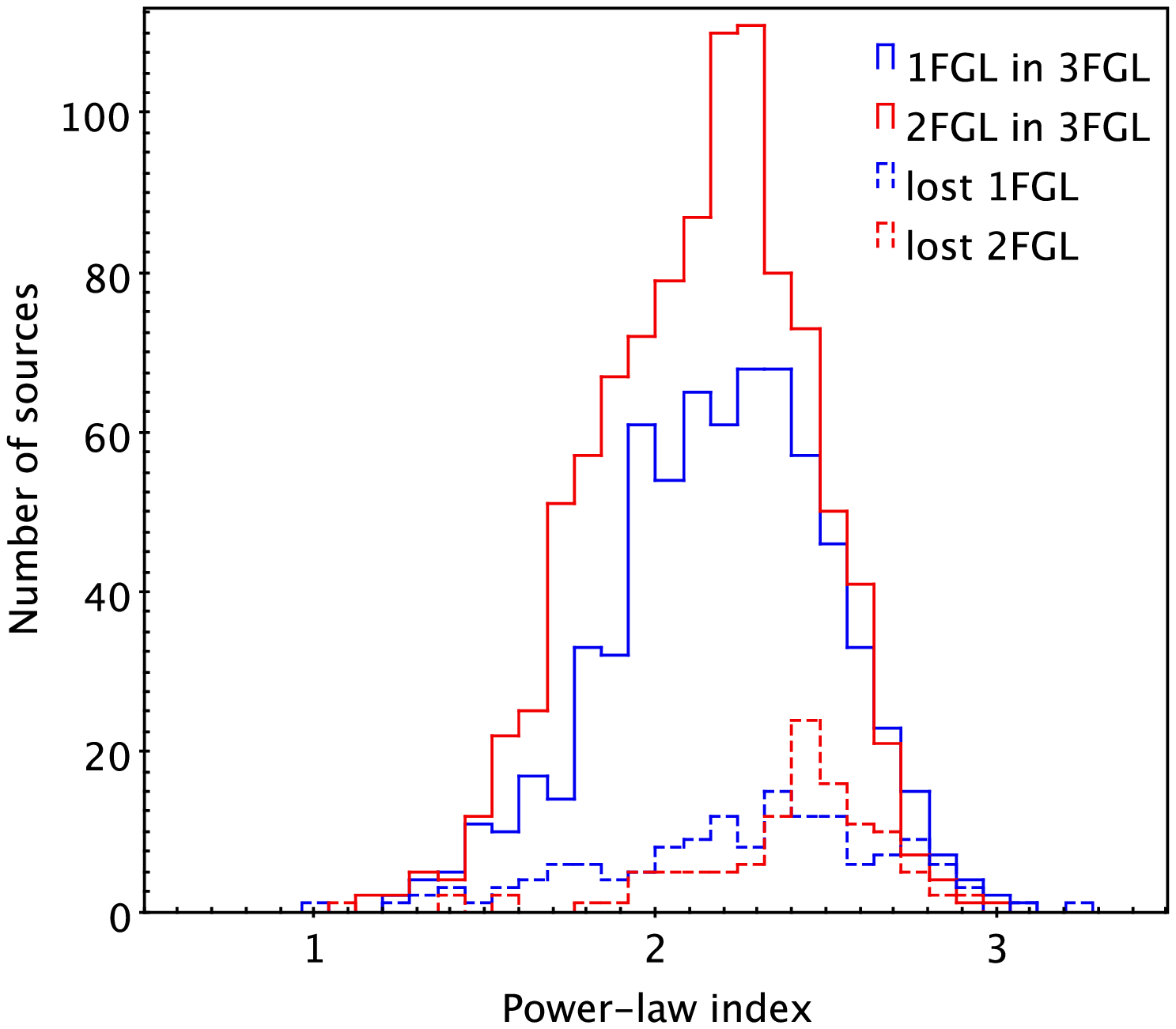}
\caption{Distributions of the power-law index of the 1FGL sources (blue solid line) and 2FGL sources (red solid line) in the 3FGL catalog and of the `lost' 1FGL sources (blue dashed line) and 2FGL sources (red dashed line). All samples include only high-latitude sources ($|b|>10\degr$).}
\label{fig:fig_spindex_InOut}
\end{figure}

The power-law indices of high-Galactic latitude ($|b| > 10\degr$) `lost' sources with power-law spectral type tend to be softer than average for their catalogs (Figure~\ref{fig:fig_spindex_InOut}).


The numbers of associated sources among the 0FGL, 1FGL, and 2FGL catalogs, and the 3FGL catalog do depend on the criteria used to define spatial coincidence (Eq.~\ref{eq:compnFGL}).  
The numbers of 1FGL -- 3FGL, 2FGL -- 3FGL and 1FHL -- 3FGL associated sources increase if we use  $\Delta < d_{99.9}$ as association criterion\footnote{Assuming a Rayleigh distribution for the source angular separations, $d_{99.9}$ is evaluated using \\ $\theta_{99.9} = 1.52\ \theta_{95}$.}.  The 193 additional associations (listed in Table~\ref{tab:2-3fgl} and corresponding 0FGL, 1FGL, 1FHL tables in the column `3FGL ($\Delta<d_{99.9}$)') represent about  5\% of the 0FGL, 1FGL, 2FGL, and 1FHL sources,  as expected when passing from $d_{95}$ to $d_{99.9}$. 
Furthermore, the improved model of the Galactic diffuse emission (\S~\ref{DiffuseModel}) used to build the 3FGL catalog together with the expected increase of the signal-to-noise ratio due to the use of 48 months of data, allowed us to obtain better localizations of the sources at positions that might be outside the 95\% confidence error regions reported in the 0FGL, 1FGL, or 2FGL catalogs. Indeed, about half of the 193 additional associations concern sources located along the Galactic plane.
Also, in the 1FGL catalog the positions of sources associated with the LAT-detected pulsars and X-ray binaries are the high-precision positions of the identified counterparts.  (These sources can be easily recognized because they have null values in the localization parameters reported in the 1FGL catalog.) Not all of these associations appear in the 3FGL catalog because they cannot be associated using $d_{95}$, but those that can be associated using $d_{99.9}$ are listed in Table~\ref{tab:2-3fgl} (and corresponding 0FGL, 1FGL, and 1FHL tables).


To study a possible reason for 0FGL, 1FGL, 2FGL, and 1FHL sources to disappear in the 3FGL catalog, we have compared the $TS$ they had when published in their respective catalogs with their values in the 3FGL $pointlike$ analysis. 
The 3FGL catalog was built, in fact, starting from 4003 seeds with  $TS > 10$ in the $pointlike$ analysis (\S~\ref{catalog_detection}). The final $gtlike$ analysis, which did not change the positions of the seeds, resulted in the 3033 sources with  $TS > 25$ that make up the 3FGL catalog. 
Therefore, possibly many seeds did not reach the threshold but can be associated with 0FGL, 1FGL, 2FGL, and 1FHL sources (using $\Delta<d_{99.9}$). These sources, marked with `T' (for `$\it{true}$') in the `3FGL Seed' column of the Table~\ref{tab:2-3fgl}, can be considered to be previously confirmed sources whose significance dropped below the threshold, either as a result of time variability, changes in the model or in the catalog analysis procedure for
 Galactic diffuse emission.
Finally we looked for those `lost' sources whose distances from an extended 3FGL source are less than $1\degr$, and these are flagged with `E' in the column `Flag' of Table~\ref{tab:2-3fgl}.

\subsubsection{Step-by-step from 2FGL to 3FGL}
\label{compare_stepbystep}


In order to understand the improvements of the 3FGL analysis with respect to 2FGL, we have considered the effects of changing the analysis, the data set, and the diffuse emission model without changing the time range (i.e., leaving it as two years). To that end we started with the 2FGL catalog and changed each of those three elements in sequence and compared each intermediate result with the previous one.
\begin{itemize}
\item The main difference between the analyses is the $Front$/$Back$ handling (\S~\ref{catalog_significance}). The comparison showed that using identical isotropic diffuse spectra for $Front$ and $Back$ events in 2FGL resulted in underestimating the low-energy fluxes of high-latitude sources. As a consequence, the corrected analysis leads to larger $TS$ values, higher photon fluxes, softer spectra, and smaller curvatures than in 2FGL. The effects are small on the scale of individual sources but collectively obvious. Quantitatively, the average difference in spectral index induced by this change was measured to be +0.05. Because that effect is due to the background, it is at the same level in $\sigma$ units ($\simeq 0.4 \sigma$) for faint and bright sources.
\item Changing data from Pass 7 (2FGL) to Pass 7 reprocessed (3FGL) results in somewhat larger $TS$, harder sources and more curved spectra (but no change of integral flux on average). The average difference in spectral index is $-$0.03. This goes in the opposite direction to (and therefore partly offsets) the difference due to the separate $Front$/$Back$ handling. However the dependence on flux is not the same. The reprocessing affects essentially all spectral indices and curvatures equally in absolute terms.
\item Finally, changing the model for Galactic diffuse emission from \texttt{gal\_2yearp7v6\_v0} used in 2FGL to \texttt{gll\_iem\_v05\_rev1} results in smaller $TS$, lower fluxes and less curved spectra (but no change of spectral index on average). Like the first point above, this background-related effect is smaller in absolute (curvature) or relative (flux) terms for brighter sources.
\end{itemize}
In conclusion, to first order the resulting net changes are not very large, consistent with the general comparison between 3FGL and 2FGL at the beginning of this section. The 3FGL sources tend to be less curved than 2FGL ones. In particular, there are fewer pathological very strongly curved sources (with $\beta=1$ and Flag 12 set) in 3FGL (41) than in 2FGL (64) even though there are more \texttt{LogParabola} spectra in 3FGL (395) than in 2FGL (336) because of the better statistics.


\begin{deluxetable}{lcrrrccccccccl}
\rotate
\tablewidth{0pc} 
\setlength{\tabcolsep}{0.18cm}
\tabletypesize{\scriptsize}
\tablecaption{2FGL Sources Not in the 3FGL Catalog \label{tab:2-3fgl}}
\tablehead{ 
  \colhead{2FGL} &
  \colhead{2FGL Assoc.\tablenotemark{(a)}} &
  \colhead{$l$\tablenotemark{(a)}} &
  \colhead{$b$\tablenotemark{(a)}} &
  \colhead{$\theta_{95}$\tablenotemark{(a)}} &
  \colhead{$\sigma$\tablenotemark{(a)}} &
  \colhead{Spec.\tablenotemark{(a)}} &
  \colhead{Var\tablenotemark{(a)}} &
 \colhead{3FGL\tablenotemark{(b)}}&
 \colhead{3FGL\tablenotemark{(c)}}&
 \colhead{$\Delta$\tablenotemark{(d)} }&
 \colhead{$\Delta\ / d_{99.9}$} &
   \colhead{3FGL\tablenotemark{(e)}} &
  \colhead{Flags\tablenotemark{(f)}} \\
 \colhead{} &  
  \colhead{}&  
  \colhead{(deg)} & 
  \colhead{(deg)} & 
  \colhead{(deg)} &  
  \colhead{} &
  \colhead{Index}& 
   \colhead{$p >$ 90\%}& 
  \colhead{($\Delta <d_{99.9}$)} & 
  \colhead{($d_{99.9} < \Delta< 1\degr$) }&  
   \colhead{(deg)} & 
   \colhead{} &
  \colhead{Seed} &   
   \colhead{} 
}
\startdata 
  J0004.2+2208 & \nodata & 108.732 & $-$39.430 & 0.194 & 5.4 & 2.49 & F & \nodata & \nodata & \nodata & \nodata & T & \nodata \\
  J0011.3+0054 & PMN J0011+0058 & 102.317 & $-$60.352 & 0.223 & 6.2 & 2.43 & T & \nodata & \nodata & \nodata & \nodata & \nodata & F\\
  J0013.8+1907 & GB6 J0013+1910 & 110.786 & $-$42.858 & 0.160 & 4.1 & 2.06 & T & \nodata & \nodata & \nodata & \nodata & \nodata & \nodata \\
  J0014.3$-$0509 & \nodata & 99.374 & $-$66.312 & 0.169 & 4.2 & 2.45 & F & J0014.3$-$0455 & \nodata & 0.322 & 0.725 & T & F\\
  J0118.6$-$4631 & \nodata & 289.226 & $-$69.872 & 0.134 & 4.5 & 1.78 & F & \nodata & \nodata & \nodata & \nodata & \nodata & \nodata \\
  J0124.6$-$2322 & \nodata & 188.135 & $-$81.611 & 0.113 & 7.5 & 2.31 & F & \nodata & J0123.7$-$2312 & 0.217 & 1.225 & \nodata & S,F\\
  J0128.0+6330 & \nodata & 126.998 & 0.922 & 0.255 & 5.0 & 2.57 & F & \nodata & J0128.4+6257 & 0.432 & 1.262 & \nodata & F\\
  J0129.4+2618 & \nodata & 133.451 & $-$35.784 & 0.333 & 4.9 & 2.56 & F & J0127.9+2551 & \nodata & 0.813 & 0.707 & T & \nodata \\
  J0158.6+8558 & \nodata & 124.201 & 23.262 & 0.180 & 4.2 & 2.52 & F & J0145.6+8600 & \nodata & 0.334 & 0.691 & T & \nodata \\
  J0214.5+6251c & \nodata & 132.251 & 1.495 & 0.134 & 4.1 & 2.26 & F & \nodata & \nodata & \nodata & \nodata & \nodata & c,F\\
  J0218.7+6208c & \nodata & 132.937 & 0.975 & 0.078 & 10.5 & 2.77 & F & J0217.3+6209 & \nodata & 0.208 & 0.753 & T & S,c,F\\
  J0219.1$-$1725 & 1RXS J021905.8$-$172503 & 191.883 & $-$67.564 & 0.148 & 4.3 & 1.92 & F & \nodata & \nodata & \nodata & \nodata & T & \nodata \\
  J0221.3+6025c & \nodata & 133.810 & $-$0.528 & 0.110 & 4.6 & 2.45 & F & \nodata & J0221.1+6059 & 0.391 & 1.451 & \nodata & c,F\\
  J0221.4+6257c & \nodata & 132.962 & 1.856 & 0.118 & 9.5 & 2.64 & F & \nodata & J0223.5+6313 & 0.316 & 1.11 & T & S,c,F\\
  J0227.2+6029c & \nodata & 134.471 & $-$0.220 & 0.097 & 6.1 & 2.38 & F & \nodata & \nodata & \nodata & \nodata & \nodata & c,F\\
 \enddata 
\tablenotetext{a}{All the values reported in these columns are from the 2FGL catalog.}
\tablenotetext{b}{Name of the 3FGL source associated with the 2FGL source with positional coincidence evaluated using $d_{99.9}$.}
\tablenotetext{c}{Closest 3FGL source having a distance  $d_{99.9} < \Delta < 1\degr$ from  the position of the 2FGL source.}
\tablenotetext{d}{In this column is reported  the angular separation ($\Delta$) between the 2FGL source and the 3FGL sources associated using $d_{99.9}$ or the closest 3FGL source. }  
\tablenotetext{e}{T: The 2FGL source and  one of the initial seeds for the 3FGL analysis have angular separation $< d_{99.9}$.}
\tablenotetext{f}{S: The 2FGL source was split/resolved into one or more seeds; c: The 2FGL source was flagged with \texttt{c}, i.e., possibly contaminated by the diffuse emission; F: the 2FGL source had analysis flags; E: The 2FGL source has a distance $<$ 1\degr from an extended 3FGL source.\\
This table is published in its entirety in the electronic edition of the Astrophysical Journal Supplements. A portion is shown here for guidance regarding its form and content.\\
Similar tables are available, only in the electronic edition, 
for lost 0FGL, lost 1FGL, and lost 1FHL sources.}
\end{deluxetable}

\section{Source Association and Identification}
\label{source_assoc_main}
    \subsection{Firm Identifications}
\label{source_assoc_firm}

As with the 2FGL and earlier LAT catalogs, we retain the distinction between associations and firm identifications.  Although many associations that we list between LAT sources and potential counterparts at other wavelengths, particularly those for AGN, have very high probability of being true, a firm identification, shown in the catalog by capitals in the Class column in Table~\ref{tab:classes}, is based on one of three criteria:
\begin{enumerate}
\item Periodic Variability.  Pulsars are the largest class in this category.  All PSR labels indicate that pulsed $\gamma$ rays have been seen from the source with a probability of the periodicity occurring  by chance of less than 10$^{-6}$.  Pulsars detected in blind searches of LAT data are indicated as `LAT PSR' in the  `ID or Assoc.' column of Table~\ref{tab:sources}; the other PSR detections are based on folding with radio or X-ray ephemerides \cite[see][]{LAT13_2PC}.  A similar chance probability requirement applies to the other set of periodic sources, the high-mass binaries (HMB).  Three of these are included in the catalog:  LS\,I +61 303 \citep{LAT09_LSI}, LS 5039 \citep{LAT09_LS5039}, and 1FGL J1018.6$-$5856 \citep{ATel.3221}.  Although not quite meeting the same chance probability, another binary (BIN) is included as an identification: Eta Carinae \citep{2012EtaCar, 2014EtaCar}.
\item Spatial Morphology.  Spatially extended sources whose morphology can be related to extent seen at other wavelengths include SNR, PWNe, and galaxies, as described in \S~\ref{catalog_extended}.  The Centaurus A  lobes and core are both marked as identified, because they are part of the same extended source, although the core itself does not show spatial extent.  
Although individual molecular clouds could in principle be included in this list, the catalog construction incorporates most known clouds into the diffuse model, and so no sources of this type are identified in the catalog. 
\item Correlated Variability.  Variable sources, primarily AGN, whose $\gamma$-ray variations can be matched to variability seen at one or more other wavelengths, are considered to be firm identifications.  Although some cases are well documented, such correlated variability is not always easily defined.   We conservatively require data in more than two energy bands for comparison.  Finding a blazar to have a high X-ray flux at the same time as a  $\gamma$-ray flare, for example, does not qualify if there is no long-term history for the X-ray emission. We include those sources whose variability properties are documented either in papers or with Astronomer's Telegrams\footnote{See \url{http://www.astronomerstelegram.org}.}.  This list does not represent the result of a systematic study.  Ongoing work will undoubtedly enlarge this list.  The one Galactic source identified in this way is nova V407 Cygni \citep{LAT10_V407Cyg}. Similarly short duration tangent gamma-ray emission observed from the classical novae, V959 Mon 2012 and V1324 Sco 2012, were not detected in the 4-year integrated analysis in the 3FGL \citep{LAT14_novae}.
\end{enumerate}

We include one exception to these rules.  The Crab PWN is listed as a firm identification even though it does not meet any of these criteria.  The well-defined energy spectrum, distinct from the Crab pulsar spectrum and matching spectra seen at both lower and higher energies provides a unique form of identification \citep{LAT10_Crab}.

In total, we firmly identify 238 out of the 3033 3FGL sources.
Among those,
143 are pulsars,
66 are active galactic nuclei (BCU, BLL, FSRQ, NLSY1, or RDG)
12 are SNR,
4 are binaries (BIN or HMB),
9 are PWN,
2 are normal galaxies, 
1 is a massive star-forming region, and
1 is a nova (Table~\ref{tab:classes}).

    \subsection{Automated Source Associations}
\label{source_assoc_automated}

Our approach for automated source association closely follows that used 
for the 2FGL, and details of the method are provided in \citet[][1FGL]{LAT10_1FGL} and \citet[][2FGL]{LAT12_2FGL}. 


In summary, we use a Bayesian approach that trades the positional
coincidence of possible counterparts with 3FGL sources against the expected 
number of chance coincidences to estimate the probability that a specific
counterpart association is indeed real (i.e., a physical association).
As for 1FGL and 2FGL, we retain counterparts as associations if they reach a posterior
probability of at least 80\%.
We apply this method to a set of counterpart catalogs for which we
calibrate the prior source association probabilities using Monte Carlo simulations of 
fake 3FGL catalogs.



Table \ref{tab:catalogs} lists the catalogs used in the automatic association procedure,
organized into four categories:
(1) known or plausible $\gamma$-ray-emitting source classes,   
(2) surveys at other frequencies,
(3) GeV sources, and
(4) identified $\gamma$-ray sources.
The first category allows us to assign 3FGL sources to object classes, while the second category reveals
multiwavelength counterparts that may suggest the possible nature of the associated
3FGL sources.
The third category allows assessment of former GeV detections of 3FGL sources, and the
fourth keeps track of all firm identifications (cf.~\S~\ref{source_assoc_firm}).
For this last category we claim associations based on the spatial overlap of the true
counterpart position with the 3FGL $99.9\%$ confidence error ellipse.

With respect to 2FGL, we updated all catalogs for which more comprehensive compilations became
available.  As for 1FGL and 2FGL, we separately consider energetic and nearby pulsars, with spin-down energy flux $\dot{E}/d^2 > 5 \times 10^{32}$ erg kpc$^{-2}$ s$^{-1}$.  We also consider millisecond pulsars (MSPs) separately.  For spin period $P$ (s) and spin-down rate $\dot{P}$, we define MSPs as pulsars satisfying $\log_{10} \dot{P} + 19.5 + 2.5 \log_{10} P < 0$.

Catalogs indicated with an asterisk in Table~\ref{tab:catalogs} have source location uncertainties greater than those for the 3FGL sources.  Catalogs indicated with a dagger have extended sources, with sizes greater than the source location uncertainty regions for 3FGL.  For these catalogs we cannot apply the Bayesian association method.  For the former catalogs we base associations on overlap of the 95\% confidence error radii.  For the latter, we require overlap between the extended source and the 95\% confidence radius (semi-major axis) of the 3FGL source.
These approaches are much less reliable than the Bayesian associations, so we do not claim any associations based on overlap in our final catalog.
We record, however, any spatial overlap with a TeV source in the FITS file version of the catalog, 
and use a special flag in our catalog (\texttt{TEVCAT\_FLAG}), distinguishing point-like (P, angular diameter $<20\arcmin$) from extended 
(E) TeV counterparts (see App.~\ref{appendix_fits_format}).
We furthermore list all unassociated 3FGL sources that are spatially overlapping with SNRs or PWNe in
Table~\ref{tab:snrext}.


\begin{deluxetable}{lrr}
\setlength{\tabcolsep}{0.04in}
\tablewidth{0pt}
\tabletypesize{\scriptsize}
\tablecaption{Catalogs Used for the Automatic Source Association Methods
\label{tab:catalogs}
}
\tablehead{
\colhead{Name} & 
\colhead{Objects\tablenotemark{a}} & 
\colhead{Ref.}
}

\startdata
High $\dot{E}/d^2$ pulsars & 213 & \citet{ATNFcatalog}\tablenotemark{b} \\
Other normal pulsars & 1657  & \citet{ATNFcatalog}\tablenotemark{b} \\
Millisecond pulsars & 137 & \citet{ATNFcatalog}\tablenotemark{b} \\
Pulsar wind nebulae & 69 & Collaboration internal \\
High-mass X-ray binaries & 114 & \citet{HMXBcatalog} \\
Low-mass X-ray binaries & 187  & \citet{LMXBcatalog} \\
Point-like SNR & 157  & \citet{SNRcatalog} \\
Extended SNR$^\dag$ & 274  & \citet{SNRcatalog} \\
O stars & 378  & \citet{OStarCatalog} \\
WR stars & 226  & \citet{vanderHucht2001} \\
LBV stars & 35  & \citet{LBVcatalog} \\
Open clusters & 2140  & \citet{OpenClustersCatalog} \\
Globular clusters & 160  & \citet{GlobClusterCatalog} \\
Dwarf galaxies$^\dag$ & 100  & \citet{DwarfGalaxies} \\
Nearby galaxies & 276  & \citet{NearbyGalaxiesCatalog} \\
IRAS bright galaxies & 82  & \citet{IRAScatalog} \\
BZCAT (Blazars) & 3060  & \citet{BZcatalog} \\
BL Lac & 1371  & \citet{AGNcatalog} \\
AGN & 10066  & \citet{AGNcatalog} \\
QSO & 129,853  & \citet{AGNcatalog} \\
Seyfert galaxies & 27651  & \citet{AGNcatalog} \\
Radio loud Seyfert galaxies & 29  & Collaboration internal \\
\hline
1WHSP & $~$1000 & \citet{1WHSP} \\
WISE blazar catalog & 7855 & \citet{WISE}\\
NRAO VLA Sky Survey (NVSS)\tablenotemark{c} & 1,773,484 &\citet{NVSScatalog} \\
Sydney University Molonglo Sky Survey (SUMSS)\tablenotemark{c} & 211,050 &\citet{SUMSScatalog} \\
Parkes-MIT-NRAO survey\tablenotemark{c} & 23277&\citet{PMNcatalog} \\
CGRaBS & 1625 & \citet{CGRaBS} \\
CRATES & 11499  & \citet{CRATES} \\
VLBA Calibrator Source List & 5776  & http://www.vlba.nrao.edu/astro/calib/vlbaCalib.txt \\
ATCA 20 GHz southern sky survey & 5890  & \citet{AT20G} \\
ATCA follow up of 2FGL unassociated sources & 424 & \citet{ATCA_2FGL} \\
ROSAT All Sky Survey (RASS) Bright and Faint Source Catalogs\tablenotemark{c} & 124,735 &\citet{RASSbright},\tablenotemark{d}\\
58 months BAT catalog &1092 & \citet{BATcatalog} \\
4$^{th}$ IBIS catalog & 723 & \citet{IBIScatalog} \\
\hline
1st AGILE catalog$^\ast$ & 47   & \citet{AGILEcatalog} \\
3rd EGRET catalog$^\ast$ & 271  & \citet{3EGcatalog} \\
EGR catalog$^\ast$ & 189  & \citet{EGRcatalog} \\
0FGL list$^\ast$ & 205  & \citet[][0FGL]{LAT09_BSL} \\
1FGL catalog$^\ast$ & 1451 & \citet[][1FGL]{LAT10_1FGL} \\
2FGL catalog$^\ast$ & 1873 & \citet[][2FGL]{LAT12_2FGL} \\
1FHL catalog$^\ast$ & 514 & \citet[][1FHL]{LAT13_1FHL} \\
TeV point-like source catalog$^\ast$ & 82 & http://tevcat.uchicago.edu/ \\
TeV extended source catalog$^\dag$ & 66   & http://tevcat.uchicago.edu/ \\
\hline
LAT pulsars & 147 & Collaboration internal \\
LAT identified & 137  & Collaboration internal \\ 
\enddata

\tablenotetext{a}{Number of objects in the catalog.}
\tablenotetext{b}{http://www.atnf.csiro.au/research/pulsar/psrcat}
\tablenotetext{c}{All-sky surveys used only in the Likelihood Ratio method, see \S~\ref{assoc:agn}}
\tablenotetext{d}{The RASS Faint Source Catalog is available from \url{ http://www.xray.mpe.mpg.de/rosat/survey/rass-fsc/}.}

\end{deluxetable}

\subsubsection{Active Galactic Nuclei associations}
\label{assoc:agn}

AGN, and in particular blazars, are the largest class of associated
sources in 3FGL at high Galactic latitudes.

In 3LAC, as in the Second LAT Catalog of AGN \citep[2LAC, ][]{LAT11_2LAC}, we added another association method 
to the automatic one described above. This is the Likelihood Ratio method (LR), frequently  used 
to assess identification probabilities for radio, infrared and optical sources \citep[e.g., ][]{deRuiter77,Pre83,ss92,Lon98,Mas01,LAT11_2LAC}. 
It is based on uniform surveys in the radio and in X-ray  bands, enabling us to search for possible counterparts among the faint radio and X-ray sources. The LR method makes use of counterpart densities (assumed spatially constant over the survey region) through the $\log N - \log S$ relation and therefore the source flux. This approach has been already used in 2LAC and we refer the reader to \S~3.2 of that paper for a comprehensive description of the method, which computes the probability that a suggested association is the `true' counterpart.\footnote{We note that the implementation of the LR method for the 2LAC associations was plagued with an error in the management of the sky coordinate precession that affected some of the associations. These false associations were also included in the active galaxy associations in 2FGL.  The error has been fixed and the 2FGL associations re-derived.  The corrected 2FGL catalog file has been delivered to the FSSC for distribution.  In the present work, comparisons with 2FGL findings are based on this corrected set of associations.} 

A source is considered as a likely counterpart of the $\gamma$-ray source if its reliability (see Eq.~4 in 2LAC) is greater than 0.8 in at least one survey.

In total, our automatic association procedure based on the Bayesian method finds 1663 3FGL sources that are associated with
AGN while the LR-based association method finds 1340 3FGL sources. For 405 sources only the Bayesian method provides an association, and for 82 sources only the LR method does so.

Overall, 3FGL includes 1745 sources associated with AGN ($58\%$ of all 3FGL sources) of which 
1145 are blazars, 
573 are candidate blazars, 
15 are radio galaxies, 
5 are Seyfert galaxies, and 
3 are other AGN.
The Seyfert galaxies are narrow-line Seyfert 1 galaxies that have been established as a
new class of $\gamma$-ray active AGN \citep{LAT09_Seyfert}.

Comparing to 2FGL we can make the following observations:
\begin{itemize}
\item The 3FGL includes 610 more sources of AGN type than the 2FGL, i.e., a  76\% increase. The fraction of new sources (not present in 2FGL) is slightly higher for hard-spectrum (i.e., $\Gamma<$ 2.2) sources than for soft-spectrum ones (i.e., $\Gamma>$ 2.2), 51\% vs. 47\% respectively, but the relative increase reaches 72\% for very hard-spectrum (i.e., $\Gamma<$ 1.8) sources. In the 3FGL, 477 counterparts are new (81 FSRQs, 146 BL~Lacs, 240 candidate blazars of  unknown type, 10 non-blazar objects); the other counterparts were present in previous AGN {\it Fermi} catalogs but not included in any of the 0FGL, 1FGL, or 2FGL catalogs  for various reasons (e.g., the corresponding $\gamma$-ray sources were not associated with AGN, had more than one counterpart or were flagged in the analysis). 
\item The fraction of counterparts of unknown type (named `bcu') has increased notably between the two catalogs (from 20\% to 28\%). The number of these sources in the 3FGL has increased by more than a factor of 2.5  relative to that in the 2FGL, becoming almost equal to that of FSRQs. This increase is mainly due to the lower probability of having a published high-quality spectrum available for these fainter sources because of the lack of  optical observing programs. In 
3LAC, sources of the `bcu' type have been divided into three sub-types depending of the multi-wavelength information available to characterize their `blazarness'. In this paper we do not propagate this sub-division and refer to the 3LAC for census.
\item The relative increase in `bcu' drives a drop in the proportions of FSRQs and BL~Lacs, which only represent 29\% and 41\% of the 3FGL respectively  (38\% and 48\% for 2FGL). The relative increase in the number of sources with respect to 2FGL is 34\% and 42\% for FSRQs and BL~Lacs respectively. 
\item Out of 825 AGN in the 2FGL, a total of 68 are missing in the full 3LAC sample, most of them due to variability effects.  A few others are present in 3FGL but with shifted positions, ruling out the association with their former counterparts.
\end{itemize}

\subsubsection{Normal galaxies}

The $\gamma$-ray emission of normal galaxies is powered by cosmic-ray interactions with interstellar gas and radiation.  They are numerous but typically faint relative to active galaxies.  The most luminous of the normal galaxies are starburst galaxies, which have very high densities of gas and massive star formation near their centers.  Less distant are normal galaxies in the local group.  As described above we searched for associations with sources in catalogs of nearby galaxies and IRAS bright galaxies (Table ~\ref{tab:catalogs}).

In the 3FGL catalog we do not find additional associations with normal galaxies relative to those reported already in  2FGL :  starburst galaxies M82 (3FGL J0955.4+6940), NGC 253 (3FGL J0047.5$-$2516) NGC 1068 (3FGL J0242.7$-$0001), and NGC 4945 (3FGL J1305.4$-$4926), and local group galaxies LMC (3FGL J0526.6$-$6825e), M31 (3FGL J0042.5+4117), and the SMC (3FGL J0059.0$-$7242e).  

Five sources in the 3FGL catalog lie within the extended source model for the LMC and are otherwise unassociated with counterparts at other wavelengths.  These sources (3FGL J0456.2$-$6924, 3FGL J0524.5$-$6937,  3FGL J0525.2$-$6614, 3FGL 0535.3$-$6559, and 3FGL J0537.0$-$7113) are classified here as `gal' based solely on the spatial coincidence with the LMC and their associations are listed as {\it LMC field}.  Their particular natures remain uncertain.  

\subsubsection{Pulsars}

Because pulsed emission can be such a clear signature, pulsars represent the largest class of firmly identified astrophysical objects in the 3FGL catalog. An extensive discussion of $\gamma$-ray pulsar properties is found in the Second {\it Fermi} Large Area Telescope Catalog of Gamma-ray Pulsars \citep[][2PC]{LAT13_2PC}.  The public catalog of LAT-detected pulsars is regularly updated\footnote{See \url{https://confluence.slac.stanford.edu/display/GLAMCOG/Public+List+of+LAT-Detected+Gamma-Ray+Pulsars}.}.  At the time of the 3FGL association analysis, this catalog had 147 pulsars (Table~\ref{tab:catalogs}).  Only 137 of the LAT-detected pulsars have associations in the 3FGL catalog, however (Table~\ref{tab:classes}).  The missing ten did not reach the $TS \ge 25$ criterion based on their average fluxes.  Three of these are PSR J0737$-$3039A \citep{LAT13_PSRJ0737}, J1640+2224, and J1705$-$1906 \citep{Hou2014}, and the remaining seven are flagged in the 2PC `spectral results' tables as being either too faint to fit, or requiring an on-peak analysis to obtain spectra.

\subsubsection{Pulsar wind nebulae}

In addition to the four pulsar wind nebulae found in 2FGL (Crab, Vela-X, MSH 15$-$52, HESS J1825$-$137), the 3FGL catalog includes seven new PWNe associations.   Five of these are firm identifications because they are spatially extended LAT sources (see Table~\ref{tbl:extended}):  HESS J1303$-$631 (3FGL J1303.0$-$6312e), HESS J1616$-$508 (3FGL J1616.2$-$5054e), HESS J1632$-$478 (3FGL J1633.0$-$4746e), HESS J1837$-$069 (3FGL J1836.5$-$0655e), and HESS J1841$-$055 (3FGL J1840.9$-$0532e).  The other two are positional associations with known PWNe: G279.8$-$35.8  (3FGL J0454.6$-$6825) and G0.13$-$0.11  (3FGL J1746.3$-$2859c).  

\subsubsection{Globular clusters}

Two globular cluster associations from the 2FGL catalog are not found in the 3FGL catalog:

\begin{enumerate}
\item 2FGL J1727.1$-$0704, previously associated with IC 1257, is found as 3FGL J1727.6$-$0654.  This source is not formally associated with the globular cluster. 
\item 2FGL J1824.8$-$2449, which was associated with NGC 6626, has been firmly identified as PSR J1824$-$2452A \citep{LAT13_PSRB1821}.  Its catalog listing is 3FGL J1824.6$-$2451.
\end{enumerate}

The number of globular clusters associated with LAT sources does continue to grow.  New associations are NGC 2608 (3FGL J0912.2$-$6452), NGC 6316  (3FGL J1716.6$-$2812), NGC 6441 (3FGL J1750.2$-$3704), NGC 6541 (3FGL J1807.5$-$4343), NGC 6717 (3FGL J1855.1$-$2243), and NGC 6752 (3FGL J1910.7$-$6000).  NGC 6752 had previously been noted as a likely LAT source by \citet{Tam2011_globular}.

\subsubsection{Supernova remnants}

Twelve SNRs are firmly identified in the 3FGL catalog as spatially extended sources (see Table~\ref{tbl:extended}). Six had previously appeared in the 2FGL catalog: IC 443, W28, W30, W44, W51C, and the Cygnus Loop.  Additions are: S147 \citep[3FGL J0540.3+2756e, ][]{LAT12_S147}, Puppis A \citep[3FGL J0822.6$-$4250e, ][]{2012LAT_PuppisA}, Vela Jr. \citep[3FGL J0852.7$-$4631e, ][]{LAT11_RXJ0852}, RX J1713.7$-$3946 \citep[3FGL J1713.5$-$3945e, ][]{LAT11_RXJ1713}, Gamma Cygni \citep[3FGL J2021.0+4031e, ][]{LAT12_extended}, and HB21 \citep[3FGL J2045.2+5026e, ][]{Reichardt13_HB21,LAT13_HB21}. 

Additionally we consider eleven unresolved 3FGL sources as being confidently associated with SNRs, based on individual studies of these SNRs in LAT data \citep[see][and references therein]{Ferrand2012_SNRCat}\footnote{\url{http://www.physics.umanitoba.ca/snr/SNRcat/}}.  These are given the `snr' designator (Table~\ref{tab:classes}).  The 2FGL sources that were designated `snr' have been added to the `spp' class in 3FGL.  Many of the 68 SNR or PWNe in this table are spatially extended sources at other wavelengths, and therefore the chance probability of an overlap with a LAT source is non-negligible.  As for previous LAT catalogs, we encourage great care in any analysis using these potential associations.

\subsubsection{Binaries}
\label{assoc:binaries}

Three HMB sources that appeared in 2FGL are also found in 3FGL: LS I +61 303 (3FGL J0240.5+6113), 1FGL J1018.6$-$5856 (3FGL J1018.9$-$5856), and LS 5039 (3FGL J1826.2$-$1450).  All were firmly identified by binary periodicity.  
We note that each of the three is a TeV emitter but that the two other binary systems that have been detected in the TeV energy regime, HESS J0632$+$057 and PSR B1259$-$63, do not have counterparts in 3FGL.
PSR B1259$-$63/LS 2883 is a bright LAT source during a small part of the 3.4-year binary period following periastron \citep{Tam_PSRB1259, LAT11_PSRB1259}.
HESS J0632$+$057 has not been detected at all.

A fourth HMB from 2FGL, Cygnus X$-$3 (2FGL J2032.1+4049) does not appear in 3FGL.  It is an intermittent LAT source and was not active enough averaged over the four years of this catalog to produce a significant detection.
Eta Carinae, which appeared in 2FGL as a possible massive star association, is included as 3FGL J1045.1$-$5941, identified as a binary system. The LAT data exhibit the known 5.5-year binary period \citep{2012EtaCar, 2014EtaCar}.  Although listed in the catalog as a nova, V407 Cygni is also a binary system.  Its flaring activity in 2010 \citep{LAT10_V407Cyg} was bright enough that it appears as 3FGL J2102.3+4547. 




\subsubsection{Multiwavelength associations}

In addition to the catalogs of classified sources, we also searched for associations with catalogs of
radio and TeV sources.
Our association procedure for AGN heavily relies on associations with radio sources as most
of the $\gamma$-ray emitting AGN are bright sources of radio emission 
(see \S~\ref{assoc:agn}).
In fact, essentially all of the radio associations we find have been classified subsequently as AGN.

We did not search for general associations with infrared, optical, or soft X-ray catalogs.  Within the LAT source error regions we would find multiple potential counterparts, most of which necessarily would be due to chance, since many of the sources in these catalogs are thermal in nature. 
We included, however, the hard X-ray catalogs INTEGRAL-IBIS and Swift-BAT, and blazar candidates extracted from the infrared WISE catalog \citep{WISE} in the automated association pipeline.  These data, when included in a study of the spectral energy distributions to evaluate their synchrotron peak frequencies and general behaviors, help in understanding the natures of the candidate counterparts. This was done especially for all the sources classified as bcu/BCU and agn.

\subsubsection{Statistics of association results}

In total we find that 1976 of the 3033 sources in the 3FGL catalog (59\%) have been associated 
with at least one non-GeV $\gamma$-ray counterpart by the automated procedures.
Table \ref{tab:assoc_results} summarizes the association results.

\begin{deluxetable}{lcccccc}
\rotate
\tablewidth{0pc} 
\setlength{\tabcolsep}{0.18cm}
\tabletypesize{\scriptsize}
\tablecaption{Statistics of Source Associations \label{tab:assoc_results}}
\tablehead{ 
\colhead{Category} &
\colhead{0FGL } &
 \colhead{1FGL } &
 \colhead{2FGL } &
 \colhead{1FHL\tablenotemark{a}} &
 \colhead{3FGL} &
}
\startdata 
Total & 205 & 1451 & 1873 & 514 &	3033 \\
Associated &  168 &  821 &  1224 & 449 & 2023 \\
Unassociated & 37 & 630 & 649 & 65 & 1010\\
New $\gamma$-ray sources\tablenotemark{b} & - & 1265 & 762 & 52 & 1312 \\
Sources associated with former LAT detections & - & 186 & 1111 & 462 & 1721 \\
Sources associated with former GeV detections\tablenotemark{c}  & 74 & 162 & 170 & 4 & 206 \\
Firmly identified sources & 31 & 65 & 124 & 60 & 238 \\
Sources associated with at least one object of known type & 153 & 623 & 952 & 385& 1398 \\
Sources which have counterparts only in the multiwavelength catalogs & - & 92 & 214 & 58 & 576 \\
\enddata 
\tablenotetext{a}{1FHL: $>$10 GeV}
\tablenotetext{b}{Non-overlapping 95\% source location confidence contours compared to previous LAT catalogs, at the level of overlapping 95\% source location confidence contours.}
\tablenotetext{c}{{\b Here only the 1AGL, 3EG, and EGR catalogs are considered.}}
\end{deluxetable} 



    \subsection{GeV and TeV Source Associations}
\label{source_assoc_tev}

Through 2014 August, 155 flaring {\it Fermi}-LAT sources were detected and promptly reported in more than 249 Astronomer's Telegrams\footnote{\url{https://www-glast.stanford.edu/cgi-bin/pub\_rapid}}. Of these sources, six are not in 3FGL. For two of these the flaring state was detected outside the time interval covered by 3FGL: S5 1044+71 (a 2FGL source classified as an FSRQ) and PMN J1626$-$2426 (in the proximity of an unassociated 2FGL source flagged as potentially contaminated by the diffuse emission). The other four are Cyg X$-$3 (an HMB associated with 2FGL J2032.1+4049; see \S~\ref{assoc:binaries}), CGRaBS J1848+3219 (an FSRQ associated with 2FGL J1848.6+3241), PKS 1124$-$186 (an FSRQ associated with 2FGL J1126.6$-$1856) and PKS 2123$-$463 (an FSRQ associated with 2FGL J2125.0$-$4632). The reason that these three FSRQs are missing from the 3FGL catalog is probably that they have average fluxes below the detection threshold.

Sources in 3FGL that are positionally associated with sources seen by
the ground-based TeV telescopes are of particular interest for
broad-band spectral studies.  As for the 2FGL catalog we studied
associations with the
TeVCat\footnote{\url{http://tevcat.uchicago.edu}} compilation of
sources detected by very-high-energy observatories.  The energy
threshold for TeVCat sources is not uniform, but it is typically
greater than 100~GeV.  We used a compilation of TeVCat sources 
prepared on 2014 October 27 that has 148 unique entries. This comprises the so-called
``Default'' and ``Newly Announced'' TeV catalogs. We note that, as in
1FGL and 2FGL, the ``Galactic Centre Ridge'' was not included for
association purposes since it represents diffuse emission over an
extended region along the Galactic plane
\citep{2006Natur.439..695A}.

This TeVCat compilation by its nature does not represent a complete
survey, and our general statistical procedure for evaluating
probabilities of chance association could not be applied.  As for 2FGL
we separately considered point-like and extended TeVCat sources
(Table~\ref{tab:catalogs}). For point-like sources the criterion for
association was overlapping 95\% source location regions (indicated by
`P' in the `TeV' column of Table~\ref{tab:sources}).  For extended
sources (LAT or TeVCat) the criterion was spatial overlap within their
respective angular extents (indicated by `E').  We note that, in
  the literature, the shapes of the extended TeV sources are usually
  approximated to a circle or to an ellipse. For the purposes of our
  association pipeline, we imposed a circular geometry on all extended
  TeV sources, setting the radius to the length of the semi-major
  axis. In the case of TeV sources whose morphologies depart
  significantly from a simple ellipse or a circle, this simplification
  of their geometry could lead to missed 3FGL associations.

In total, 124 3FGL sources have TeV counterparts.  Of the 148 TeV
sources considered, 117 have 3FGL associations and, out of these, 6
TeV sources have multiple 3FGL associations. Five of these are
extended TeV sources and one is the Crab, which is associated with
both the synchrotron and the inverse-Compton Crab 3FGL sources (3FGL
J0534.5+2201s/3FGL J0534.5+2201i). The TeV sources HESS J1018$-$589,
Westerlund 2, HESS J1458$-$608 and MGRO J2019+37 have two 3FGL
associations each while Westerlund 1 has three 3FGL associations.

We note that the TeV source HESS J1018$-$589 has two components, denoted
A (a point source) and B (extended emission). Of the two
3FGL associations, 3FGL~J1018.9$-$5856, a LAT high-mass binary, lies
closer to location A and 3FGL~J1016.3$-$5858, a LAT pulsar, lies closer to
location B.

Table~\ref{tab:ext_tev_assoc} shows the associations between extended
TeVCat sources and 3FGL catalog sources.  Some of these, designated
with \texttt{e} appended to their source names, were explicitly
modeled as extended sources corresponding to H.E.S.S. sources (see
\S~\ref{catalog_extended}).

Out of the 58 TeV AGN, 57 have associated sources in 3FGL. Only HESS
J1943+213, tentatively classified as a high-synchrotron peaked (HSP) blazar
\citep{HESS_J1943p213, 2012A&A...539A.128L}, does not have a
3FGL association. This presumed blazar is unique in the TeV sky in
that it shines through the Galactic plane. We note that the VERITAS
source VER~J2016$+$371 is positionally associated with
3FGL~J2015.6$+$3709 although the VERITAS source is probably a PWN
\citep{2014ApJ...788...78A} and the LAT source is associated with an
FSRQ of unknown redshift.


The Milagro source, MGRO~J2031$+$41, which comprises
TeV~J2032$+$4130, was also included as a separate source in the TeV
list that was used to evaluate the associations with 3FGL because
it is postulated that its emission is due to more than one source \citep{Milagro07}.
Due to the large extent
of MGRO~J2031$+$41 \citep[$1\fdg8$,][]{2012ApJ...753..159A}, it has positional
overlap with eight 3FGL sources in addition to 3FGL~J2032.2$+$4126. We
have 
listed just one LAT source 3FGL~J2028.6$+$4110e, the Cygnus Cocoon, as being associated with
MGRO~J2031$+$41.

Due to its large extent ($2\fdg6$), the Milagro Geminga source
\citep{Milagro09} has positional overlap with two 3FGL
sources, 3FGL~J0633.9$+$1746 (the Geminga pulsar) and the
unidentified source 3FGL~J0626.8$+$1743. We have associated only
the Geminga pulsar with the Milagro source.
\citet{Milagro09} postulate that the Milagro
emission could be due to a pulsar-driven wind associated with
Geminga.

The TeV sources Boomerang and SNR~G103.3$+$02.7 have positional
overlap at TeV energies. They are each positionally coincident with
the same two 3FGL sources, the unidentified source 3FGL J2225.8+6045
and the pulsar, 3FGL J2229.0+6114. We have associated
SNR~G103.3$+$02.7 with 3FGL~J2225.8$+$6045 and Boomerang, classified
as a PWN at TeV energies, with 3FGL J2229.0+6114, the LAT pulsar.

Relative to 2FGL eight TeV sources are newly associated with LAT sources. All
but one of these sources (HESS~J1641$-$463) had already been detected
at TeV energies when 2FGL and 1FHL were released. None of these eight
sources, however, have counterparts in those catalogs:

\begin{enumerate}
\item The H.E.S.S. Galactic Center source is associated with 3FGL~J1745.6$-$2859c. The corresponding source 2FGL~J1745.6$-$2858, had a large enough position offset
  that it was not considered associated with the TeV source.
  The Galactic center remains, however, a particularly complex region whose detailed study is beyond the scope of this paper.

\item Three HSP blazars (SHBL~J001355.9$-$185406, 1ES~0229+200 and
  1ES~0347$-$121) are associated with 3FGL sources.

\item The shell SNR Tycho is associated with the faint LAT source
  3FGL~J0025.7+6404. Although the center of the TeV emission is
  offset by $0\fdg12$ from the LAT source, the
  relatively large uncertainty of the LAT position indicates sufficient positional overlap for association. 

\item The TeV PWN HESS~J1809$-$193 is another new TeV association in
  3FGL. Discovered at TeV energies in 2007, it is a relatively bright
  TeV source with an integral flux 14\% that of the Crab Nebula
   in the same energy band \citep{2007A&A...472..489A}.

\item The  unidentified TeV sources HESS~J1626$-$490 and
  HESS~J1641$-$463 are also new TeV associations for 3FGL.

\end{enumerate}


Thirty-one TeV sources  have no counterparts in 3FGL; one of
these, the unidentified extended TeV source HESS~J1857$+$026, has an
association in 1FHL (1FHL~J1856.9$+$0252). It is the only TeV source
to have a 1FHL association but none in 2FGL or
3FGL. The other TeV sources having no counterparts in 3FGL comprise 10
other unidentified TeV sources, 10 PWNe, 2 binaries, 4 shell-type SNRs,
2 SNR-molecular cloud associations, 1 composite SNR (i.e., 9 out of the
19 SNRs in TeVCat) and the HSP blazars discussed earlier.

We note that two TeV sources, HESS~J1634$-$472 and SNR W49B, were each
associated with LAT sources in 2FGL and in 1FHL but do not have
associations in 3FGL. A source coincident with W49B has been
detected by the LAT, 3FGL~J1910.9$+$0906 but this does not have
positional overlap with the TeV detection of W49B.

The TeV source, HESS~J1427$-$608 had an association in 2FGL but was not
associated with a LAT source in 1FHL and does not have an association
in 3FGL.


\begin{deluxetable}{ll}
\setlength{\tabcolsep}{0.04in}
\tablewidth{0pt}
\tabletypesize{\scriptsize}
\tablecaption{Associations of 3FGL with Extended TeV Sources\label{tab:ext_tev_assoc}}
\tablehead{
\colhead{TeVCat Name\tablenotemark{a}} &
\colhead{3FGL Name}
}
\startdata
Boomerang & J2225.8+6045, J2229.0+6114\\
CTA 1 & J0007.0+7302\\
CTB 37A & J1714.5$-$3832\\
Geminga & J0633.9+1746\\
HESS J1018$-$589 & J1016.3$-$5858, J1018.9$-$5856\\
HESS J1303$-$631 & J1303.0$-$6312e\\
HESS J1356$-$645 & J1356.6$-$6428\\
HESS J1458$-$608 & J1456.7$-$6046, J1459.4$-$6053\\
HESS J1503$-$582 & J1503.5$-$5801\\
HESS J1507$-$622 & J1506.6$-$6219\\
HESS J1614$-$518 & J1615.3$-$5146e\\
HESS J1616$-$508 & J1616.2$-$5054e\\
HESS J1626$-$490 & J1626.2$-$4911\\
HESS J1632$-$478 & J1633.0$-$4746e\\
HESS J1640$-$465 & J1640.4$-$4634c\\
HESS J1708$-$443 & J1709.7$-$4429\\
HESS J1718$-$385 & J1718.1$-$3825\\
HESS J1745$-$303 & J1745.1$-$3011\\
HESS J1800$-$240 & J1800.8$-$2402\\
HESS J1804$-$216 & J1805.6$-$2136e\\
HESS J1809$-$193 & J1810.1$-$1910\\
HESS J1825$-$137 & J1824.5$-$1351e \\
HESS J1834$-$087 & J1834.5$-$0841\\
HESS J1837$-$069 & J1836.5$-$0655e\\ 
HESS J1841$-$055 & J1840.9$-$0532e\\
HESS J1848$-$018 & J1848.4$-$0141\\
HESS J1858+020 & J1857.9+0210\\
IC 443 & J0617.2+2234e\\
Kookaburra (Rabbit) & J1418.6$-$6058\\
Kookaburra PWN & J1420.0$-$6048\\
MGRO J1908+06 & J1907.9+0602\\
MGRO J2019+37 & J2021.1+3651, J2017.9+3627\\
MGRO J2031+41 & J2028.6+4110e\\
MSH 15$-$52 & J1514.0$-$5915e\\
RX J0852.0$-$4622 & J0852.7$-$4631e\\
RX J1713.7$-$3946 & J1713.5$-$3945e\\
SNR G292.2$-$00.5 & J1119.1$-$6127\\
Terzan 5 & J1748.0$-$2447\\
TeV J2032+4130 & J2032.2+4126\\
Vela X & J0833.1$-$4511e\\
VER J2019+407 & J2021.0+4031e\\
W 28 & J1801.3$-$2326e\\
W 51 & J1923.2+1408e\\
Westerlund 1 & J1648.3$-$4611, J1650.3$-$4600, \\
 & J1651.5$-$4626\\
Westerlund 2 & J1023.1$-$5745, J1024.3$-$5757\\
\enddata
\tablenotetext{a}{From \url{http://tevcat.uchicago.edu}.}
\end{deluxetable}

    \subsection{Properties of Unassociated Sources}
\label{source_assoc_unassoc}

Among the 3033 sources in the 3FGL catalog, 2033 have associations or identifications with known astrophysical objects. Although that number is greater than the total number of sources in the 2FGL catalog, 1010 (33\%) of the 3FGL sources remain unassociated. 
Among these unassociated sources are many that were found in previous LAT catalogs, indicating that some persistent mysteries remain despite extensive efforts to find associations over the past few years. The continued prevalence of unassociated sources is expected, as the improvement in sensitivity with four years of flight data and improvements to the characterization of backgrounds have allowed {\it Fermi}-LAT to probe the $\gamma$-ray sky to unprecedented depths. As a result, direct comparison to previous releases is difficult.

\begin{figure}[!ht]
\plotone{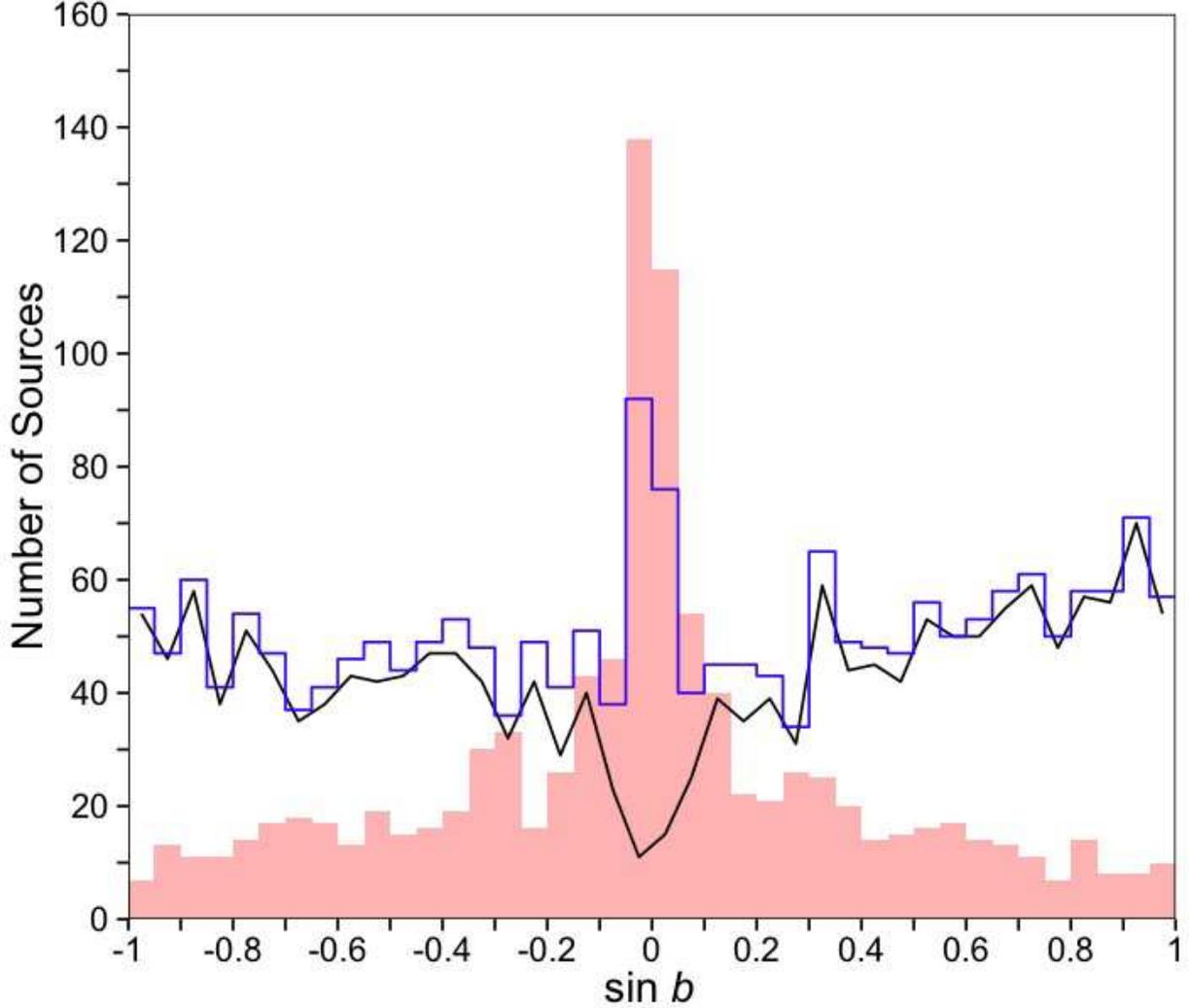}
\caption{Distributions in Galactic latitude $b$ of unassociated sources (shaded red region), all associated sources (blue histogram) and
all active galaxy source classes (black line). Binned in $\sin b$, an isotropic distribution would be flat.}
\label{fig:sinb}
\end{figure}

The distribution of unassociated sources on the sky is compared in Figure~\ref{fig:sinb} to the distribution of the associated sources. 
The plot reveals some important features that should be kept in mind when considering unassociated 3FGL sources. 
Of the 992 unassociated sources in the 3FGL,
334 fall within the Galactic plane ($|b| < 5\degr$). 
This leaves 658 unassociated at $|b| > 5\degr$ sources for an average density of $ 1.75 \times 10^{-2}$ sources deg$^{-2}$.
While this  density reflects the finite angular resolution and sensitivity of the instrument, some of the shortfall of  associations is attributable to the fact that not all areas of the sky have been mapped uniformly at other wavelengths. 
For instance,  a 3269 deg$^2$ ``overlap region'' 
in the North Galactic Cap intensively covered by five radio catalogs GB6 (6~cm) FIRST (20~cm), NVSS (20~cm), WNSS (92~cm), and VLSSr (4~m), as well as by the Sloan Digital Sky Survey (SDSS DR9) optical survey \citep{Kimball2008}, contains only 31 unassociated sources.   The corresponding density of unassociated sources is
$9.48 \times 10^{-3}$ sources deg$^{-2}$, nearly half of the overall average.

Within the Galactic plane, the unassociated source population is a combination of both Galactic and extragalactic source classes. Outside the plane, the LAT-detected AGN source density is 0.045~deg$^{-2}$.   By extrapolation, this implies that there should be 160 detectable AGN within the 10-degree band surrounding the Galactic plane (not accounting for incompleteness of AGN catalogs at low latitudes). 
Only 56 sources are associated with active galaxies in this region.  At low latitudes the LAT detection threshold is higher, and catalogs of active galaxies are incomplete, but extrapolation from higher latitudes suggests that fewer than $\sim$100 of the unassociated sources in the region $|b| < 5\degr$ could be active galaxies.

Of the remaining unassociated sources, we expect most to be Galactic objects. If we use the fractions of Galactic associated sources in this latitude range as a guide, we find that nearly half should be pulsars (47\%), and most of the remainder should be SNRs (44\%), with only a small number of other Galactic sources (9\%). These yet-to-be-detected pulsars may never be seen to pulse: part of the neutron star population will have magnetic and rotation axis orientations causing $\gamma$-ray emission to be spread over a large part of the rotation phase \citep{Hou2014}. This, combined with a low $\gamma$-ray flux in the presence of high background, renders blind periodicity searches insensitive to pulsations \citep{Dormody2011}. That many of these sources are likely to be previously undetected SNRs is also reasonable. \citet{LAT12_1FGLUnassoc} recognized that the distribution of unassociated LAT sources near the Galactic plane matched the scale height appropriate for Population I objects, such as the SNR parent population of massive stars.

It should be emphasized that a substantial fraction of the unassociated sources (40\%) has at least one analysis flag set (\S~\ref{catalog_analysis_flags}). We find that 57\% of the sources with at least one flag have $|b| < 5\degr$, which reflects the complexity of the Galactic diffuse emission (\S~\ref{catalog_diffusemodel} and \S~\ref {catalog_ism}). Because of the difficulties in source detection against the bright diffuse background along the Galactic plane, unassociated sources with analysis flags set should be considered with caution.
That difficulty is acute in the Galactic bulge (within $5\degr$ of the Galactic center), in the Vela ($l \sim 268\degr$) and Cygnus ($l \sim 80\degr$) regions.
The immediate vicinity of the Galactic center is particularly uncertain, with 7 sources within $1\degr$, on top of bright diffuse emission. We did not attempt in the framework of 3FGL to devise models dedicated to those regions, so the source positions and characteristics there are not as reliable as in the extragalactic sky.

We also note a number of clusters of unassociated sources, mostly near the Galactic plane. Obvious ones are near ($l,b$) = (133.5,+1), (340,$-$2). Since there is no specific search for extended sources in 3FGL, many of those clusters are probably extended sources (supernova remnants, pulsar wind nebulae, star formation regions). The extended sources which are currently declared (\S~\ref{catalog_extended}) are all relatively bright, with a median significance of $32\sigma$. This leaves a lot of room for fainter extended sources.

\section{Galactic Source Number Counts}
\label{source_counts}

\def\Lgamma{L_\gamma}
\def\Lgammamin{L_{\gamma,min}}
\def\Lgammamax{L_{\gamma,max}}
\def\Sgamma{S_\gamma}
\def\Sgammamin{S_{\gamma,min}}
\def\Sgammamax{S_{\gamma,max}}
\def\Lunits{ph~s$^{-1}$\ }
\def\perLunits{/ph~s$^{-1}$}
\def\Lergunits{erg\ ~s$^{-1}$\ }
\def\Sunits{ph~cm$^{-2}$~s$^{-1}$\ }
\def\kpc3{kpc$^{3}$}
\def\perkpc3{kpc$^{-3}$}

The 3LAC companion paper discusses briefly the source number counts of extragalactic sources. 
Here we address the Galactic source number counts, following the analysis for the 1FHL catalog
which was based on the method described by  \citet{Strong2007_source_populations}, and which  addressed energies above 10 GeV. 
For 3FGL we use energies above 1~GeV.
Photon fluxes over that energy range ($F_{35}$ of Table~\ref{tab:sources}) are more accurate than over the full band, as explained in the 2FGL paper.
The much larger number of sources in 3FGL compared to 1FHL  means that the entire analysis is more robust. 

The motivation for performing a  Galactic source population analysis is firstly to
obtain estimates of the global source characteristics, i.e., space density and luminosity function, 
secondly to estimate the contribution from sources below the detection threshold to the Galactic `diffuse' emission,
and thirdly to generate templates of this emission, to be incorporated into  diffuse emission models for future source catalogs.
Source population analysis also puts the detected sources in the context of the total source content of the Galaxy.

The method is guided by properties of known sources such as pulsars but does not attempt physical modeling of the sources; the approach is essentially geometrical, 
nevertheless the analysis reveals the basic properties of the source population.
We refer to the modeling of the source population(s) as {\it population synthesis}.
Since the   population synthesis includes all sources down to arbitrarily low flux levels (for a given model),
it can also be used to study the flux limit of the actual catalog, 
and assess how the observed  source number counts are affected by the detection procedure. This serves as a consistency check on other methods of assessing the detection threshold.

An essential principle is to use the fact that low-latitude sources probe the high-luminosity, low space densities at large distances,  while the high-latitude sources constrain the  low-luminosity, high space density nearby objects.
This is because high-luminosity sources are rare but visible to large distances which are only sampled in the Galactic plane, while low-luminosity sources are common but only visible when nearby, so dominate outside the plane.
 These samples are complementary and allow the full luminosity function to be estimated.

\subsection{Source Population Synthesis}

The population synthesis and subsequent analysis is performed using the GALPLOT software, which is publicly available\footnote{\url{http://sourceforge.net/projects/galplot}.}.
Let $\Lgamma$ be the luminosity of a source in \Lunits in some energy range.
We use photon luminosities since they are most directly related to the  detectability of sources and detection thresholds. 
The luminosity function at Galactocentric distance $R$ and distance from Galactic plane $z$
is the space density of sources per unit luminosity $\rho(\Lgamma,R,z)$.  
The shape of the luminosity function is assumed independent of position, i.e.,  $\rho(\Lgamma,R,z)$ is separable in $\Lgamma$ and $(R,z)$.
After \citet{Strong2007_source_populations} we assume that the luminosity function depends on luminosity as $\Lgamma^{-\alpha}$ for  $\Lgammamin< \Lgamma< \Lgammamax$ and is zero outside these
limits.
The total space density of sources is  $\rho(R,z)=\int\rho(\Lgamma,R,z)\
d\Lgamma$, 
which we normalize to the value $\rho_\odot$ at $(R, z) = (R_\odot, 0)$. 
For a source of luminosity $\Lgamma$ at distance $d$ the flux is $\Sgamma=\Lgamma/4\pi d^2$.  
The differential source number counts
are defined as $N(\Sgamma)$ sources per unit flux  
over the area of sky considered.   At lower $\Sgamma$, both the luminosity function and the spatial boundaries
influence $N(\Sgamma)$.
For this analysis the  source fluxes are binned in log($\Sgamma$) so that the plotted distributions are proportional to $\Sgamma N(\Sgamma)$; we use 5 bins per decade of $\Sgamma$, appropriate to the statistics available.
We use standard Monte Carlo techniques to sample  $\rho(\Lgamma,R,z)$ throughout  the  Galaxy,
using oversampling to reduce statistical fluctuations if necessary.
We use the sources generated from such simulations to form simulated catalogs extending below the 3FGL flux limit and compare the flux distributions with the observations.
  

In the present work we do not explicitly account for the source detection efficiency in 3FGL, but simply compare the predictions with the data mindful
of the range of the estimated detection threshold.
Since the detection efficiency and threshold depend on direction, mainly because of the Galactic diffuse emission, and also exposure variations, accounting for this would require considerably more study than possible in the present work, and is not required for the scope here.

A large number of models was generated and compared with the data; for this paper we  choose one reference model which is found to reproduce the data satisfactorily, but which  is not  unique. 
This suffices to illustrate  plausible properties of the Galactic source population.
A complete study of source number counts, optimizing the model over all parameters and considering spatial distributions and more sky regions, or using information about particular source classes, is beyond the present paper, and is foreseen in a future work. 

\subsection{Model and Comparison With Data}


Our reference model for the luminosity function has  $\rho_\odot= 100$  kpc$^{-3}$,
 and an $\Lgamma^{-1.8}$ dependence on luminosity 
in the range $2 \times 10^{34} - 2 \times 10^{39}$ \Lunits  above 1~GeV.   
The luminosity law is discussed in \citet{Strong2007_source_populations};  the chosen  power-law index 1.8 is larger than expected  for normal pulsars (1.5 or less) or MSPs, but here we wish to encompass both of these source types, and also other sources like SNR,  with a single power-law function, for simplicity. 
This index is required to fit  $N(\Sgamma)$  at both low and high latitudes; the steep slope ensures enough low-luminosity sources to match high-latitude number counts.

The distribution  in Galactocentric distance is based on the model of  \citet{Lorimer2006_pulsar_population} for the distribution of pulsars, taken as representative of Galactic sources.
We adopt an exponential scale height
of 500~pc, guided by that of pulsars; the source number count distribution $N(\Sgamma)$ depends only weakly on the scale height.
This distribution peaks near  $R=4$ kpc and falls to zero at $R=0$;
it was chosen for illustration and has not been optimized for the 3FGL source number counts.
The spectrum of sources was taken as a power-law with exponential cutoff, index 1.4 and cutoff energy 3.2~GeV, with correlated dispersion in these parameters as found for pulsars. 
This is representative of pulsars but is not critical for this work since we use integral photon fluxes  above 1~GeV as the basis for the analysis of  $N(\Sgamma)$.
The spectrum is  used here only when comparing with the spectrum of interstellar emission, to estimate the contribution of sources to the  diffuse emission (section 6.3).


\begin{figure*}[!ht]
\begin{center}
\includegraphics[width=8.0cm]{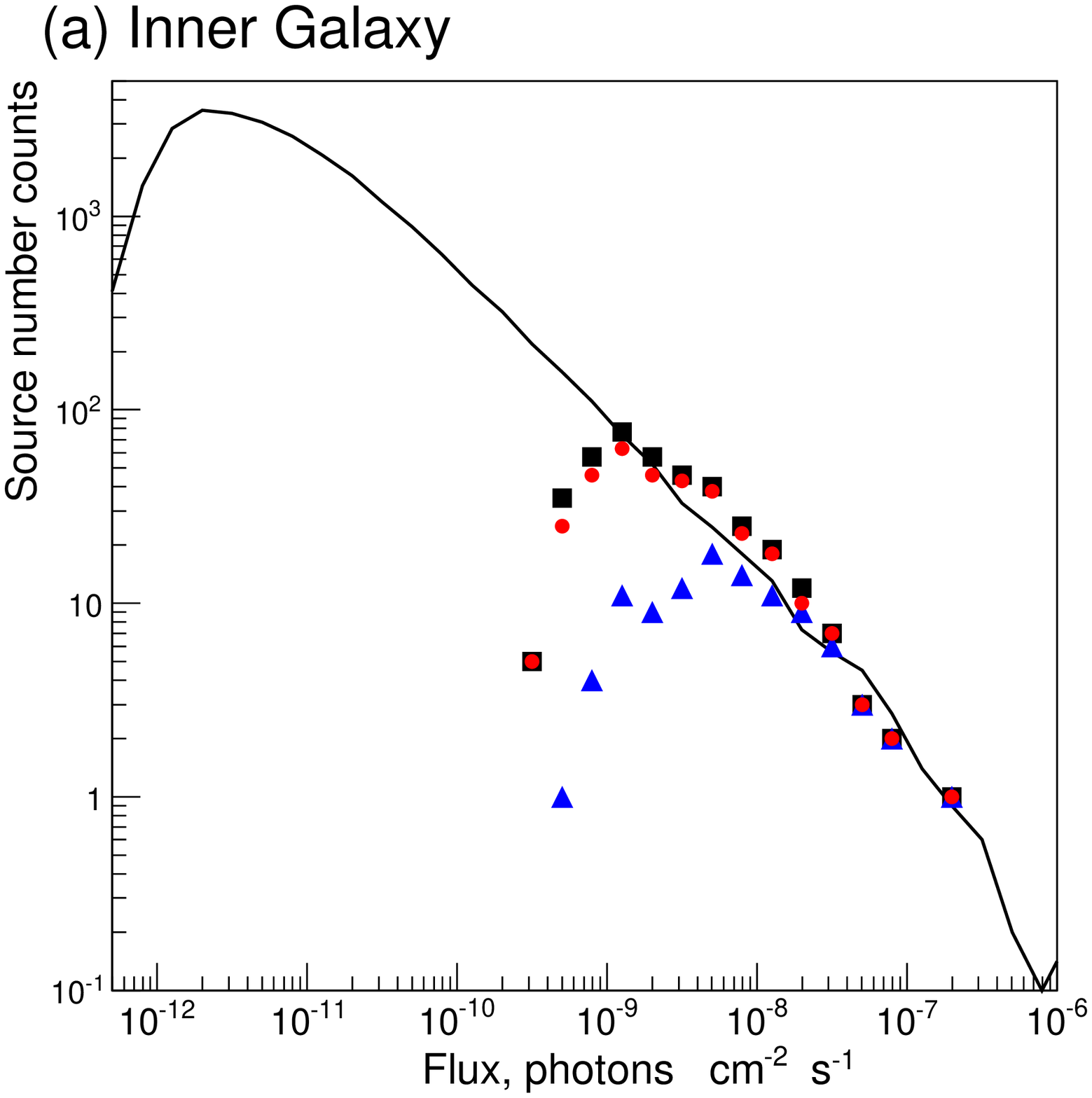}
\includegraphics[width=8.0cm]{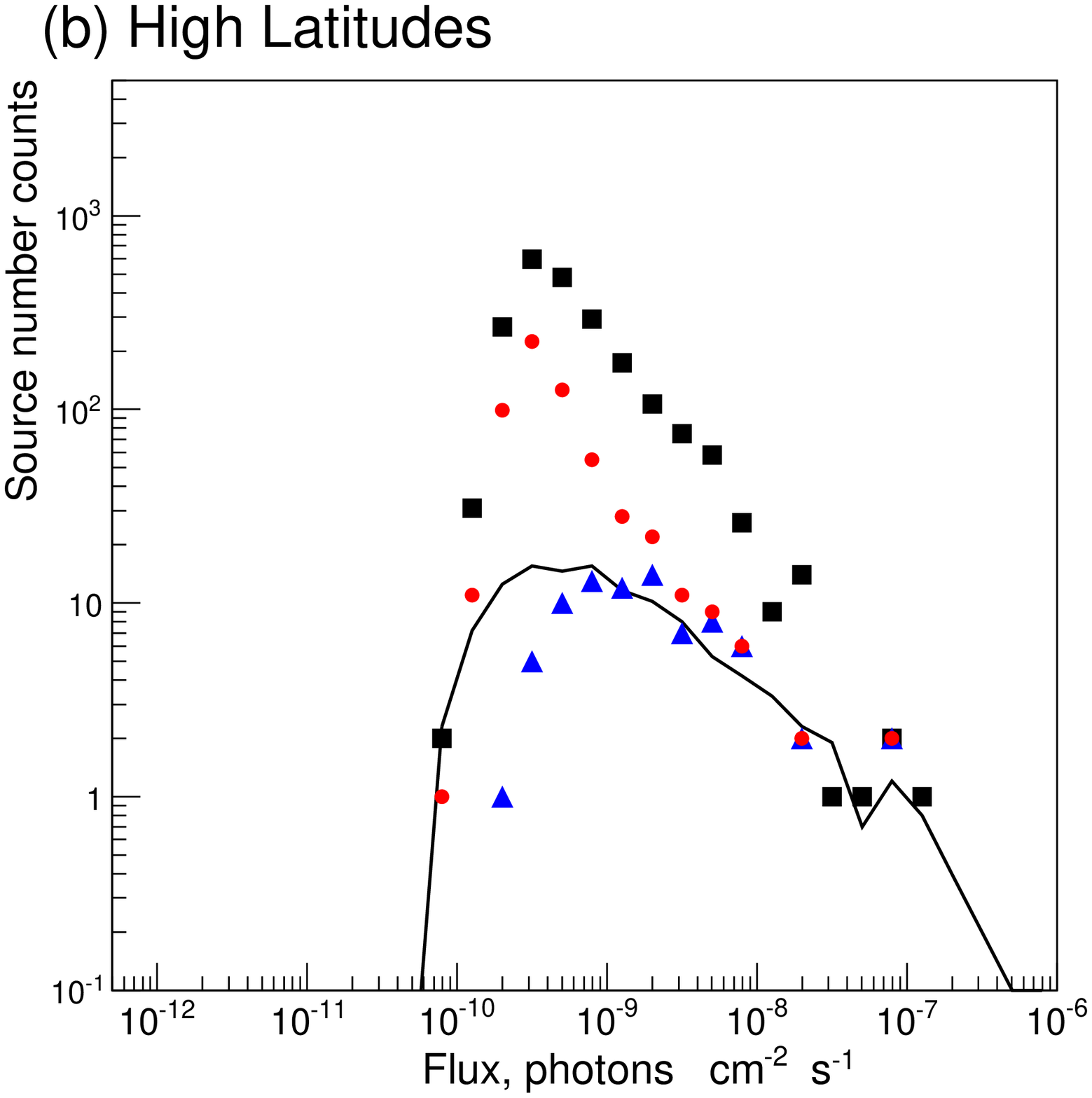}
\includegraphics[width=8.0cm]{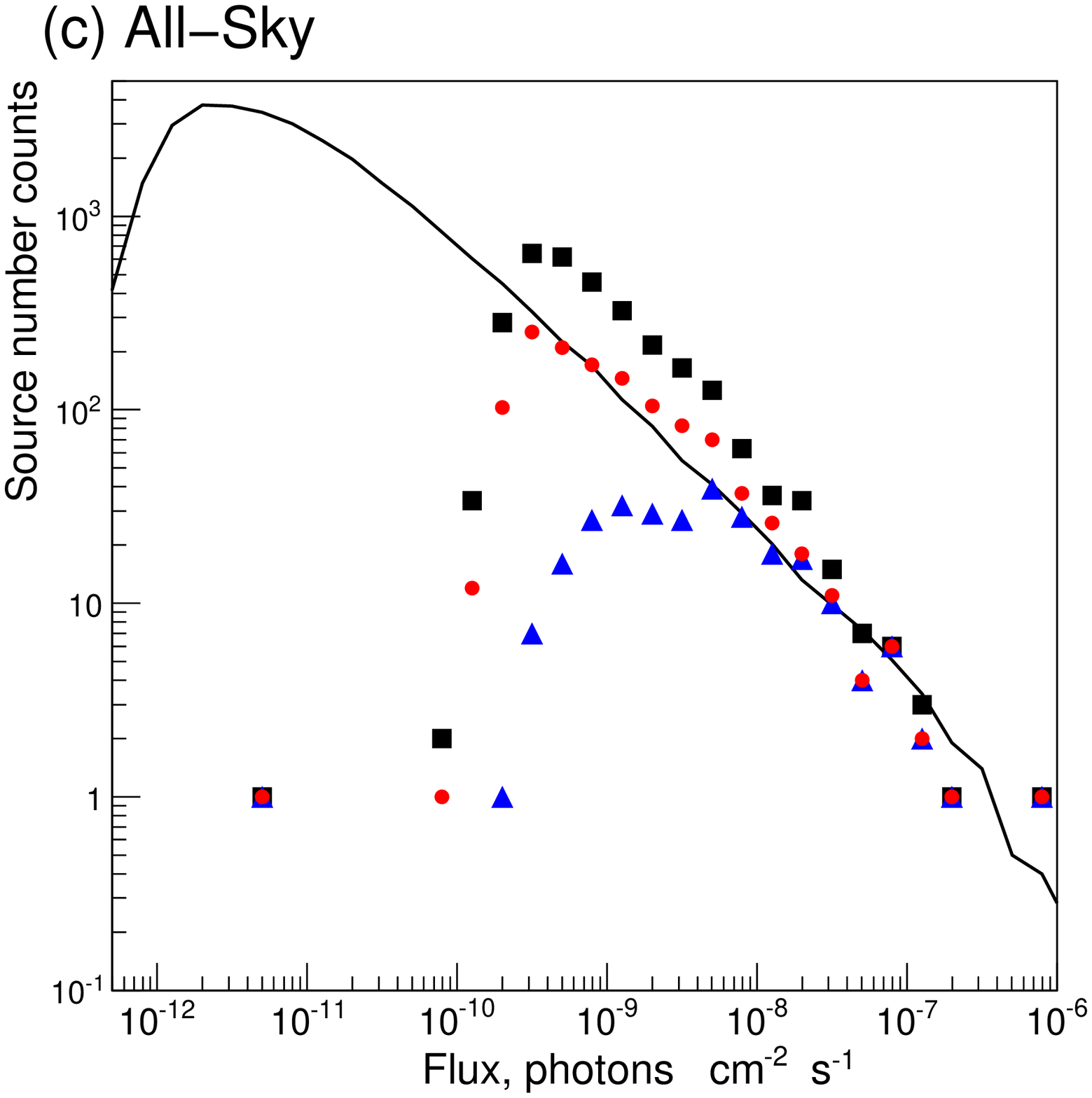}
\end{center}
\caption{\label{PopulationGal} 
Dependence of source number counts (number of sources per 0.2 dex) on source photon flux $S$ above 1 GeV.
The markers are source number counts from the 3FGL catalog; blue triangles are identified and associated Galactic sources, red circles are identified and associated Galactic, and unassociated sources, and black squares are all sources including extragalactic (for reference). 
The curves are from the reference model described in the  text.
(a) inner Galaxy  ($|b|<10\degr$, $300\degr<l<60\degr$); (b) high latitudes ($|b|>10\degr$, all longitudes); (c) all-sky.
}
\end{figure*}

Figure~\ref{PopulationGal}   compares the simulated $N(\Sgamma)$ with the observed flux distributions of 3FGL sources at  low latitudes in the inner Galaxy  ($|b|<10\degr$, $300\degr<l<60\degr$)
 and high latitudes  ($|b|>10\degr$), as well as for the full sky for reference.
The predictions agree reasonably with Galactic plus  unassociated sources  at low latitudes, and with Galactic associated sources at high latitudes (where unassociated sources are probably mainly AGN).
At high latitudes it is important that the model does not over-predict  Galactic plus  unassociated sources, and this condition is satisfied.
The reference model is consistent with the low-latitude source number counts, having the observed dependence on flux above the source detection threshold; the slope
reflects the spatial distribution (independent of the shape of the luminosity function)
above $10^{-8}$ ph cm$^{-2}$ s$^{-1}$, while the distributions for both the model and observed source number counts flatten at lower fluxes, following the luminosity function. 

Some of the  unassociated sources in the inner Galaxy may be AGN; using high-latitude AGN and unassociated sources (which are probably mainly AGN)  and scaling by solid angle, we estimate about 30 AGN above the threshold for the inner Galaxy.
This can be  compared to 254 Galactic identified plus associated sources in the inner Galaxy for the same threshold, so that the AGN contribution is negligible, especially considering that most of these should be identified/associated AGN and hence excluded from our sample.
For comparison there are 33  (identified/associated) AGN in this region, similar to our estimate from high latitudes,  so the selection of Galactic classes certainly avoids the presence of more than a few AGN among the unassociated sources.

Pulsars, including MSPs, have a range of luminosities $10^{32}-10^{37}$ \Lergunits \citep{LAT13_2PC}, corresponding to about   $3 \times 10^{34}-3 \times 10^{39}$ \Lunits
(taking a mean energy 2~GeV = $3.2 \times 10^{-3}$ erg). 
This is consistent with the range we have found from our  $N(\Sgamma)$ analysis at low and high latitudes, although we are not assuming anything about the physical nature of the sources.

In this model there are $2.9\times10^4$ sources in the Galaxy (in the luminosity range considered), with a total luminosity $>1$~GeV of $2 \times 10^{40}$ ph~s$^{-1}$, or about $6 \times 10^{37}$ erg\ ~s$^{-1}$. 
3FGL  contains about 266 Galactic sources (identified plus associated
in Table~\ref{tab:classes})
so that the LAT detects about  1\% of the sources in the Galaxy;
allowing for a significant number of unassociated  sources at low latitudes being Galactic, a larger number is certainly included in 3FGL.
Figure~\ref{PopulationGal} shows that the distribution of simulated sources (in the reference model) continues down to fluxes $\sim$100 times below the detection threshold, the cutoff being due to the finite spatial extent of the Galaxy.
The ratio of total flux below threshold to above threshold is about 0.25, which gives an estimate of the contribution of the undetected sources  to the `diffuse' emission (see below).

Using the work of \cite{Watters2011_pulsars} we can  estimate the number of $\gamma$-ray pulsars in the Galaxy; they give a pulsar birthrate of 1 per 59 years, which corresponds to $1.7\times10^4$  pulsars up to age 1 Myr. This is consistent with our model;
we  include other classes of sources, but this shows that our value is plausible.


\subsection{Contribution of Undetected Sources to Diffuse Galactic Emission}



Judging from the turnover in the inner Galaxy  $N(\Sgamma)$ data, the detection threshold there is  about $1  \times 10^{-9}$ ph cm$^{-2}$ s$^{-1}$, and we adopt this for the following estimates.
For the inner Galaxy, the total  flux from sources is $3  \times 10^{-6}$ ph cm$^{-2}$ s$^{-1}$, with   $2.4  \times 10^{-6}$ \Sunits above threshold,  $0.6  \times 10^{-6}$ \Sunits below threshold.
So 20\% of the total  source flux is below threshold, and the ratio of flux below/above threshold is 25\%.
We can use the measured diffuse spectrum directly, comparing with the contribution from sources relative to interstellar emission: for the inner Galaxy as defined here,  this gives 12\% from  sources above threshold, 3\% from sources below threshold, at 1~GeV.
These estimates are clearly model-dependent, in particular the adopted luminosity function gives a large number of low-luminosity sources, but they are certainly of the correct order,
since varying the models within the range consistent with the data  does not change the estimates greatly; details are beyond the scope of this paper.

For comparison with our estimates, \cite{Watters2011_pulsars} used physical modeling of young pulsars, and estimated their contribution to diffuse emission as 2.8\%, however using all-sky averages and for only 6 months of LAT data taking.
A study of the MSP contribution to the Galactic emission, for energies above 100~MeV, has been given by  \cite{Gregoire2013_MSPs}; they find the contribution is at the few percent level.

Population synthesis can be used to estimate the increase in the number of sources with improved detection limits; in this model, reducing the threshold by a factor 2 would yield about twice as many sources at low latitudes.

Finally we consider the global picture.
The luminosity of the interstellar emission from cosmic-ray interactions is about  $1 \times 10^{41}$ \Lunits  or   $4 \times 10^{38}$ \Lergunits for energies above 1~GeV
\citep{Strong2010_Galactic_luminosity}
so that sources have about  20\% of the interstellar  luminosity. 
This is also an estimate of the contribution of all Galactic  sources to the total  Galactic $\gamma$-ray intensity, averaged over the sky.

\section{Conclusions}
\label{conclusions}

The third {\it Fermi} LAT catalog is the deepest-yet in the 100~MeV--300~GeV energy range.  The increased sensitivity relative to the 2FGL catalog is due to both the longer time interval (4 years vs. 2 years for 2FGL) and the use of reprocessed Pass 7 data, which provides a narrower PSF above 3~GeV.  The 3FGL catalog also benefits from higher-level improvements in the analysis, including an improved model for Galactic diffuse emission, and a refined method for source detection.   

The 3FGL catalog includes 3033 sources.  The sources are detected ($TS > 25$) based on their average fluxes in the 4-year data set; 647 of the sources are found to be significantly variable on monthly timescales.  We flag 78 (2.6\%) of the sources as potentially being related to imperfections in the model for Galactic diffuse emission; the character \texttt{c} is appended to their names.  An additional 572 (18.9\%) are flagged in the catalog for less serious concerns, e.g., for the spectral model having a poor fit or for being close to a brighter source.  Of the 3033 sources in the catalog, 238 (7.8\%) are considered identified, based on correlated variability or (for 25 of the identified sources) correlated angular sizes with observations at other wavelengths.  Of the remainder, we find likely lower-energy counterparts for 1786 sources (59.6\%).  The remaining 992 sources (32.7\%) are unassociated.

The identified and associated sources in the 3FGL catalog include many Galactic and extragalactic source classes.  The largest Galactic source class continues to be pulsars, with {\b 143} known $\gamma$-ray pulsars and 24 candidates.  Other Galactic source classes have continued to grow; fifteen globular clusters are now associated with LAT sources.  Our analysis of Galactic source counts, informed by a model for the luminosity function, suggests that at 1 GeV $\sim$3\% of the Galactic diffuse emission is due to unresolved Galactic sources.  Blazars remain the largest class of extragalactic source, with more than 1100 identified or associated with BL Lac or FSRQ active galaxies.  Non-blazar classes of active galaxies are also found, including a Seyfert galaxy (Circinus galaxy), a compact steep spectrum radio source (3C 286) and several radio galaxies. The populations of active galaxies in 3FGL are considered in more detail in the companion 3LAC catalog.  

\acknowledgments

The {\it Fermi}-LAT Collaboration acknowledges generous ongoing support
from a number of agencies and institutes that have supported both the
development and the operation of the LAT as well as scientific data analysis.
These include the National Aeronautics and Space Administration and the
Department of Energy in the United States, the Commissariat \`a l'Energie Atomique
and the Centre National de la Recherche Scientifique / Institut National de Physique
Nucl\'eaire et de Physique des Particules in France, the Agenzia Spaziale Italiana
and the Istituto Nazionale di Fisica Nucleare in Italy, the Ministry of Education,
Culture, Sports, Science and Technology (MEXT), High Energy Accelerator Research
Organization (KEK) and Japan Aerospace Exploration Agency (JAXA) in Japan, and
the K.~A.~Wallenberg Foundation, the Swedish Research Council and the
Swedish National Space Board in Sweden.

Additional support for science analysis during the operations phase is gratefully
acknowledged from the Istituto Nazionale di Astrofisica in Italy and the Centre National d'\'Etudes Spatiales in France.

This work made extensive use of the ATNF pulsar  catalog\footnote{http://www.atnf.csiro.au/research/pulsar/psrcat}  \citep{ATNFcatalog}.  This research has made use of the NASA/IPAC Extragalactic Database (NED) which is operated by the Jet Propulsion Laboratory, California Institute of Technology, under contract with the National Aeronautics and Space Administration, and of archival data, software and online services provided by the ASI Science Data Center (ASDC) operated by the Italian Space Agency.

This research has made use of Aladin\footnote{http://aladin.u-strasbg.fr/}, TOPCAT\footnote{http://www.star.bristol.ac.uk/\~mbt/topcat/} and APLpy, an open-source plotting package for Python\footnote{http://aplpy.github.com}. The authors acknowledge the use of HEALPix\footnote{http://healpix.jpl.nasa.gov/} \citep{Gorski2005}.

{\it Facilities:} \facility{Fermi}.

\bibliography{3FGL_Catalog_arXiv3}

\appendix

\section{Diffuse model adjustments}
\label{appendix_diffuse_params}

\begin{figure*}
   \centering
   \begin{tabular}{cc}
                      
     \includegraphics[bb=10 36 489 344,clip,width=0.527\textwidth]
                     {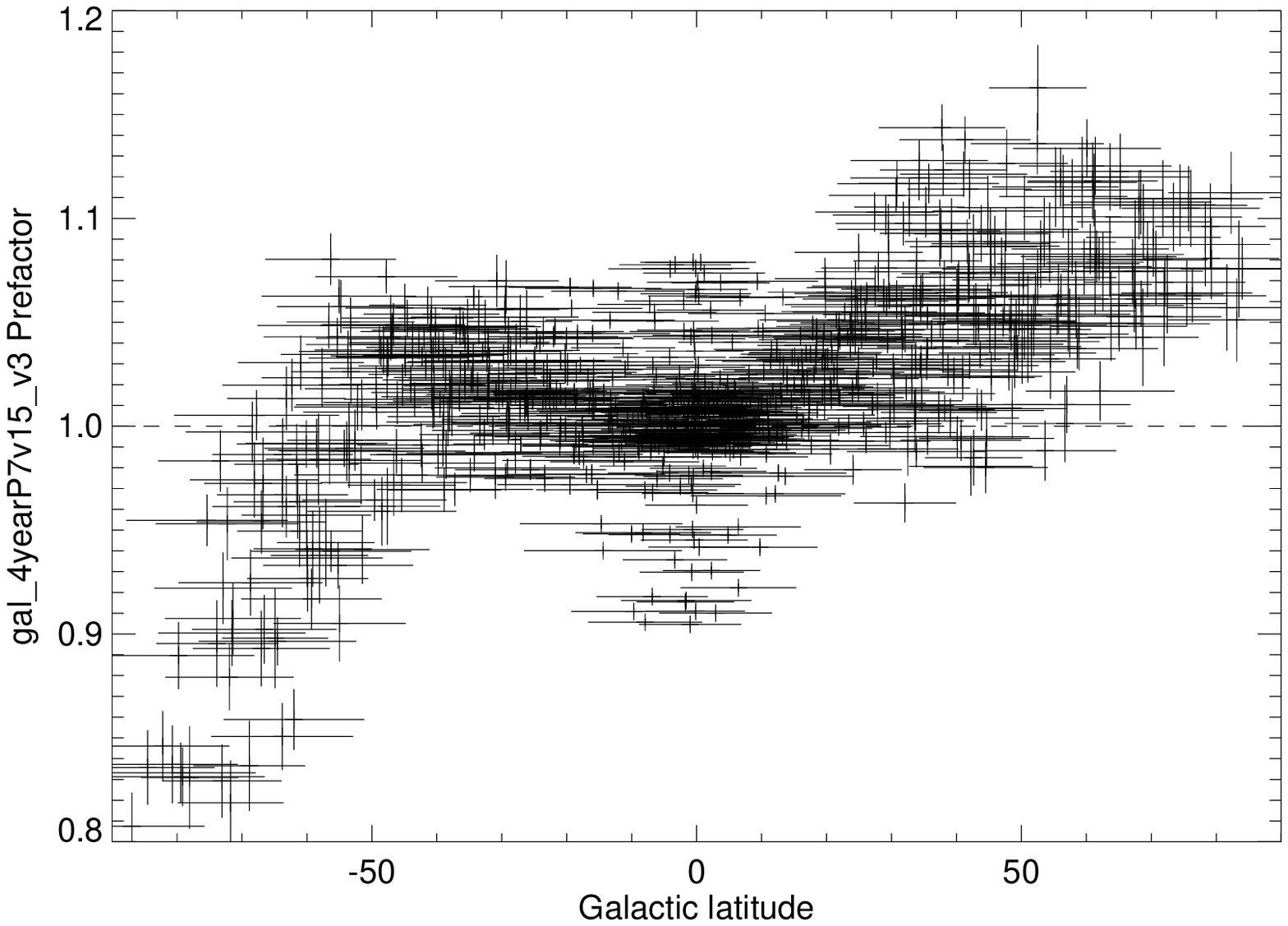} & \hspace{-0.5cm}
     \includegraphics[bb=60 36 489 344,clip,width=0.472\textwidth]
                     {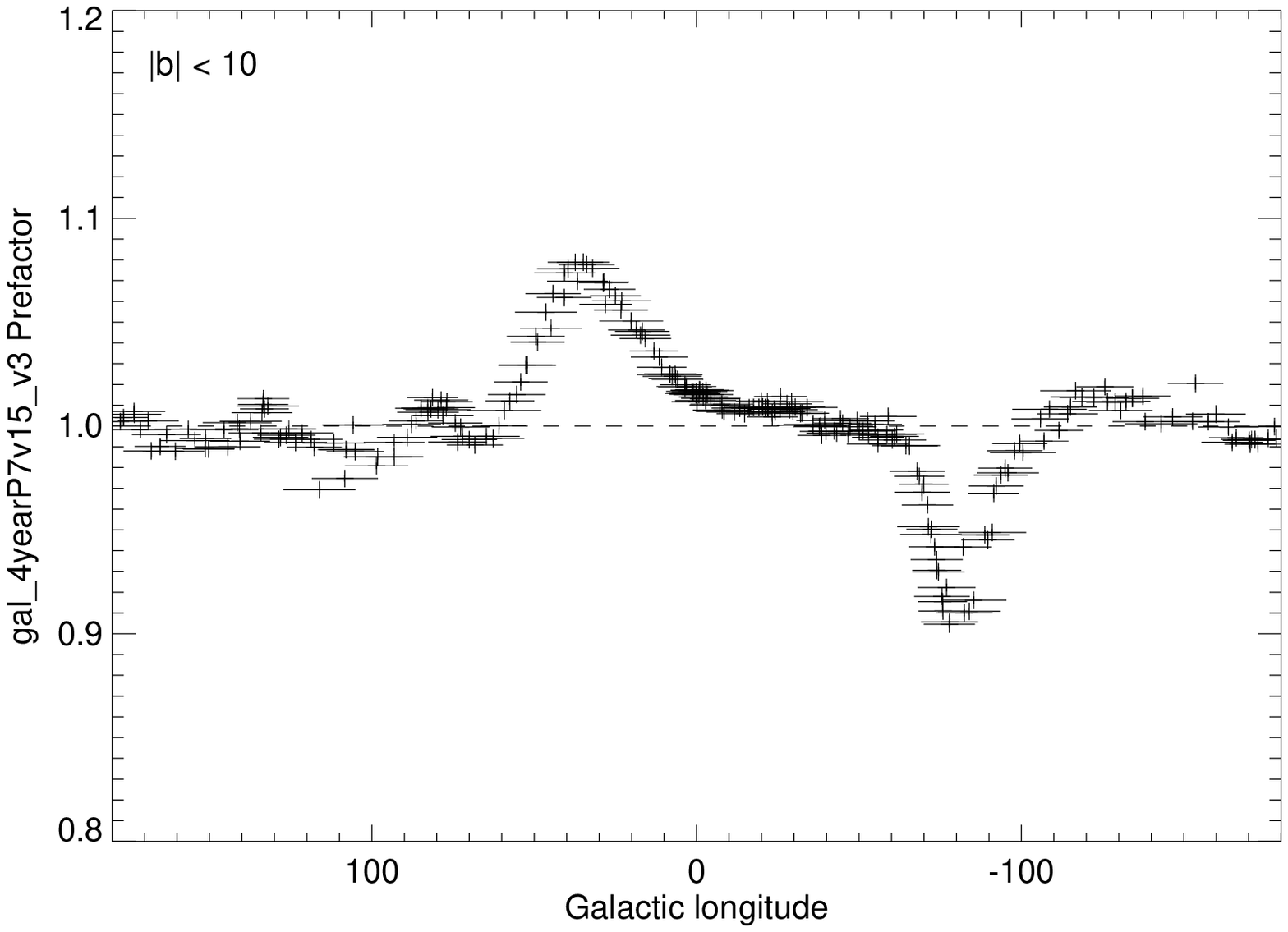} \\
     \includegraphics[bb=10 36 489 344,clip,width=0.527\textwidth]
                     {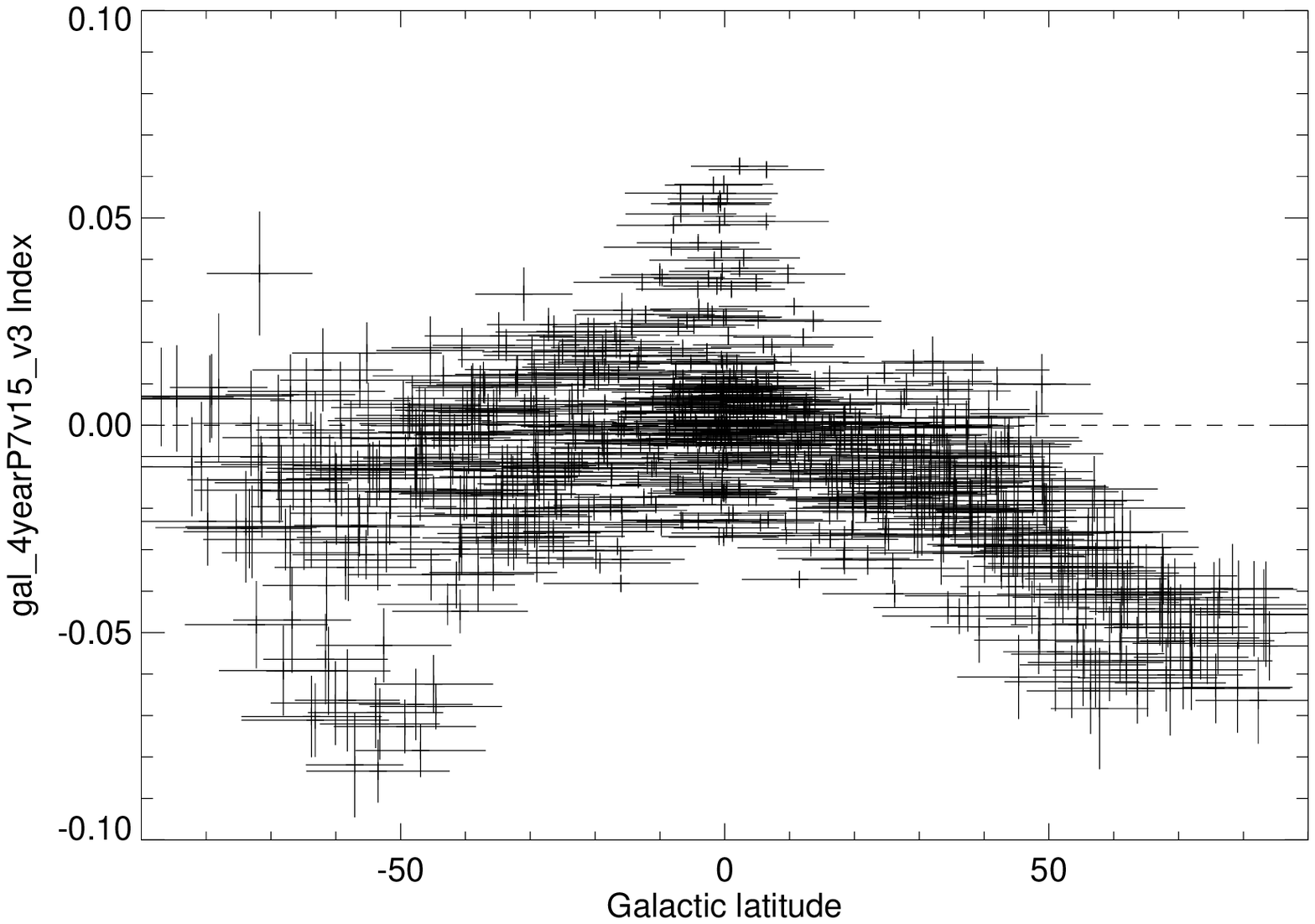} & \hspace{-0.5cm}
     \includegraphics[bb=60 36 489 344,clip,width=0.472\textwidth]
                     {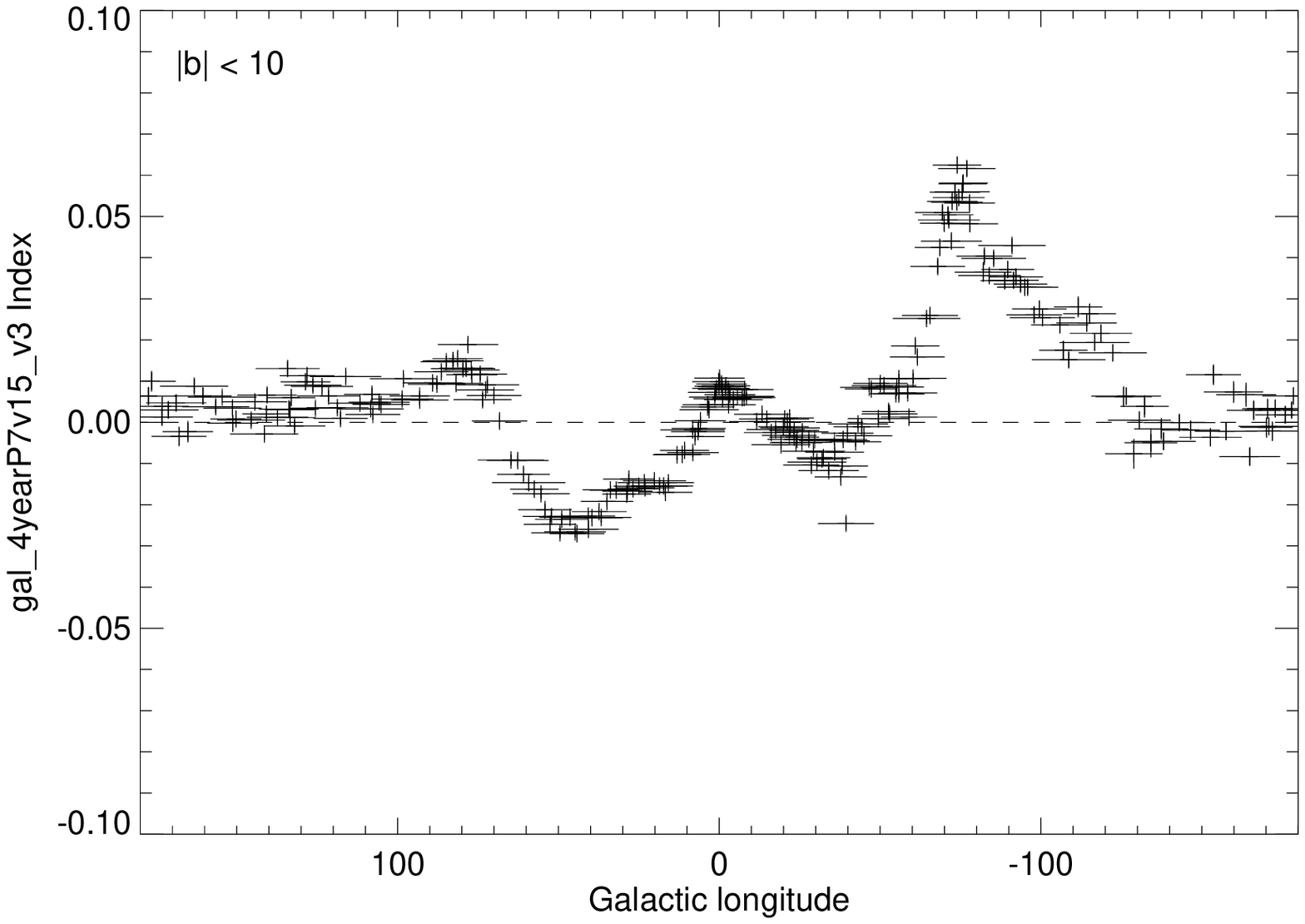} \\
      \includegraphics[bb=10 0 489 344,clip,width=0.527\textwidth]
                     {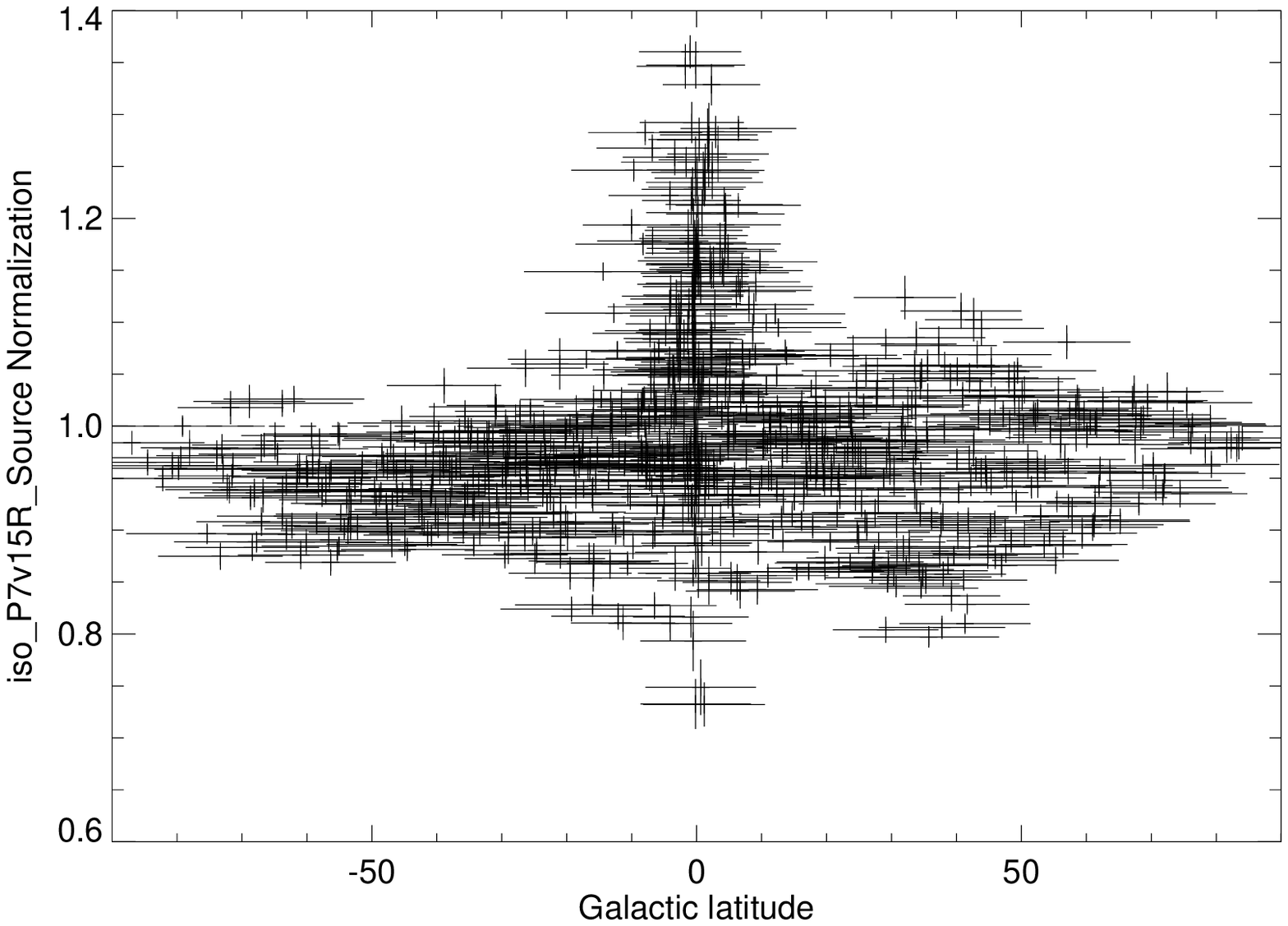} & \hspace{-0.5cm}
      \includegraphics[bb=60 0 489 344,clip,width=0.472\textwidth]
                     {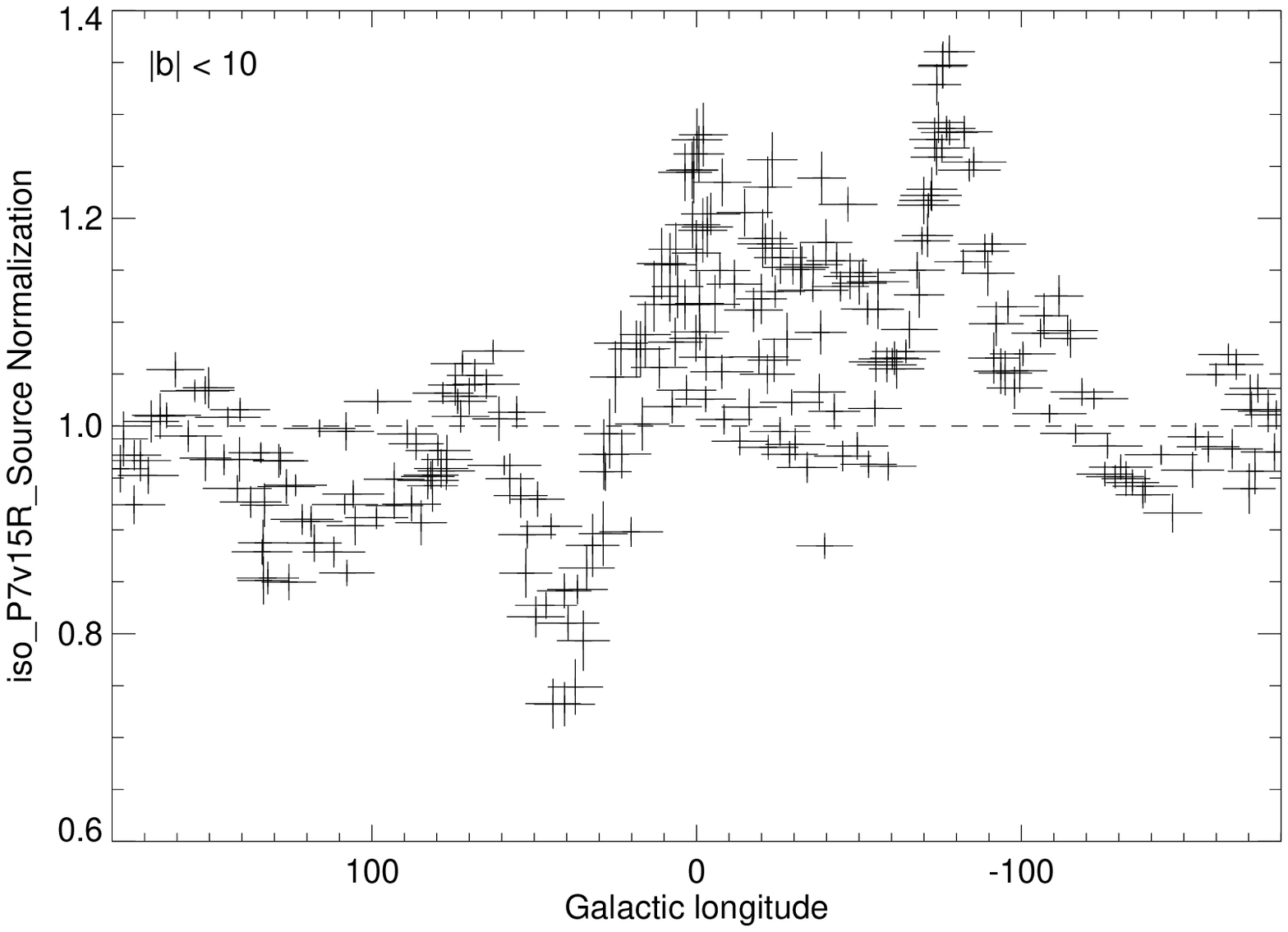}
   \end{tabular}
   \caption{Diffuse model parameters in each RoI. The horizontal error bar is the RoI radius. The vertical error bar is the statistical error from the fit. The vertical scale is the same in the left and right plots for a given parameter.
            Top: Normalization of the Galactic diffuse component (at 500~MeV).
            Center: Spectral index of the power-law correction to the Galactic diffuse component (positive means a harder model).
            Bottom: Normalization of the isotropic component.
            Left: All values as a function of Galactic latitude.
            Right: Galactic plane only as a function of Galactic longitude.
           }
   \label{fig:diffparams}
\end{figure*}

\begin{deluxetable}{lccl}
\setlength{\tabcolsep}{0.04in}
\tablewidth{0pt}
\tabletypesize{\scriptsize}
\tablecaption{LAT 3FGL FITS Format: ROIs Extension\label{tab:ROIs}}
\tablehead{
\colhead{Column} &
\colhead{Format} &
\colhead{Unit} &
\colhead{Description}
}
\startdata
ROI\_num & I & & ROI number (cross-reference to main table) \\
RAJ2000 & E & deg & Right Ascension of ROI center \\
DEJ2000 & E & deg & Declination of ROI center \\
GLON & E & deg & Galactic Longitude of ROI center \\
GLAT & E & deg & Galactic Latitude of ROI center \\
Radius & E & deg & ROI radius (unbinned mode) or half-side (binned mode) \\
PARNAM$i$\tablenotemark{a} & E & \nodata & Value of diffuse model parameter $i$ \\
Unc\_PARNAM$i$ & E & \nodata & $1\sigma$ error on PARNAM$i$ \\
\enddata
\tablenotetext{a} {Two columns (value and error) for each diffuse model parameter. The parameter name is given by the \texttt{PARNAM$i$} keyword in the extension header.}
\end{deluxetable}

In \S~\ref{catalog_significance} we noted that the diffuse emission model has three free parameters in each RoI. We report their values in the ROIs extension of the catalog (Table~\ref{tab:ROIs}) and we show in Figure~\ref{fig:diffparams} how they vary over the sky.
The first thing to notice is that the amplitude of the variations is relatively small. Overall the Galactic normalization does not vary by more than 20\%, and in the Galactic plane (i.e., where it is the dominant component) it does not vary by more than 10\%. The slope of the power-law correction does not exceed 0.1 (positive or negative) and inside the plane it does not exceed 0.05. The isotropic normalization does not vary by more than 40\%, and outside the Galactic plane (i.e., where it is the dominant component) it does not vary by more than 20\%. This indicates that the diffuse model is quite accurate.
Nevertheless, the statistical precision of the data is so good that the deviations are formally very significant. Leaving those parameters free allows releasing some of the tension that exists locally between the data and the diffuse model.

At high latitudes (left-hand plots) the isotropic component is stable (as it should) but the Galactic normalization shows a clear North-South effect. The model is too high in the South but lacks emission in the North. The middle plot also indicates that the model is somewhat too hard particularly in the North. The group of points where the model is too hard in the South is around (RA,Dec) = (+10\degr,$-$60\degr).

At low latitudes (right-hand plots) the error bars on the Galactic diffuse parameters are very small and the parameter values are very correlated between an RoI and its neighbor. This is because the distance between the centers of neighboring RoIs in the plane (a few degrees) is much smaller than their diameter (15 to 20\degr). The model appears to be too low and somewhat too hard East of the Galactic center (around longitude +35\degr) whereas it is too high and too soft just West of the Carina region (around longitude $-$80\degr). Outside those two regions the fitted Galactic model is very close to the original one. Inside the plane the isotropic component is a minor contributor and it tends to fluctuate a lot.

\begin{figure*}
   \centering

      \includegraphics[bb=0 365 598 475,clip,width=\textwidth]
                     {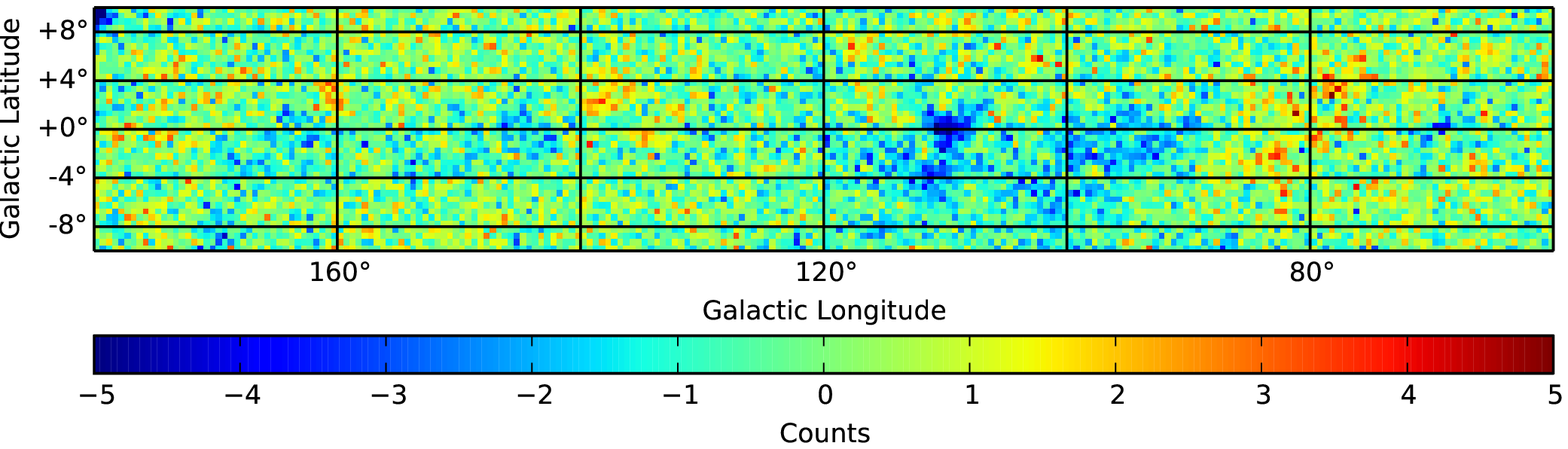}
      \includegraphics[bb=0 365 598 475,clip,width=\textwidth]
                     {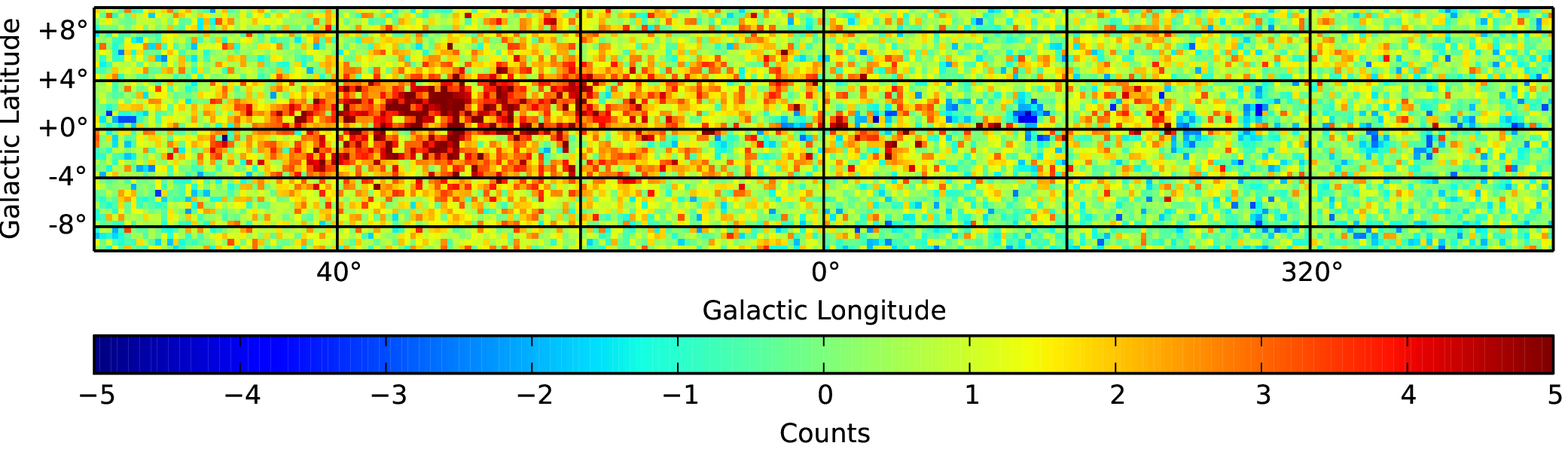}
      \includegraphics[bb=0 305 598 475,clip,width=\textwidth]
                    {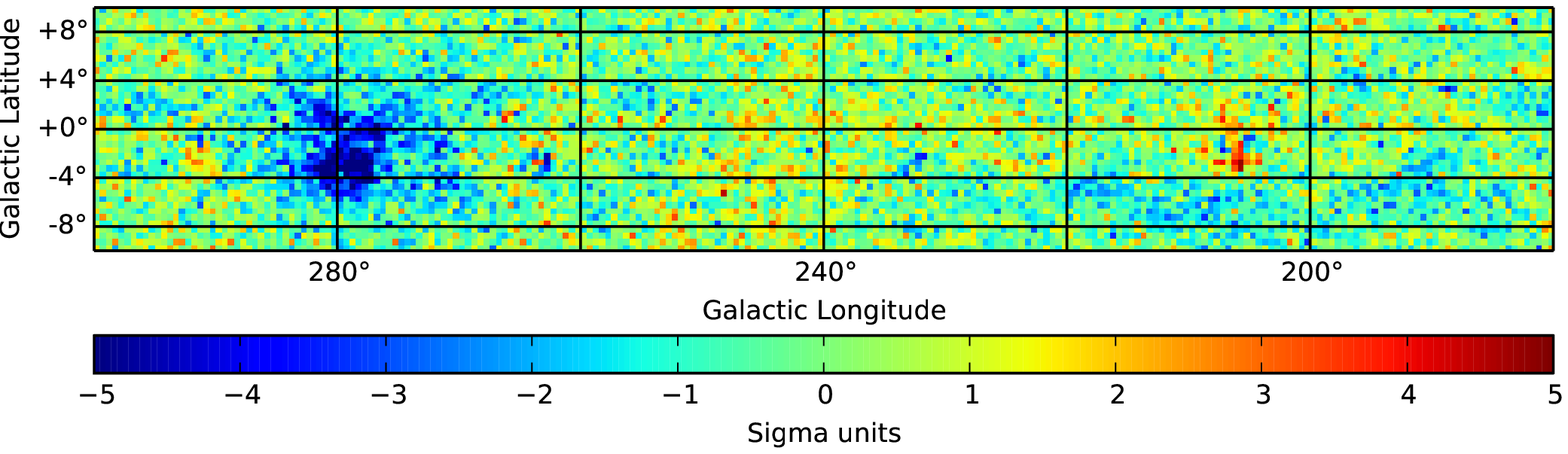}
   \caption{Residuals when setting the diffuse model normalizations to 1 and no power-law correction, integrated from 100~MeV to 100~GeV and expressed in sigma units over $0\fdg5$ pixels.
            Top: Positive Galactic longitudes from the anticenter to Cygnus.
            Center: Galactic ridge.
            Bottom: Negative Galactic longitudes from Carina to the anticenter.
           }
   \label{fig:diffresiduals}
\end{figure*}

Figure~\ref{fig:diffresiduals} is another illustration of the same effect. It shows the residuals between the data and the full sky model (original diffuse model + 3FGL sources, without any free parameter), integrated from 100~MeV to 100~GeV. The figure is restricted to the Galactic plane because at the scale shown here ($0\fdg5$ pixels) nothing comes out clearly at high latitude. The units are sigma units (statistical deviations). The model was originally computed with 0\fdg1 pixels in order to model the sources accurately. The main features are the data excess East of the Galactic center and the data deficit around $-$80\degr, which correspond to the features in Figure~\ref{fig:diffparams}.


\section{Description of the FITS Version of the 3FGL Catalog}
\label{appendix_fits_format}

The FITS format version of the 3FGL catalog\footnote{The file is available  from the $Fermi$ Science Support Center, http://fermi.gsfc.nasa.gov/ssc} has four binary table extensions.  The extension {\tt LAT\_Point\_Source\_Catalog Extension} has all of the information about the sources, including the monthly light curves (Table~\ref{tab:columns}).  

The extension {\tt Hist\_Start} lists the Mission Elapsed Time (seconds since 00:00 UTC on 2000 January 1) of the start of each bin of the monthly light curves.  The final entry is the ending time of the last bin.  

The extension {\tt GTI} is a standard Good-Time Interval listing the precise time intervals (start and stop in MET) included in the data analysis.  The number of intervals is fairly large because on most orbits ($\sim$95~min) $Fermi$ passes through the South Atlantic Anomaly (SAA), and science data taking is stopped during these times.  In addition, data taking is briefly interrupted on each non-SAA-crossing orbit, as $Fermi$ crosses the ascending node.  Filtering of time intervals with large rocking angles, other data gaps, or operation in non-standard configurations introduces some more entries.  The GTI is provided for reference and would be useful, e.g., for reconstructing the precise data set that was used for the 2FGL analysis.

The extension {\tt ExtendedSources} (format unchanged since 2FGL) contains information about the 25 spatially extended sources that are modeled in the 3FGL catalog, including locations and shapes.

\begin{deluxetable}{lccl}
\setlength{\tabcolsep}{0.04in}
\tablewidth{0pt}
\tabletypesize{\scriptsize}
\tablecaption{LAT 3FGL FITS Format: LAT\_Point\_Source\_Catalog Extension\label{tab:columns}}
\tablehead{
\colhead{Column} &
\colhead{Format} &
\colhead{Unit} &
\colhead{Description}
}
\startdata
Source\_Name & 18A & \nodata & Official source name 3FGL JHHMM.m+DDMM \\
RAJ2000 & E & deg & Right Ascension \\
DEJ2000 & E & deg & Declination \\
GLON & E & deg & Galactic Longitude \\
GLAT & E & deg & Galactic Latitude \\
Conf\_68\_SemiMajor & E & deg & Long radius of error ellipse at 68\% confidence \\
Conf\_68\_SemiMinor & E & deg & Short radius of error ellipse at 68\% confidence \\
Conf\_68\_PosAng & E & deg & Position angle of the 68\% long axis from celestial North, \\
& & & positive toward increasing RA (eastward) \\
Conf\_95\_SemiMajor & E & deg & Long radius of error ellipse at 95\% confidence \\
Conf\_95\_SemiMinor & E & deg & Short radius of error ellipse at 95\% confidence \\
Conf\_95\_PosAng & E & deg & Position angle of the 95\% long axis from celestial North, \\
& & & positive toward increasing RA (eastward) \\
ROI\_num & I & \nodata & ROI number (cross-reference to ROIs extension) \\
Signif\_Avg & E & \nodata & Source significance in $\sigma$ units (derived from Test Statistic) \\
& & & over the 100~MeV to 300~GeV band \\
Pivot\_Energy & E & MeV & Energy at which error on differential flux is minimal \\
Flux\_Density & E & cm$^{-2}$ MeV$^{-1}$ s$^{-1}$ & Differential flux at Pivot\_Energy \\
Unc\_Flux\_Density & E & cm$^{-2}$ MeV$^{-1}$ s$^{-1}$ & $1\sigma$  error on differential flux at Pivot\_Energy \\
Spectral\_Index & E & \nodata & Best fit photon number power-law index:  for LogParabola spectra, \\
& & & index at Pivot\_Energy; for PL(Super)ExpCutoff spectra, low-energy index\\
Unc\_Spectral\_Index & E & \nodata & $1\sigma$ error on Spectral\_Index \\
Flux1000 & E & cm$^{-2}$ s$^{-1}$ & Integral photon flux from 1 to 100~GeV \\
Unc\_Flux1000 & E & cm$^{-2}$ s$^{-1}$ & $1\sigma$ error on integral photon flux from 1 to 100~GeV \\
Energy\_Flux100 & E & erg cm$^{-2}$ s$^{-1}$ & Energy flux from 100~MeV to 100~GeV obtained by spectral fitting \\
Unc\_Energy\_Flux100 & E & erg cm$^{-2}$ s$^{-1}$ & $1\sigma$  error on energy flux from 100~MeV to 100~GeV \\
Signif\_Curve & E & \nodata & Significance (in $\sigma$ units) of the fit improvement between power-law \\
& & & and either LogParabola (for ordinary sources) or PLExpCutoff (for pulsars) \\
& & & A value greater than 4 indicates significant curvature \\
SpectrumType & 18A & \nodata & Spectral type (PowerLaw, LogParabola, PLExpCutoff, PLSuperExpCutoff) \\
beta & E & \nodata & Curvature parameter ($\beta$ of Eq.~\ref{eq:logparabola}) for LogParabola; NULL for other spectral types \\
Unc\_beta & E & \nodata & $1\sigma$ error on $\beta$ for LogParabola; NULL for other spectral types\\
Cutoff & E & MeV & Cutoff energy ($E_c$ of Eq.~\ref{eq:expcutoff}) for PL(Super)ExpCutoff; NULL for other spectral types \\
Unc\_Cutoff & E & MeV & $1\sigma$ error on cutoff energy for PL(Super)ExpCutoff; NULL for other spectral types \\
Exp\_Index & E & \nodata & Exponential index ($b$ of Eq.~\ref{eq:expcutoff}) for PLSuperExpCutoff; NULL for other spectral types \\
Unc\_Exp\_Index & E & \nodata & $1\sigma$ error on exponential index for PLSuperExpCutoff; NULL for other spectral types \\
PowerLaw\_Index & E & \nodata & Best fit power-law index; equal to Spectral\_Index if SpectrumType is PowerLaw \\
Flux30\_100 & E & cm$^{-2}$ s$^{-1}$ & Integral photon flux from 30 to 100~MeV (not filled) \\
Unc\_Flux30\_100 & 2E & cm$^{-2}$ s$^{-1}$ & $1\sigma$ lower and upper error on integral photon flux from 30 to 100~MeV (not filled) \\
nuFnu30\_100 & E & erg cm$^{-2}$ s$^{-1}$ & Spectral energy distribution between 30 and 100~MeV (not filled) \\
Sqrt\_TS30\_100 & E & \nodata & Square root of the Test Statistic between 30 and 100~MeV (not filled) \\
Flux100\_300 & E & cm$^{-2}$ s$^{-1}$ & Integral photon flux from 100 to 300~MeV \\
Unc\_Flux100\_300 & 2E & cm$^{-2}$ s$^{-1}$ & $1\sigma$ lower and upper error on integral photon flux from 100 to 300~MeV\tablenotemark{a} \\
nuFnu100\_300 & E & erg cm$^{-2}$ s$^{-1}$ & Spectral energy distribution between 100 and 300~MeV \\
Sqrt\_TS100\_300 & E & \nodata & Square root of the Test Statistic between 100 and 300~MeV \\
Flux300\_1000 & E & cm$^{-2}$ s$^{-1}$ & Integral photon flux from 300~MeV to 1~GeV \\
Unc\_Flux300\_1000 & 2E & cm$^{-2}$ s$^{-1}$ & $1\sigma$ lower and upper error on integral photon flux from 300~MeV to 1~GeV\tablenotemark{a} \\
nuFnu300\_1000 & E & erg cm$^{-2}$ s$^{-1}$ & Spectral energy distribution between 300~MeV and 1~GeV \\
Sqrt\_TS300\_1000 & E & \nodata & Square root of the Test Statistic between 300~MeV and 1~GeV \\
Flux1000\_3000 & E & cm$^{-2}$ s$^{-1}$ & Integral photon flux from 1 to 3~GeV \\
Unc\_Flux1000\_3000 & 2E & cm$^{-2}$ s$^{-1}$ & $1\sigma$ lower and upper error on integral photon flux from 1 to 3~GeV\tablenotemark{a} \\
nuFnu1000\_3000 & E & erg cm$^{-2}$ s$^{-1}$ & Spectral energy distribution between 1 and 3~GeV \\
Sqrt\_TS1000\_3000 & E & \nodata & Square root of the Test Statistic between 1 and 3~GeV \\
Flux3000\_10000 & E & cm$^{-2}$ s$^{-1}$ & Integral photon flux from 3 to 10~GeV \\
Unc\_Flux3000\_10000 & 2E & cm$^{-2}$ s$^{-1}$ & $1\sigma$ lower and upper error on integral photon flux from 3 to 10~GeV\tablenotemark{a} \\
nuFnu3000\_10000 & E & erg cm$^{-2}$ s$^{-1}$ & Spectral energy distribution between 3 and 10~GeV \\
Sqrt\_TS3000\_10000 & E & \nodata & Square root of the Test Statistic between 3 and 10~GeV \\
Flux10000\_100000 & E & cm$^{-2}$ s$^{-1}$ & Integral photon flux from 10 to 100~GeV \\
Unc\_Flux10000\_100000 & 2E & cm$^{-2}$ s$^{-1}$ & $1\sigma$ lower and upper error on integral photon flux from 10 to 100~GeV\tablenotemark{a} \\
nuFnu10000\_100000 & E & erg cm$^{-2}$ s$^{-1}$ & Spectral energy distribution between 10 and 100~GeV \\
Sqrt\_TS10000\_100000 & E & \nodata & Square root of the Test Statistic between 10 and 100~GeV \\
Variability\_Index & E & \nodata & Sum of 2$\times$log(Likelihood) difference between the flux fitted in each time \\
& & & interval and the average flux over the full catalog interval; a value greater \\
& & & than 72.44 over 48 intervals indicates $< $1\% chance of being a steady source \\
Signif\_Peak & E & \nodata & Source significance in peak interval in $\sigma$ units \\
Flux\_Peak & E & cm$^{-2}$ s$^{-1}$ & Peak integral photon flux from 100~MeV to 100~GeV \\
Unc\_Flux\_Peak & E & cm$^{-2}$ s$^{-1}$ &  $1\sigma$ error on peak integral photon flux \\
Time\_Peak & D & s (MET) & Time of center of interval in which peak flux was measured \\
Peak\_Interval & E & s & Length of interval in which peak flux was measured \\
Flux\_History & 48E & cm$^{-2}$ s$^{-1}$ & Integral photon flux from 100~MeV to 100~GeV in each interval (best fit from \\
& & &  likelihood analysis with spectral shape fixed to that obtained over full interval)\\
Unc\_Flux\_History & $2 \times 48$E & cm$^{-2}$ s$^{-1}$ &  $1\sigma$ lower and upper error on integral photon flux in each interval \\
& & &  added in quadrature with 2\% systematic component \\
Extended\_Source\_Name & 18A & \nodata & Cross-reference to the ExtendedSources extension for extended sources, if any \\
0FGL\_Name & 18A & \nodata & Name of corresponding 0FGL source, if any \\
1FGL\_Name & 18A & \nodata & Name of corresponding 1FGL source, if any \\
2FGL\_Name & 18A & \nodata & Name of corresponding 2FGL source, if any \\
1FHL\_Name & 18A & \nodata & Name of corresponding 1FHL source, if any \\
ASSOC\_GAM1 & 18A & \nodata & Name of likely corresponding 1AGL source \\
ASSOC\_GAM2 & 18A & \nodata & Name of likely corresponding 3EG source \\
ASSOC\_GAM3 & 18A & \nodata & Name of likely corresponding EGR source \\
TEVCAT\_FLAG & A & \nodata & P if positional association with non-extended source in TeVCat \\
& & & E if associated with a more extended source in TeVCat, N if no TeV association \\
ASSOC\_TEV & 24A & \nodata & Name of likely corresponding TeV source from TeVCat \\
CLASS1 & 5A & \nodata & Class designation for associated source; see Table~\ref{tab:classes} \\
ASSOC1 & 26A & \nodata & Name of identified or likely associated source \\
ASSOC2 & 26A & \nodata & Alternate name of identified or likely associated source \\
Flags & I & \nodata & Source flags (binary coding as in Table~\ref{tab:flags})\tablenotemark{b} \\
\enddata
\tablenotetext{a} {Separate $1\sigma$ errors are computed from the likelihood profile toward lower and larger fluxes. The lower error is set equal to NULL and the upper error is derived from a Bayesian upper limit if the $1\sigma$ interval contains 0 ($TS < 1$, see \S~\ref{catalog_flux_determination}).}
\tablenotetext{b} {Each condition is indicated by one bit among the 16 bits forming \texttt{Flags}. The bit is raised (set to 1) in the dubious case, so that sources without any warning sign have \texttt{Flags} = 0.}

\end{deluxetable}

\end{document}